\documentclass[12pt,preprint]{aastex}

\shorttitle{Circumnuclear Dust in Nearby Galaxies}
\shortauthors{Martini, Regan, Mulchaey, \& Pogge}

\slugcomment{ApJS Accepted [21 Oct 2002]}

\begin{document}

\newcommand{\um}{\mu{\rm m}}
\newcommand{\hst}{{\it HST\,\,}}
\newcommand{\eg}{{\rm e.g.\,}}
\newcommand{\kms}{{\rm \,km\,s^{-1}}}
\newcommand{\etal}{{\rm et al.\,}}


\title{Circumnuclear Dust in Nearby Active and Inactive Galaxies. 
I. Data\altaffilmark{1}}

\altaffiltext{1}{Based on observations with the
NASA/ESA {\it Hubble Space Telescope} obtained at the the Space Telescope
Science Institute, which is operated by the Association of Universities for
Research in Astronomy, Incorporated, under NASA contract NAS5-26555.}

\author{Paul Martini \altaffilmark{2}, 
Michael W. Regan \altaffilmark{3} 
John S. Mulchaey \altaffilmark{2}, 
Richard W. Pogge\altaffilmark{4}, 
}

\altaffiltext{2}{Carnegie Observatories, 813 Santa Barbara St.,
Pasadena, CA 91101-1292, martini@ociw.edu, mulchaey@ociw.edu}

\altaffiltext{3}{Space Telescope Science Institute, 3700 San Martin Drive, 
Baltimore, MD 21218, mregan@stsci.edu}

\altaffiltext{4}{Department of Astronomy, Ohio State University,
140 W. 18th Ave., Columbus, OH 43210, pogge@astronomy.ohio-state.edu}

\begin{abstract}

The detailed morphology of the interstellar medium (ISM) in the central 
kiloparsec of galaxies is controlled by pressure and gravitation. The 
combination of these forces shapes both circumnuclear star formation and 
the growth of the central, supermassive black hole. 
We present visible and near-infrared {\it Hubble Space Telescope} images 
and color maps of 123 nearby galaxies that show the distribution of the 
cold ISM, as traced by dust, with excellent spatial resolution. 
These observations reveal that nuclear dust spirals are found in the majority 
of active and inactive galaxies and they possess a wide range in coherence, 
symmetry, and pitch angle. 
We have used this large sample to develop a classification system 
for circumnuclear dust structures. 
In spite of the heterogeneous nature of the complete sample, we only find 
symmetric, two-arm nuclear dust spirals in galaxies with large scale bars 
and these dust lanes clearly connect to dust lanes along the leading 
edges of the large scale bars. Not all dust lanes along large scale bars 
form two arm spirals, however, and several instead end in nuclear rings. 
We find that tightly wound, or low pitch angle, nuclear dust spirals are 
more common in unbarred galaxies than barred galaxies. 
Finally, the extended narrow line region in several of the active galaxies 
is well-resolved. The connection between the ionized gas and circumnuclear 
dust lanes in four of these galaxies provides additional evidence that a 
significant fraction of their extended narrow line region is ambient gas 
photoionized {\it in situ} by the active nucleus. 
In a future paper, we will use our classification system for circumnuclear 
dust to identify differences between active and inactive galaxies, as well 
as barred and unbarred galaxies, in well-matched subsamples of these data. 

\end{abstract}


\section{Introduction}

The kinematics of the cold interstellar medium (ISM) largely determines the 
future star formation and black hole growth in galaxies. 
In the circumnuclear region of most galaxies, corresponding to approximately 
the central kiloparsec, both pressure and gravitational forces can dominate 
these kinematics. In the absence of velocity information, observations of dust 
lanes in absorption from visible and near-infrared (NIR) images can provide a 
snapshot of the cold ISM. 

Before the advent of \hst, high spatial resolution observations of dust in the 
circumnuclear region were only possible for relatively nearby galaxies, 
including our own \citep{morris96}, Andromeda \citep{mcelroy83,ciardullo88}, 
and several other relatively nearby galaxies where circumnuclear dust 
is apparent on photographic plates in, for example, the Carnegie Atlas of 
Galaxies \citep{sandage94}. 
Large single-band \hst programs have vastly expanded the number of galaxies 
with high spatial resolution observations \citep[resolutions less than 50 
parsecs;][]{malkan98,carollo98,ferruit00}. These observations have revealed a 
wealth of detail in the circumnuclear structure of spiral galaxies
and show that this nuclear structure is frequently quite different from 
the large scale morphology in any given galaxy. Multiband observations 
of smaller samples of active galaxies clearly show that much of this 
circumnuclear structure is due to nuclear dust \citep{elmegreen98,quillen99}, 
although mostly material that is only a factor of a few higher density than 
the ambient ISM \citep{regan99,martini99}. 

One of the most common dust features found in the circumnuclear region 
are nuclear dust spirals, which exhibit a wide range of intrinsic shapes. 
These include symmetric, two-arm spirals, one-arm spirals, and chaotic, 
multiarm spirals \citep{elmegreen98,laine99,regan99,martini99}. 
A number of theoretical models exist that can give rise to coherent dust 
structures in the circumnuclear region, although in all cases some differential 
rotation is required to form the spiral structure. Galaxies with strong, 
large scale bars form shock fronts along the leading edges of their bar 
\citep{prendergast83,athanassoula92} and hydrodynamic simulations show that 
this inflowing material can form symmetric and two-arm or ``grand design'' 
spiral structure in the circumnuclear region \citep{englmaier00,patsis00}. 
\citet{maciejewski02} recently modeled the circumnuclear region of barred 
galaxies and found grand design nuclear spirals similar to those found by 
\citep{englmaier00}. While the nuclear spirals in \citet{englmaier00} only 
corresponded to weak perturbations, the gas velocity field of the spirals 
in \citet{maciejewski02} had a large negative divergence, characteristic of 
strong shocks and similar to the shocks in the large-scale bar. Significant 
inflow is likely to be associated with these spirals shocks, as shown by 
\citet{fukuda98}. 

More chaotic, or ``flocculent'' spirals can form in the circumnuclear region 
from pressure forces, or acoustic turbulence, in the ISM \citep{elmegreen98,
montenegro99}. Detailed studies of the LINERs NGC\,4450 and NGC\,4736 by 
\citet{elmegreen02} show that the azimuthal Fourier transform spectra 
have a power law slope of -5/3, characteristic of turbulence. These structures 
therefore appear to be formed by the same turbulent processes that create 
structure in the ISM of galaxies at larger radii, but have been sheared by the 
presence of some differential rotation in the circumnuclear region 
and given a spiral appearance. Recent hydrodynamic simulations of the central 
ISM also naturally produce a filamentary, spiral structure in surface 
density with pressure and gravitational forces \citet{wada99,wada01}. 
Both these analytic and numerical studies suggest some inflow can occur 
due to turbulence alone. 

The nuclear dust spirals on these small scales 
(they are generally a kiloparsec or less in length) are distinct from their 
host galaxy spiral arms both spatially and in composition. 
Nuclear spirals rarely extend out of the region dominated by the 
central bulge. Observations of Seyfert galaxies over many kiloparsecs show 
that while the grand design nuclear spirals appear to connect to the 
dust lanes along the large scale bar, most types of nuclear spirals do not 
connect to the large-scale spiral arms \citep{pogge02}. 
These nuclear spirals also appear to only be enhancements in the dust surface 
density and not sufficiently overdense to be self-gravitating. 
Only a small number of nuclear dust spirals have associated star formation 
\citep[\eg NGC\,4321, Coma D15:][]{knapen95,caldwell99} and NIR 
observations do not show enhancements in the stellar surface density 
associated with the spirals \citep{martini01}. In addition, density waves 
tend to get stronger with increasing radius, whereas 
the nuclear spirals get weaker with increasing radius 
\citep{montenegro99,elmegreen02}.

One of the primary motivations for most of these multiband observations 
was to study the circumnuclear environment of active galaxies in order to 
determine the processes that removed angular momentum from host galaxy 
material and fueled the central, supermassive black hole. Recent 
\hst programs have provided strong evidence that all galaxies with a spheroid 
component have a supermassive black hole \citep{richstone98,ferrarese00,
gebhardt00,ferrarese01,gebhardt01}. 
This result has added new urgency to the question of how nuclear activity is 
fueled and why it is not currently fueled in all galaxies. 
Nuclear spiral structure was proposed as a mechanism for fueling nuclear 
activity because it was both discovered in nearly all of the observations of 
active galaxies and that these structures could be formed by shocks 
\citep{regan99,martini99}. 

Both bar-driven inflow and acoustic turbulence are theoretically viable 
mechanisms for the removal of angular momentum. In the case of inflow 
from large-scale bars, the gas and dust loses essentially all of its 
angular momentum in the shock front at the leading edge of the bar. For 
acoustic turbulence, the shocks in the ISM caused by the turbulence could 
dissipate sufficient energy and angular momentum to lead to significant 
inflow \citep{elmegreen02}. However, aside from the absence of kinematic 
data to demonstrate that inflow does occur, the selective investigation
of active galaxies resulted in relatively few comparable observations of 
inactive galaxies. Such observations are needed to determine if there is an 
excess of nuclear spiral structure, or other morphological features, in 
active galaxies over inactive galaxies of similar Hubble types and 
luminosities. We have obtained \hst observations of a significantly larger 
sample of both active and inactive galaxies to perform this analysis and 
present the data and a morphological study of each of the galaxies in our 
sample. In a future paper \citep{martini02} we compare the circumnuclear 
environment of active and inactive galaxies for a well-matched subsample 
of these data, as well as discuss some interesting results on differences in 
the circumnuclear environments of barred and unbarred galaxies.

\section{Sample Selection and Data Processing}

 \subsection{Sample Selection}

This galaxy sample was initially formed from the 91 Seyfert galaxies that meet 
the Revised Shapely-Ames (RSA) catalog magnitude requirement ($B_T < 13.4$ mag) 
discussed by \citet{maiolino95}, minus those with $v > 5000 \kms$ and 
axial ratios $R_{25} > 0.35$, to insure the circumnuclear region could be 
studied at high spatial resolution with minimal inclination effects. 
The resulting sample comprises 67 Seyfert galaxies, of which eight were GTO 
targets. Inactive galaxies with approximately the same velocity, inclination, 
Hubble type, and luminosity were selected for each of 67 these Seyferts.  
Additional Seyfert and non-Seyfert galaxies of particular interest were 
then added to increase the sample of nearby galaxies. 
From this input catalog, 104 galaxies were observed as part of a NICMOS 
snapshot program (SN 7330; PI Mulchaey). 
The galaxies observed with NICMOS, minus those that already had unsaturated, 
WFPC2 F606W images from \hst, then formed the input catalog of 88 galaxies 
for our WFPC2 snapshot program (SN 8597; PI Regan) and 76 of these galaxies 
have been observed to date. 
To these galaxies we have added the NICMOS observations of Seyfert 2s 
(GO 7867; PI Pogge) from the CfA Survey \citep{huchra92}, in order to place 
them on a common classification scheme, and the RSA Seyferts observed by 
NICMOS GTO programs, to create the final sample of 123 galaxies. 
Table~\ref{tbl:basic} lists the most common name(s) for each galaxy and 
provides the \hst program that obtained the visible and NIR 
images. The WFPC2 filter used for the visible-wavelength 
image is also listed; the F160W filter (hereafter $H$) was used for all 
of the NICMOS observations. 

Comparison of the circumnuclear dust properties of active and inactive 
galaxies requires reliable classifications. While the active 
galaxies are clearly active, the question of just what constitutes a 
inactive galaxy is much more dependent on the quality of previous 
observations. The inactive galaxies in this sample are generally bright and 
nearby, and therefore have been observed with sufficient sensitivity to 
detect at least a bright active nucleus, but genuine nuclear activity at 
very low levels usually can not be ruled out. For example, three of the 
35 inactive RSA galaxies observed for the original NICMOS snapshot program 
were later reclassified as Seyfert~2 galaxies by the Palomar Survey 
\citep{ho97b} and a comparable number were classified as LINERs. In 
Table~\ref{tbl:morph} we list the type of nuclear activity present in 
each galaxy from Table~\ref{tbl:basic}, as well as references to the
original work and most recent classification. 
For the inactive galaxies, we list references to any surveys that could 
have detected some level of nuclear activity if it were present. 

The question of whether or not a galaxy possesses a large-scale bar 
presents a problem comparable to the presence of nuclear activity. While 
all of the galaxies have been observed as part of the RC3 
\citep{devaucouleurs91} and have visible-wavelength observations, 
many galaxies that do not appear to have bars based on visible-wavelength data 
do have them when studied in the NIR \citep{mcleod95,mulchaey97b}. 
A large fraction of the galaxies in this sample were observed in the NIR, 
although there are nevertheless many apparently unbarred galaxies that do not 
have NIR observations.  
The classification of these unbarred galaxies are not as 
reliable as the unbarred galaxies that do have NIR observations. 
Nevertheless, aside from a few striking exceptions such as NGC\,1068 
\citep{scoville88,thronson89}, very few galaxies that show no evidence for 
a bar at visible wavelengths are strongly barred in the NIR 
\citep{mulchaey97b}. 
Table~\ref{tbl:morph} lists the bar classification for each galaxy, 
along with a flag to represent the quality of the bar classification. 

\begin{figure}
\epsscale{0.75}
\plotone{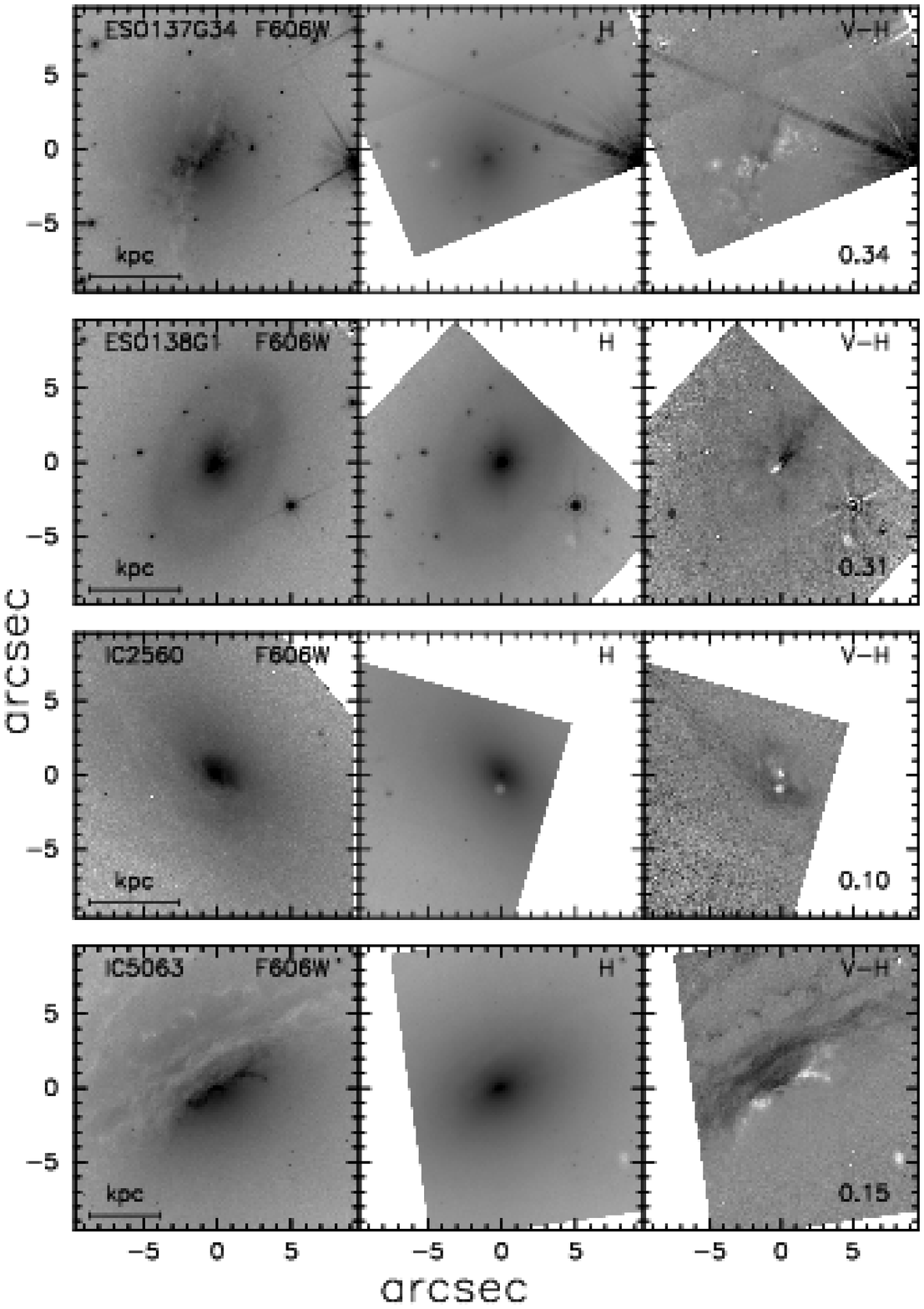}
\caption{$V$ and $H$ images and $V-H$ color maps of the galaxies, one galaxy 
per row. The $V$ image was obtained through the WFPC2 filter named in the upper 
right of the left panel. The bar in the lower left of the left panel shows 
the projected spatial scale at the distance of the galaxy (for $H_0 = 75 \kms$
Mpc$^{-1}$). The middle panel shows the NICMOS F160W image. Those galaxies 
with smaller images were observed with Nicmos Camera~1; most galaxies were 
observed with Nicmos Camera~2. The right panel shows the $V-H$ colormap, 
where dark on the greyscale corresponds to red and light to blue. All three 
panels are on a logarithmic greyscale, have a $20''$ field of view, and 
are oriented such that North is up and East is to the left. 
\label{fig:cmaps} }
\end{figure}

Table~\ref{tbl:morph} also lists the distance to each of these galaxies 
used to compute the projected spatial scale of the images shown in 
Figure~\ref{fig:cmaps}. As the Catalog of Nearby Galaxies 
\citep{tully88} was the single largest source for these distances, we have 
adopted a Hubble constant of $H_0 = 75 \kms$ Mpc$^{-1}$. For galaxies that did 
not have a previously published distance, we have used the relationship 
derived by \citet{yahil77} to estimate the distance. 

 \subsection{Initial Processing}

The majority of the visible-wavelength images were obtained as part of 
the snapshot program SN 8597. These consist of two exposures of length 
140s and 400s obtained through the F606W filter and with the galaxy centered 
on the PC chip. 
The image pairs were processed through the standard calibration pipeline and 
then combined with an XVista script as described in \citet{pogge02}. 
The remaining visible wavelength images were obtained from the \hst archive 
and constitute either image pairs similar to those obtained for SN 8597, or 
single exposures. These later observations required additional steps to remove 
the cosmic ray contamination and to process these images we have developed a 
new algorithm, which is described in the next section. 
 
The NIR images were all processed by the standard NICMOS pipeline 
developed at STScI. Many of the earliest images were obtained while this 
pipeline was still under development and were later recalibrated with 
new procedures and reference files as they became available. The basic 
details of the NICMOS reduction steps are described in \citet{regan99} 
for SN 7330 and \citet{martini99} for GO 7867. The galaxies not observed 
by either of these programs were obtained from the STScI archive and 
processed by the best reference and calibration files. None of the 
images were processed with CALNICB, instead the final shift and add 
step and any cosmic ray cleaning was performed in 
IRAF\footnote{IRAF is distributed by the National Optical Astronomy 
Observatories, which are operated by the Association of Universities for 
Research in Astronomy, Inc., under cooperative agreement with the National 
Science Foundation.}. 

 \subsection{Cosmic Ray Removal}

Techniques for the removal of cosmic rays fall into two general families. 
The first relies on multiple images of the same field, such as those obtained 
as part of our snapshot program. Here cosmic rays are identified as 
an anomalously large intensity increase in a given pixel and can be rejected 
when the images are combined. The second method is for the case where only a 
single image is available. In this latter case cosmic rays are detected 
by the deviation of a given pixel by a certain threshold over the 
variation in the neighboring pixels and the value of this pixel is then 
replaced with some value calculated from the neighboring pixels. 

For the galaxies from the \hst archive with only single observations, 
the majority of which were from snapshot program SN 5479 \citep{malkan98}, 
we developed a new method of cosmic ray detection and removal based on 
detailed knowledge of the \hst point spread function (PSF) and the fact that 
cosmic rays produce image artifacts which are usually sharper than the PSF. 
To identify cosmic rays, the original image is convolved with a PSF 
model generated with the TINYTIM software \citep{krist99} 
and then this convolved image is divided back into the original image 
to form $I/(I \otimes P)$, where $I$ is the original image, $P$ is the PSF, 
and $\otimes$ is the convolution operator. 
Pixels values greater than two in this divided image were empirically 
determined to nearly always correspond to bad pixels and these pixels were 
flagged and replaced by the average pixel value in the immediate vicinity in 
the original image. 
Cosmic rays were typically found to contaminate neighboring pixels 
within approximately one pixel of those identified by this technique. We 
therefore smoothed an initial mask of the flagged pixels with a 3x3 boxcar 
kernel to generate a final mask of all contaminated pixels. Each of these 
masked pixels was then replaced by the mean value of a 5x5 pixel box centered
on the pixel, minus any other masked pixels within this region. 

This technique is based on the ``structure map'' concept detailed in 
\citet{pogge02}, where a structure map is defined as: 
\begin{equation}
S = \left[\frac{I}{I \otimes P}\right]\otimes P^{t}
\end{equation}
where $S$ is the structure map and $P^{t}$ is the transpose of the model PSF, 
$P^{t}(x,y)=P(-x,-y)$. 
The cosmic ray detection image differs from the structure map as the original 
image is simply divided by a convolved version of itself, and not then 
convolved again by the transpose of the PSF. 
The image used for cosmic ray detection is similar to the second ``correction 
image'' formed in typical implementations of Richardson-Lucy (R-L) image 
restoration \citep{richardson72,lucy74}, although formally this latter image 
is $(I \otimes P^t)\otimes P$. 
In practice, image anomalies such as cosmic rays or saturated stars needed to 
be removed from the galaxies we studied in \citet{pogge02} as the structure 
maps overemphasized these features to the point of obliterating neighboring 
physical structures of interest. In several instances we also cleaned images
by hand or used the method of Laplacian edge 
detection described by \citet{vandokkum01}. 

 \subsection{Color maps}

The visible-wavelength images were calibrated with the standard calibration 
header keywords to approximately correspond to either ground-based $V$ 
or $I$ band, depending on the filter. 
All of the NIR observations were obtained through the NICMOS F160W filter, 
which is close to, but slightly broader than, a ground-based $H$ filter. 
The majority of these observations were obtained with NICMOS Camera 2, although 
22 observations were obtained with NICMOS Camera 1. 
These NIR observations were converted to $H$ magnitudes per 
square arcsecond using the standard NICMOS photometric calibration. 
The visible and NIR images, hereafter referred to as the $V$ and 
$H$ images, respectively, were then rebinned onto a common plate scale of 
$0.02''$ per pixel and differenced to form $V-H$ color maps.  
Figure~\ref{fig:cmaps} shows these $V$, $H$, and $V-H$ images for all 
123 galaxies. 

\begin{figure}
\epsscale{0.7}
\plotone{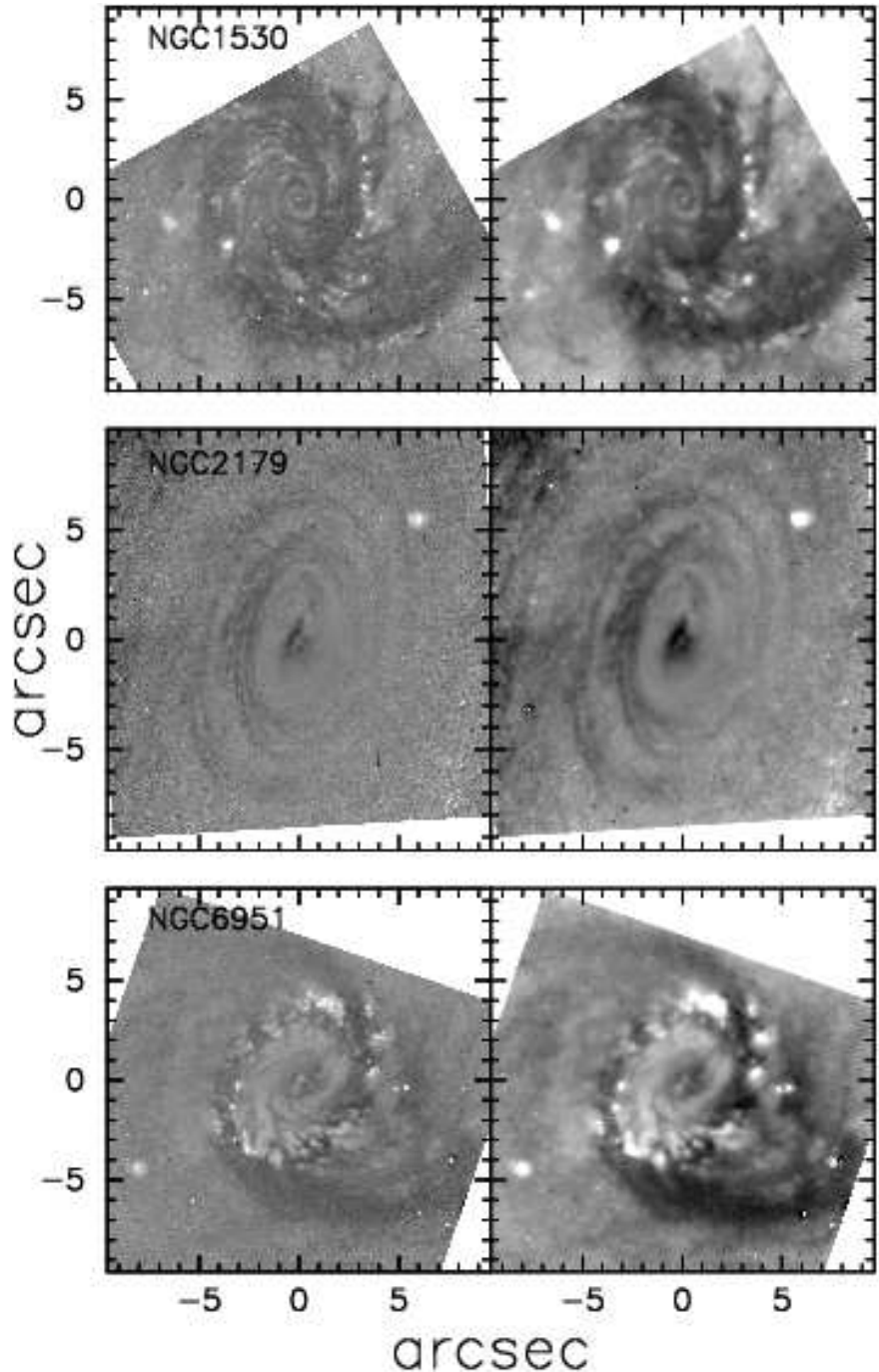}
\caption{Demonstration of the effects of image convolution for three galaxies 
with nuclear dust spirals: NGC\,1530 (GD), NGC\,2179 (TW), and NGC\,6951 (LW). 
The classification codes are described in \S\ref{sec:morph}. 
The left image shows the $V-H$ colormap from Figure~\ref{fig:cmaps}, the 
right image shows the same colormap constructed from a $V$ image 
convolved with the $H$ PSF and the $H$ image convolved with the $V$ PSF. 
Although the contrast of the dust lanes is poorer in the convolved images, the 
objects are still recognizeable as members of these three different 
nuclear spiral classes. This figure further shows that the dust features in 
the unconvolved images are not spurious features caused by the angular 
resolution mismatch of the two filters. 
\label{fig:convtest} }
\end{figure}

The angular resolution difference of $\sim 2.6$ between the diffraction limit 
of \hst at the effective wavelengths of the F606W and F160W filters results 
in very different spatial resolutions in the two filters. 
This effect is particularly striking for the active galaxies with 
bright nuclei, such as NGC\,1068 or NGC\,4151. The nuclei of these galaxies 
have striking red rings due to the prominence and larger radius of the airy 
diffraction rings in the NIR data. 
One way to mitigate the effects of these artifacts is to convolve 
the respective images by the PSF of the other in order to achieve a common 
angular resolution. However, the large mismatch in angular resolution between
these frames effectively means that the excellent angular resolution of 
the $V$ observations is significantly degraded. In fact, while the 
angular resolution mismatch is quite striking in the active nuclei, in 
general it does not effect the interpretation of dust morphology in the 
circumnuclear environment. This is because the $V$ data are approximately 
a factor of five more sensitive to dust extinction (for a standard Milky 
Way reddening law) than the $H$ data and thus nearly all of the information 
on the dust morphology is contained in the $V$ frames and the $H$ data 
mostly provides a measure of the true stellar light distribution in the 
absence of dust. To prove this point, and confirm that the angular resolution 
mismatch between $V$ and $H$ does not introduce spurious dust features which 
could impact on our analysis, we have created a series of color maps with 
the $V$ and $H$ data convolved to the same angular resolution and compared 
these convolved color maps with the unconvolved frames. A subset of these
comparison frames are shown in Figure~\ref{fig:convtest} for galaxies with 
particularly striking circumnuclear dust morphology.  
While these panels clearly show that the PSF-matching process smooths out 
dust features, it neither erases them completely nor leads to spurious 
features. 

\section{Classification}

The wide variety of different circumnuclear dust structures in 
Figure~\ref{fig:cmaps} illustrates striking similarities and differences 
to the larger scale morphology of galaxies. Most of these color maps show 
spiral lanes of dust and many of these nuclear spirals bear close 
resemblance to the grand design and multiarm spirals of varying pitch 
angle common to large scale spiral structure many kiloparsecs in extent. 
However, while such dust spirals appear to be quite common, many galaxies 
with well-ordered large scale structure have very chaotic circumnuclear 
environments, \eg NGC\,3351. 
In addition, while circumnuclear spiral structure of some form 
is present in most these galaxies, it is much less frequently as ordered and 
symmetric as large scale spiral structure. 
Perhaps most importantly, the dust structures in the circumnuclear region 
only rarely appears to have any associated star formation, 
whereas only a small fraction of all large scale spirals (\eg NGC\,7377 and 
IC\,5063) are purely defined by dust, rather than star formation. 
The NIR images also do not show evidence for any enhancement 
in the stellar surface density associated with the spirals, whereas 
large-scale, grand design \citep{elmegreen84} and flocculent \citep{thornley96} 
spirals show some enhancements in the NIR. 
Because of these physical and morphological differences, 
particularly in the spiral structure, we have decided to 
develop a separate classification system for circumnuclear dust, rather 
than attempt to continue \citet{malkan98}'s extension of the Hubble 
classification scheme to smaller scales. 

While the morphological differences between circumnuclear dust spirals and 
large scale, predominantly stellar spirals merit a different classification 
system, their similarities lead naturally to a parallel approach. 
\citet{sandage94}, in their introduction to the Carnegie Atlas of Galaxies, 
describe how the philosophy behind the Hubble system is an amalgam of the 
extreme empiricist and rationalist approaches to classification in its  
synthesis of observational data with physical insight and intuition. 
We endeavored to use a similar approach in the development of criteria for 
classification of nuclear dust structures. 
This is largely because the circumnuclear dust classification is based on the 
Hubble system, as well as the fact that some physical intuition (or prejudice) 
is unavoidable. 
Nevertheless, we have developed this classification system with reference 
to the various morphological phenomena in the circumnuclear dust and have 
avoided reference to nuclear activity and any host galaxy properties 
(with the exception of inclination). 
This approach was adopted to avoid bias imposed by the large scale structure 
that could obscure genuinely new characteristics of the circumnuclear region. 
In the following section, and in greater detail in a future paper, we 
discuss the relation between these different classes and nuclear and host 
galaxy properties. 
The presence of several connections between the circumnuclear classifications 
and both host and nuclear properties does suggest that real and distinct 
physical phenomena are connected to these different classes and we are 
not simply imposing order on chaos. 

 \subsection{Morphological Classes} \label{sec:morph}

In our previous studies of circumnuclear structure in active galaxies, 
we noted that all of them are rich in dust structures and in most cases this 
dust takes the form of spiral features with greater or lesser degrees of 
coherence. The samples studied by \citet{regan99} 
and \citet{martini99} showed that circumnuclear dust structures have 
varying degrees of symmetry, number of spiral arms, and coherence, but were 
too small to search for the morphological trends necessary to establish a 
classification scheme for circumnuclear dust structures. In their 
recent morphological study of active and HII galaxies, \citet{malkan98} did
provide new classifications for many galaxies based on much smaller-scale 
structure than those used for the original RSA or RC3 classification, however 
in most cases these classifications were still largely based on the 
large-scale structure, rather than the dust morphology in the 
circumnuclear environment at spatial scales less than a kiloparsec. 
The color maps presented here more clearly show the dust morphology at 
radii as small as several resolution elements from the nucleus in all but the 
brightest active galaxies. 
These observations clearly show trends in the circumnuclear morphology that 
are similar, but not identical, to the larger-scale morphology and the most 
obvious similarity is in the nuclear spiral dust structure. 
This structure ranges from symmetric, two-arm spirals that appear similar to
the beautiful grand-design spirals NGC\,1566 and NGC\,5194, 
to multiarm spirals similar to NGC\,2336, to spirals with only a few scattered 
arcs of dust. 
A significant fraction of the galaxies have no circumnuclear dust spirals
and instead simply have more amorphous dust structures or have no 
dust structure in their circumnuclear region at all. 
We have divided the galaxies with clear evidence for spiral structure into 
four different classes, created a fifth class for those galaxies with chaotic 
circumnuclear dust, and a sixth class for galaxies that have no evidence of 
nonuniform dust. Each of these classes are described below and an 
example of each class is shown in Figure~\ref{fig:class}. 

  \subsubsection{Grand Design Nuclear Spirals (GD)} 

This class is defined by the presence of a symmetric, two arm dust spiral 
such as shown in NGC\,5643 (shown in Figure~\ref{fig:class}) and 
NGC\,1300. 
These spirals appear similar to the grand design spirals found in 
the disks of many galaxies, yet these nuclear spirals are on scales less than 
a kiloparsec, rather than the many kiloparsecs of their larger namesakes. 
In addition, these arms are only traced by dust and do not exhibit
the enhanced star formation and stellar surface density characteristic
of galaxy-scale spiral density waves.
The lack of star formation and absence of a detectable NIR  
surface density enhancement suggests that grand design nuclear spirals either 
do not trace significant enhancements in the dust surface density or they 
have sufficient shear to inhibit star formation. 
Estimates of the dust surface density increase in the arms by \citet{martini99} 
supports the former scenario, although this does not constrain the amount of 
shear in the dust lanes. 

Many galaxies with grand design nuclear spirals have additional 
circumnuclear dust, often in the form of spiral arm fragments, in the interarm 
region between the grand design spiral arms (\eg NGC\,4303, NGC\,6814). 
We still classified galaxies as grand design nuclear spirals provided 
that the symmetric, two-arm structure was more prominent than the more 
irregular features in their interarm region. 
A common feature of many of these spirals (and the additional types we define 
below) is that the dust structures on one half of the circumnuclear disk are 
not as distinct as those in the other half, \eg NGC\,6890. 
As the division between the high and low contrast halves of the 
circumnuclear disk always coincides with the position angle of the host galaxy, 
we attribute this change in contrast to the inclination of the galaxy. 
In several cases, this effect is so strong that only a hint of the second arm 
is apparent, yet we still classify a galaxy as a grand design nuclear spiral 
if a 180 degree rotation of the first arm would clearly 
align with any structure on the other side of the disk, as is the case 
for NGC\,2639. 

  \subsubsection{Tightly Wound Nuclear Spiral (TW)}

Many galaxies with clear nuclear dust spirals lack either the clear symmetry 
of grand design spirals, two dust lanes, or both. 
We have separated these galaxies into two classes based upon the pitch angle 
of the nuclear spiral features. 
We classified spirals with small pitch angles as tightly wound (TW) and 
spirals with larger pitch angles as loosely wound (LW). The dividing line in 
pitch angle corresponds roughly to Hubble type Sab. 
Tightly wound nuclear spirals generally maintain their coherence for one or 
more complete rotations about the nucleus, \eg NGC\,2985 
(Figure~\ref{fig:class}) and NGC\,4030. 
As is the case for grand design and other nuclear spirals, these arms 
do not appear to trace circumnuclear star formation, nor are they apparent in 
NIR surface photometry. 

  \subsubsection{Loosely Wound Nuclear Spiral (LW)} 

All of the remaining galaxies with coherent nuclear spirals, but with larger 
pitch angle (comparable to the pitch angles exhibited by Hubble types Sb 
and later), are grouped into the class of loosely wound nuclear spirals. 
While these arms exhibit similar coherence as the tightly wound nuclear 
spirals, there are always several clear spiral arms in loosely wound 
nuclear spirals, \eg NGC\,6951 (Figure~\ref{fig:class}) and NGC\,4941, 
whereas the tightly wound nuclear spirals often have such sufficiently small 
pitch angles that it is difficult to discern the number of spiral arms present. 
Although the LW nuclear spirals are well separated, all of the spiral arms 
in a given galaxy appear to have comparable pitch angles and imply the same 
sense of rotation. Another common feature of LW spiral arms is that they are 
nearly always too loosely wound to make a complete rotation about the nucleus, 
whereas the tightly wound nuclear spirals typically make one or more 
complete rotations. This is not an obvious physical difference between the 
two types, however, as often signal-to-noise (\eg NGC\,3486), the NICMOS 
field of view (\eg NGC\,7392), or larger scale structures in the host 
galaxy (\eg NGC\,6951) limit the apparent radial extent of the LW spirals. 

  \subsubsection{Chaotic Nuclear Spiral (CS)}

Many galaxies exhibit evidence for one or more spiral arm fragments or dust 
arcs that suggest spiral structure, yet they lack the coherent structure 
over a large range in radius exhibited by the three classes described above. 
We classified these galaxies as chaotic nuclear spirals because they still 
show evidence for some spiral structure, though more reminiscent of the 
spiral features in irregular galaxies than in any of the traditional spiral 
galaxy classes. As in the case of the other types of nuclear dust spirals, 
the dust lanes also do not have star formation, which is not the case for 
the spiral arm fragments seen in Irregulars. 
The chaotic spirals include dust spirals that are broken into many fragments 
(\eg NGC\,6221) or do not fill the entire circumnuclear disk (\eg NGC\,2460; 
Figure~\ref{fig:class}). 
In order to be included in this class, however, they do need to possess at 
least two spiral arms that suggest the same sense of rotation. For example, 
NGC\,3368 has at least two spiral arm fragments with the same sense of 
rotation, whereas NGC\,4260, which we do not classify as a chaotic nuclear 
spiral, has at best one arm fragment. 
Instead, we classify NGC\,4260 as a member of the chaotic class defined in 
the next subsection.  

  \subsubsection{Chaotic Circumnuclear Dust (C)}

These galaxies have clearly discernible, and often considerable, 
circumnuclear dust, yet without unambiguous 
evidence for spiral structure. While there are a few galaxies that
suggest spiral features, such as NGC\,6946 (Figure~\ref{fig:class}) 
or NGC\,2336, in both of 
these cases there are features of nearly equal size and contrast that 
could be due to spiral arms in a disk with the opposite sense of rotation. 
This class is in effect a catch-all for any galaxies that show no evidence for 
spiral structure, yet a larger sample could merit the division of this 
class into several distinct groups. 
For example, a number of these galaxies have straight dust lanes that 
completely cross the nucleus, such as IC\,5267 and NGC\,3627. 
Other galaxies are at relatively high inclination and have 
several straight dust lanes crossing the nucleus on one half, \eg NGC\,4939, 
NGC\,5033 and NGC\,5970. If these galaxies were not so highly inclined 
relative to the line of sight, they might be classified as nuclear spirals. 
Several other galaxies with high inclination simply have a large cloud of 
dust on what is presumably the near side of the disk, \eg NGC\,4258 and 
NGC\,6300, rather than many lanes of dust. 

  \subsubsection{No Structure (N)} 

The final class corresponds to galaxies with no dust structure in the 
circumnuclear region. 
Galaxies in this class, \eg NGC\,1398 (Figure~\ref{fig:class}) and NGC\,357, 
do not show any dust structures within at least the 
central few hundred parsecs. 
For most of these galaxies there is also no evidence for nonuniform dust at 
larger radii. 
However, the case of NGC\,628 illustrates the complication caused by the 
large distance distribution of the full sample. 
In this galaxy the central region is essentially featureless, 
yet outside of approximately 200 parsecs radius there is significant spiral 
dust structure. 
Nevertheless we have classified this galaxy as having no structure as the 
well-resolved central region is apparently free of dust lanes. 

 \subsection{Notes on the Classification Process}

The nuclear classifications for each galaxy are provided in the last 
column of Table~\ref{tbl:nc}. Each galaxy was classified independently by 
two of us (PM and MR) after we agreed on the classification scheme described 
above and a set of prototypes for each class.
The galaxies given above as examples of each class include the prototypes 
used for the classification. 
There were disagreements for nearly half of the galaxies. This high rate of 
disagreement is some cause for concern given the subjective nature of this 
process, yet upon closer scrutiny nearly all of the sources of conflict were 
between similar classes, such as between the chaotic spirals and chaotic 
or between chaotic spirals and loosely wound spirals. 
A blind reclassification of all of the sources of conflict led 
to resolution of most of the cases, which we attribute to our increased 
familiarity with the classification scheme. 
The remaining conflicts were resolved in binding arbitration (provided by JM). 

Galaxies with $R_{25} > 0.35$ ($\sim 70$ degrees) were considered too 
highly inclined for a reliable classification. 
They have been labeled as high inclination (HI) systems in Table~\ref{tbl:nc}. 
There is a great deal of dust in these high inclination systems and 
occasionally nuclear spiral structure is visible, \eg NGC~1320. 
For these galaxies we have provided an additional classification after the HI 
in the table, although most just show chaotic dust structures. For 
inclinations greater than $R_{25} = 0.30$ the fraction of galaxies classified 
as chaotic starts to increase noticeably and therefore only galaxies with 
$R_{25} < 0.30$ are included in the detailed analysis described in Paper~II. 

Many of these galaxies have high contrast dust features present on 
half of the color map and little or no structure on the opposite half. 
This appears to be due to the inclination of the host galaxy and can be 
used, as noted by \citet{hubble43}, to break the degeneracy between the 
near and far sides of a galaxy. 
We assumed that lower contrast or the absence of dust features on one half 
of a circumnuclear disk was due to inclination alone and classified 
these cases as if similar dust structures were visible on both sides of the 
disk. For several galaxies, such as NGC\,3393 or NGC\,3458, the 
center of the galaxy fell near the edge of the NICMOS image and therefore 
only half of the galaxy is present in the color map. In these instances, and 
in others where the signal-to-noise of the NIR image was 
significantly worse than the visible image, we also used the WFPC2 ($V$) 
image to aid in the classification. 
In the Appendix we provide more detail on the classifications for individual 
galaxies. 

\section{Circumnuclear Dust Structure}

\subsection{Nuclear Activity}

\begin{figure}
\epsscale{0.85}
\plotone{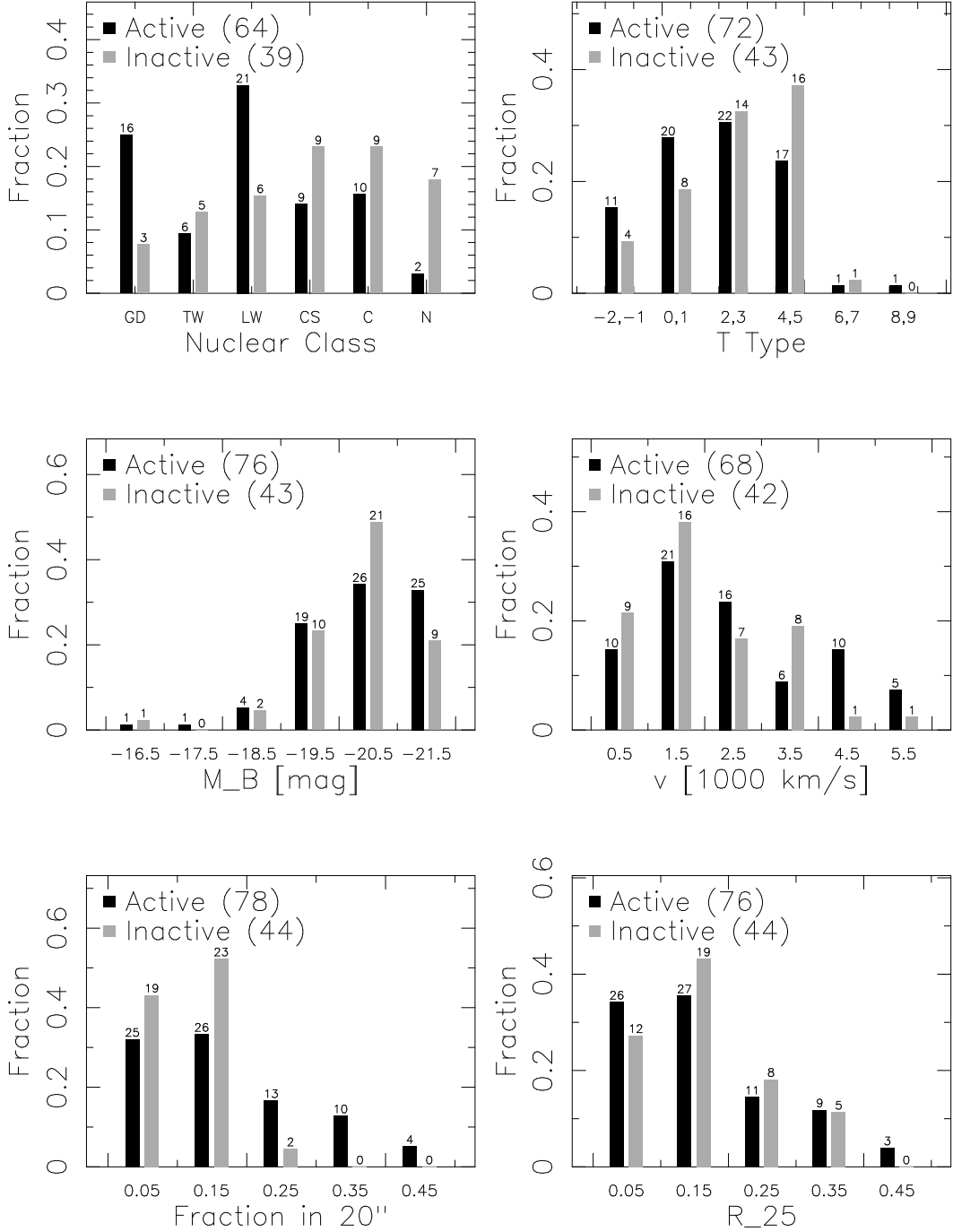}
\caption{
Distribution of the nuclear and host galaxy properties of the full sample 
of active and inactive galaxies observed with \hst.
The upper left histogram shows the distribution of these galaxies into
the six nuclear classes defined in \S\ref{sec:morph}. 
The histogram bars are normalized to the total number of galaxies of each 
type and the number of galaxies in each class is given above the histogram 
bar. 
The remaining five histograms show the distribution of the Hubble type, 
$B$ luminosity, heliocentric velocity, size, and inclination of the active 
and inactive galaxies.  The total number of galaxies in each panel does not
equal 123 as galaxies with $R_{35} > 0.30$ were not include in the first 
panel and several individual galaxies fall outside the range of each of 
the bins shown.
\label{fig:agnhist} }
\end{figure}

In Figure~\ref{fig:agnhist} we plot histograms of the fraction of 
active and inactive galaxies in each of the classes described above, as well 
as histograms of their distribution in heliocentric velocity ($v$), 
$B$ luminosities ($M_B$), Hubble types (T), size (fraction of $D_{25}$ within 
the 20$''$ field of view), and inclination ($R_{25}$). 
Each set of histograms is normalized such that a given bin reflects the 
fraction of the total active or inactive sample within that bin. 
The number on top of each histogram then gives the actual number of galaxies 
in that bin. 
The total number of galaxies in each panel does not add up to the 123 galaxies 
in the entire sample as there are typically several galaxies which fall 
outside the range of each histogram. 

The first histogram in Figure~\ref{fig:agnhist} shows that nuclear spiral 
structure is a very common feature in active galaxies as 52 of the 
64 active galaxies (with $R_{25} < 0.3$) in this sample, or 80\%, fall into 
one of the four classes of nuclear spiral structure defined above. 
Nuclear spirals of some form are also found in the majority of the inactive 
galaxies we observed: 23 of the 39 inactive galaxies, or approximately 60\% 
of the full sample. 
This sample size greatly exceeds the number studied in our previous work, in 
which we suggested that nuclear spirals could remove sufficient angular 
momentum to fuel nuclear activity.  

The remaining five histograms in the figure illustrate some of the potential 
biases that could affect the relative distributions of the active and inactive 
galaxies. From the distribution of host galaxy types it is apparent that 
the active galaxies are slightly biased toward an earlier mean Hubble type 
than the inactive galaxies. 
The distribution in $M_B$ and $v$ shows that the active galaxies are on 
average also brighter and more distant than the inactive galaxies. The large 
number of distant active galaxies is partially due to the inclusion of the 
CfA Seyferts, many of which are even more distant than the upper limit of 
$6000 \kms$ for the last bin shown in Figure~\ref{fig:agnhist}. 
The lowest two panels show the size and inclination distributions. 
The active and inactive samples clearly have different size distributions. 
The angular size of the active galaxies are systematically smaller than the 
inactive galaxies, which shows that their greater mean distance more than 
offsets the fact that they are in brighter, presumably larger galaxies. 
The inclination distributions are similar for both samples. 
These many differences between the host galaxy properties of the active and 
inactive galaxies preclude any physical interpretation of the differences 
in the distributions of their circumnuclear dust classifications. 
In a future paper we will create subsamples of active and inactive 
galaxies without differences in their host galaxy properties and investigate 
the differences in their circumnuclear dust morphology in detail. 

In several AGN the spiral nuclear dust lanes also appear to have segments 
illuminated by ionization cones emerging from the active nucleus.
The most striking example is Mrk 573 in which the grand design nuclear 
spiral appears to make two rotations about the nucleus, with 4 segments 
illuminated in emission in a double biconical form \citep{quillen99}.
Several other galaxies also show dust lanes in absorption that become 
emission in a cone-shaped region centered on the nucleus, 
particularly Mrk\,1066 and NGC\,788. 
This may also be the case in NGC\,3393, which has spectacular S-shaped 
emission filaments inside the ionization cone \citep{pogge97,cooke00}. 
Unfortunately, the $H$ observation missed the nucleus and 
there is only a $V-H$ color map of one of the ionization cones, but the 
filaments on the ends of this emission become dust lanes at larger radii 
and outside the ionization cone. Studies of both Mrk\,573 
\citep{tsvetanov92,pogge95,capetti96} and NGC\,3393 \citep{cooke00} debate 
the relative importance of bow shocks and anisotropic nuclear ionization in the 
formation of these structures. 
Following \citet{quillen99}, who studied 4 galaxies, we find additional 
evidence that in at least some of these galaxies a significant fraction of 
the gas in the extended narrow-line regions is ambient gas photoionized 
{\it in situ} by the active nucleus.

  \subsection{Bars}

\begin{figure}
\epsscale{0.85}
\plotone{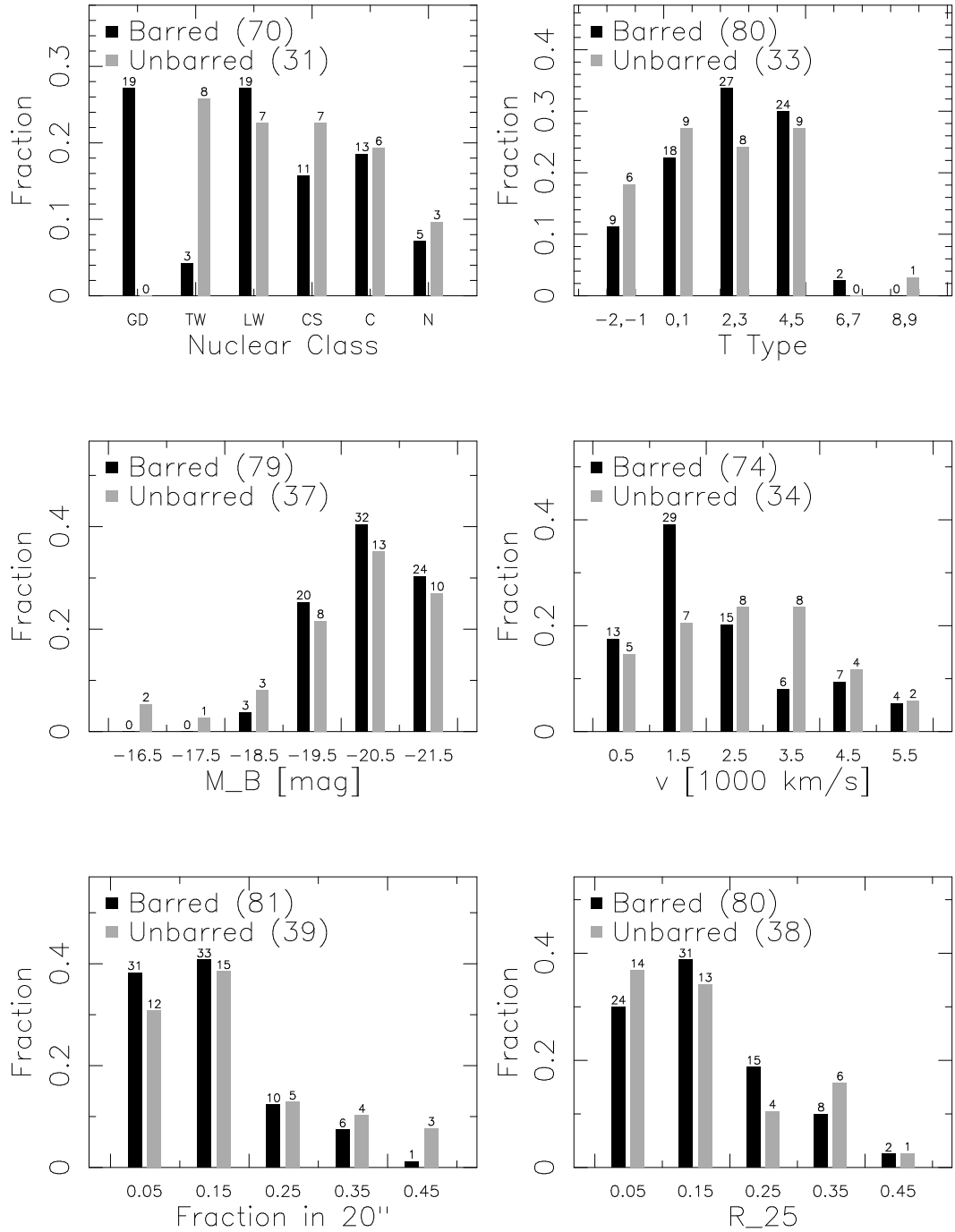}
\caption{Same as Figure~\ref{fig:agnhist} for the barred and unbarred galaxies. 
\label{fig:barhist} }
\end{figure}

Figure~\ref{fig:barhist} shows a similar set of histograms for the barred and 
unbarred galaxies in the sample. The distribution of host properties 
for the barred and unbarred galaxies are much more similar than for the 
active and inactive galaxies, although the unbarred samples are slightly 
earlier Hubble type, fainter, and more distant than the barred galaxies. 
Given these similarities, the distribution of the barred and unbarred galaxies 
into the different circumnuclear dust classes can provide real physical 
insight into the interaction of a large scale bar and the circumnuclear ISM, 
and these distributions do show some striking trends. 
The most prominent trend is that grand design nuclear spirals are only 
present in barred galaxies. There are 19 barred galaxies, or 27\% or the 
sample, with grand design nuclear spirals, yet no unbarred galaxies. 
Given the fraction of unbarred galaxies in the sample, we would expect there 
to be eight unbarred galaxies with grand design nuclear spirals if this form 
of circumnuclear structure were independent of the presence of a large scale 
bar. 

Straight dust lanes in strong, large-scale bars have been observed in galaxies 
such as NGC\,1300 or NGC\,7479. 
Simulations of barred galaxies show that shock fronts develop in the 
interstellar medium in strong bars \citep{prendergast83,athanassoula92}. 
These shocks can extend inwards as a grand-design spiral, depending on the 
properties of the potential and ISM, and cause dissipation when gas flows 
across them \citep{maciejewski02}. 
Simulations of gas flow in bars have shown that the angle between the bar and 
the dust lanes depends on the strength of the bar, where the dust lanes in 
the strongest bars are parallel to the bar's semimajor axis 
\citep{athanassoula92}. 
These dust lanes have the greatest shear, and consequently both the largest 
inflow and no associated star formation. 
Dust lanes in weaker bars \citep[\eg which on average correspond to type 
``SAB'' in the RC3 catalog, although with significant scatter][]{buta01} have 
less shear and some bars, such as in NGC\,1566 \citep{comte82}, have star 
formation in their dust lanes. 
Several of the galaxies (NGC\,4253, NGC\,5135, NGC\,5383, and NGC\,7130) 
in this sample with dust lanes and grand design 
nuclear spirals also have star formation associated with the large scale 
bar's dust lanes, which indicates that grand design nuclear dust spirals 
are not only found in strong bars. 
There are also barred galaxies with straight dust lanes that do not form 
grand design nuclear spirals, such as NGC\,4314 and NGC\,6951, where the 
bar dust lanes end in rings. 
Estimates or direct observations of gas flow along the dust lanes in bars 
have been made for NGC\,1097 \citep{quillen95} and NGC\,7479 \citep{regan97}. 
Recent, higher-resolution simulations have shown that the infalling material 
may become sheared by differential rotation within the Inner Lindblad 
Resonance of the bar or smoothed by a high sound speed \citep{englmaier00}. 
This differential rotation creates a symmetric, two-arm spiral like the grand 
design nuclear spiral class. 

\citet{englmaier00} used their models to demonstrate how the pitch angle
and coherence of the nuclear dust spirals depends on the central mass 
concentration and the pressure in the ISM, which is parametrized by a 
single value for the sound speed in their simulations. 
Models with greater central concentrations have more tightly wound arms, as 
do models with lower sound speeds. 
There is a wide range of pitch angle among the grand design nuclear spirals 
shown in Figure~\ref{fig:cmaps}, which offer the opportunity to test the 
predictions of these simulations in detail. For example, the grand 
design nuclear spiral arms in NGC\,1530 or NGC\,5643 have larger pitch 
angles than those in NGC\,1300 or Mrk\,573. As noted above, the observations 
of grand design nuclear spirals show that their circumnuclear disks are 
generally filled with finer dust structures in regions that are not 
dominated by the large, two-arm spiral. 
The presence of these smaller, flocculent spiral structures mixed with the 
dominant, two-arm spiral is a clear difference between the observations of 
grand-design spirals and the simulations of \citet{englmaier00}, who only 
produce the two-arm spiral. This difference may be due to the use of 
a single sound speed to characterize the ISM in their simulations as 
filamentary structure is seen in multiphase ISM models \citep{wada99,wada01}. 
Filamentary structure is also seen in multiphase ISM models which include a 
weak bar potential \citep{wada01b}, where the introduction of a multiphase 
ISM causes smooth spiral shocks to break down into parsec-scale, fine 
structure. 

A second trend apparent in the nuclear class distribution is that the 
tightly wound nuclear spiral class appears to be much more common 
in unbarred galaxies than barred galaxies. There are eight unbarred galaxies 
with tightly wound nuclear spirals, or 26\% of the sample, 
yet only three barred galaxies, or 4\% of the sample, have tightly 
wound spirals. 
As barred galaxies outnumber unbarred galaxies by approximately three to one,  
and there was no obvious selection bias in favor or against circumnuclear 
structure in galaxies with or without large scale bars, this also appears to 
be real and not the spurious result of a combination of 
selection effects. 

 \subsection{Nuclear Rings}

In our sample of 123 galaxies we find 14 galaxies with a ringlike 
structure in the $20''$ fields shown in Figure~\ref{fig:cmaps}. 
Eight are cases of strong nuclear rings that have both current star formation 
and may also have an older stellar population. 
In these eight galaxies the basic morphology of the star formation and dust is 
the same: the star formation is clumpy and the star formation occurs 
outside of the dust ring.
Five of the galaxies show the ring almost equally on both sides of the galaxy
(NGC\,1667, NGC\,2273, NGC\,3351, NGC\,4314, and NGC\,6951) and three galaxies 
(NGC\,1672, NGC\,5427, and NGC\,6890) show the star formation clearer on one 
side of the galaxy and the dust clearer on the other side of the galaxy.
In all the cases, the dust morphology is clumpy with the dust fragments
implying a spiral-like shape. The sense of the spiral pattern in the dust 
is in the same sense as the rotation of the nuclear ring, assuming that 
the gas in the bar dust lanes curves into the rings \citep{piner95}.
There also seems to be a tendency for the most active region of star
formation to be between the dust ring and where the bar dust lanes
joins the nuclear ring.

Four additional galaxies appear to have old stellar rings that are visible 
in both 
the $V$ and $H$ images. Two of these rings do not appear to have any 
associated dust: ESO138-G1 and Mrk\,477, and two have dust lanes immediately 
interior to the ring: NGC\,1300 and NGC\,3081. The final two galaxies 
have bright star formation rings, one a very small nuclear starburst 
ring in NGC\,864 and the other the larger starburst ring in NGC\,7469. 
The properties of the galaxies with rings are summarized in 
Table~\ref{tbl:ring} and additional information for several of these rings, 
many of which were previously studied from the ground by \citet{buta93}, 
are provided in the Appendix.

\section{Summary}

We have presented \hst observations of circumnuclear dust structure in a large 
sample of nearby galaxies. The color maps which trace this dust structure show 
that the nuclear dust spirals reported previously are the most common 
morphological feature in the circumnuclear region of spiral galaxies, 
irrespective of the presence or absence of an active nucleus or a large-scale 
bar. We have defined a set of four different classes of nuclear spiral 
structure that together encompass the main morphological features of 
nuclear spiral structure. There are in addition many galaxies that show 
no evidence of nuclear spiral structures and these galaxies have either 
very chaotic circumnuclear dust or no clearly discernible dust structure in 
their circumnuclear environment. 

While the entire sample is too heterogeneous for a detailed examination 
of the frequency of nuclear spiral structure in active and inactive 
galaxies beyond the statement that it is a common feature of both 
types, there are several results, including two clear differences between 
barred and unbarred galaxies, that should not be affected by the various
selection effects in this sample: 

\noindent 1. Only galaxies with large scale bars exhibit grand design 
nuclear spiral structure. Where high-quality data on larger scales is 
available, the two arms of the grand design nuclear spiral connect 
to the dust lanes on the leadings edges of the large scale bar. 
As these cases demonstrate that some large scale bars can remove sufficient 
angular momentum to transport matter to within at least tens of parsecs 
of the nucleus, they provide qualitative support to the hypothesis that 
some large scale bars can directly fuel nuclear activity. 
However, these dust lanes are also found in galaxies that are not known to 
harbor AGN. 

\noindent 2. Tightly wound nuclear spiral arms are more likely to be 
found in unbarred galaxies than barred galaxies. 

\noindent 3. Several active galaxies with prominent ionization cones, 
most notably Mrk\,573, Mrk\,1066, and NGC\,788, show that the emission 
line gas is the same material that forms the spiral dust lanes in the 
circumnuclear region. These dust lanes, traced by their absorption of stellar 
light, become bright emission regions where they cross the opening angle 
of the ionization cone from the nucleus. 

In a future paper we will compare a subsample of active galaxies to an 
inactive, control sample matched in Hubble type, luminosity, heliocentric 
velocity, size, and inclination to determine if there are any differences in 
the circumnuclear dust structures between active and inactive galaxies. 
We will also employ a similar sample-matching strategy to verify these 
differences between the circumnuclear dust structures of barred and unbarred 
galaxies to extend the results described above. 

\acknowledgements 

Support for this work was provided by NASA through grant numbers 
SN-7330, GO-7867, and SN-8597 from the Space Telescope Science Institute, 
which is operated by the Association of Universities for Research in Astronomy, 
Inc., under NASA contract NAS5-26555.  
PM was supported by a Carnegie Starr Fellowship. We acknowledge several 
suggestions from the referee that have improved this manuscript, and thank 
Witold Maciejewski for some helpful comments. 
This research has made use of the NASA/IPAC Extragalactic Database (NED) which 
is operated by the Jet Propulsion Laboratory, California Institute of 
Technology, under contract with the National Aeronautics and Space 
Administration. 

\appendix 

\section{Notes on Individual Objects}

Below we provide details on why we applied a particular classification 
to each galaxy, whether or not the final classification was very uncertain, 
and note other interesting features in the circumnuclear dust structure. 

\noindent
ESO137-G34 (HI) -- This Seyfert 2 is too highly inclined for an accurate 
classification, but it appears to possesses a great deal of chaotic, 
circumnuclear dust including a dust lane extending across the images. 
The emission studied by \citet{ferruit00} is also clearly visible in the 
color map. 

\noindent
ESO138-G1 (C) -- The most prominent feature of this galaxy is the smooth 
stellar ring, first observed by \citet{ferruit00},  with a radius of 
approximately 1.1 kpc. There are no young star clusters or other structures  
in the ring. The bright, blue emission to the immediate 
southeast of the nucleus was studied by \citet{ferruit00} who argue that 
it is either scattered nuclear light from the Seyfert 2 nucleus or 
produced by young, hot stars. 

\noindent
IC2560 (LW) -- The loosely wound nuclear dust spiral within the central 
500 parsecs appear to form two, resolved dust arms that connect to the large 
scale bar. The white circle immediately south of the nucleus is the 
NIC2 coronographic spot. 

\noindent
IC5063 (C) -- Although the circumnuclear dust structure is classified as 
chaotic, the straight dust lanes crossing to the north of the nucleus 
suggest that they could have a spiral morphology if viewed at a smaller 
inclination. Host galaxy dust may be responsible for its Seyfert 2 
classification as the peak nuclear brightness in $V$ and $H$ are not 
coincident. There are also pronounced ionization cones extending to the 
east and west of the nucleus.  

\noindent
IC5267 (C) -- There is a straight, north-south dust lane that crosses within
$1''$ of the nucleus and extends for over a kiloparsec. This dust lane and 
a neighboring feature with the same orientation give this galaxy its 
chaotic classification. These dust lanes appear similar to those expected 
in a nearly edge on galaxy, yet this galaxy is not highly inclined to the 
line of sight. 

\noindent
Mrk334 (LW) -- While there is clear circumnuclear spiral morphology in this 
galaxy, its distance and bright nuclear source make it difficult to resolve 
structure within a few hundred parsecs of the nucleus. 

\noindent
Mrk461 (LW) -- The coherence and pitch angle of the nuclear dust spirals 
make this a good example of a loosely wound nuclear spiral, although the 
contrast of the arms relative to the interarm region is not large. The nuclear 
spiral arms are more obvious on the near side than the far side of the galaxy. 

\noindent
Mrk471 (GD) -- There are straight dust lanes along this galaxy's large scale 
bar galaxy visible in both the $V$ image and color map and they enter the 
circumnuclear region approximately perpendicular to the nuclear bar 
candidate discussed by \citet{martini01}. The arms appear to curve inwards 
into a grand design nuclear spiral at these smaller scales, but the large 
distance of this galaxy precludes a detailed examination of the circumnuclear 
structure.

\noindent
Mrk477 (LW) -- This is one of the most distant galaxies in the sample and it 
is difficult to resolve much structure in the circumnuclear region. 
The loosely wound classification is based on much larger spatial scales than 
used for most of the galaxies and may not be representative of the circumnuclear 
structure. 

\noindent
Mrk573 (GD) -- The grand design, nuclear dust spiral in this Seyfert is 
particularly striking as these dust lanes are lit up in emission where they 
cross the ionization cone from the active nucleus. The inner part of these 
spiral arms also appear to trace the leading edges of the nuclear bar, as 
discussed by \citet{quillen99} and \citet{martini01}.  

\noindent
Mrk1066 (LW) -- The peak brightness in the $V$ and $H$ images are slightly 
offset, which suggests that dust is obscuring the direct line of sight 
to the active nucleus. At larger scales these is a multiarm, flocculent 
spiral. The inner region of these dust lanes appears to be photoionized by the 
the AGN. 

\noindent
Mrk1210 (TW) -- The tightly wound nuclear dust spiral is very regular and the 
individual dust lanes can be traced for over a full rotation about the 
nucleus.  

\noindent
NGC214 (LW) -- There is a dramatic change in the dust structure between 
the central kiloparsec and larger scales. The circumnuclear 
region is relatively smooth except for a few dust lanes, while at larger 
scales the arms are much more fragmented, fill the galaxy disk, and 
have some associated star formation. The position angle of the ring-shaped 
transition region between the circumnuclear disk and larger scales is 
offset approximately 30 degrees from the position angle of the host galaxy. 

\noindent
NGC357 (N) -- There is essentially no resolved circumnuclear dust in the 
central kiloparsec. While the nucleus is near the edge of the $H$ image, the 
$V$ image shows that the surface density is also smooth in the regions not 
visible in the color map. A few extremely low contrast dust features are 
present immediately south of the nucleus, but they are negligible compared 
to the contrast in any galaxy not in the N class. 

\noindent
NGC404 (C) -- This LINER is one of the nearest galaxies in the sample and the 
$H$ image has a mottled appearance due to the resolved giant stars. 
This complicates the identification of dust lanes and therefore the 
nuclear classification, but the dust distribution still appears chaotic. 
The $V-I$ color map shown in \citet{pogge00} also supports this classification 
and is less affected by resolved stars. 

\noindent
NGC628 (N) -- While there is no evidence for dust structure within 
$\sim 200$ pc of the nucleus, there are clear nuclear dust spirals at larger 
radii. If this galaxy were a factor of two or more distant, the empty 
central region might have been sufficiently poorly resolved to merit 
classification as a loosely wound or chaotic spiral. 

\noindent
NGC788 (LW) -- There are some emission line regions in the central few 
hundred parsecs, and these emission structures are coincident with where 
the dust lanes cross the ionization cone. 

\noindent
NGC864 (LW) -- There is a remarkably small circumnuclear starburst ring 
with a radius of slightly less than 100 parsecs. The blue light from this 
starburst appears to dominate the visible light from the nucleus. 

\noindent
NGC1068 (CS) -- This extremely well-studied galaxy has numerous short 
dust lanes in a spiral pattern, but these arms do not have the coherence of the 
loosely wound nuclear spirals. We therefore classify this as a chaotic spiral. 

\noindent
NGC1144 (TW) -- The large distance of this galaxy makes it difficult to 
classify the circumnuclear structure. On the observed scales, down to 
a limiting resolution of about 500 parsecs, it appears to be a tightly wound 
spiral.  

\noindent
NGC1241 (GD) -- This is a well defined, if inclined, example of a grand 
design dust spiral. These arms appear to fragment somewhat in the central 
few hundred parsecs, although the main two arms are still visible down to 
these smallest scales. 

\noindent
NGC1275 (C) -- There is a great deal of star formation in the circumnuclear 
region of this galaxy, as well as no coherent pattern to the dust. 

\noindent
NGC1300 (GD) -- This very well known, strongly barred galaxy is one of the 
prototypes of the grand design nuclear dust spiral class. The dust lanes 
clearly connect to the dust lanes along the leading edges of the bar studied 
by \citet{lindblad96}. 

\noindent
NGC1320 (HI/TW) -- Although it is highly inclined, a tightly wound nuclear dust 
spiral is still clearly visible. 

\noindent
NGC1398 (N) -- There is no dust structure at all and even the color map is 
extremely constant. The nucleus is only barely visible as a slightly 
bluer point on the otherwise flat color map. 

\noindent
NGC1530 (GD) -- The classification for this galaxy is somewhat questionable. 
At large scales, down to approximately 500 parsecs, there is a grand design 
nuclear spiral and some additional spiral arm fragments filling the rest of 
the circumnuclear disk. At smaller radii these arm fragments appear to have 
increasing contrast relative to the grand design spiral arms to the extent that 
it appears like a tightly wound spiral at the smallest radii. 

\noindent
NGC1638 (N) -- There is no evidence for any dust structures within the 
central kiloparsec of this galaxy. Even at larger scales, there only appear to 
be a few small star formation regions with some associated dust. 

\noindent
NGC1667 (LW) -- The star forming ring at approximately two kiloparsecs 
is composed of many spiral arm fragments intermixed with young 
star clusters. Inside of the ring the circumnuclear disk is substantially 
smoother and the few spiral dust lanes present do not fill the 
disk. 

\noindent
NGC1672 (LW) -- The ring of star formation is crossed by many dust lanes.  
These dust lanes extend inwards into a loosely wound nuclear dust spiral 
that is significantly more prominent on the near side of the galaxy. 
The peak nuclear brightness is offset between the $V$ and $H$ images, which 
suggests that the Seyfert 2 classification may be due to dust obscuration. 

\noindent
NGC1961 (TW) -- This peculiar galaxy has a multiarm, tightly wound 
nuclear spiral. There appear to be a few star clusters to the north of the 
nucleus, which may be the far side of the galaxy based on the contrast 
in the circumnuclear dust. 

\noindent
NGC2146 (C) -- The many chaotic dust lanes in this galaxy are parallel to 
the semimajor axis of the host galaxy. While there are several sites of star
formation, there is no obvious nucleus in the circumnuclear region. 

\noindent
NGC2179 (TW) -- This is an excellent examples of a tightly wound nuclear 
dust spiral. It appears to begin as a two arm spiral near the nucleus, but 
then at larger scales the two dust lanes merge into a one arm spiral that 
remains coherent for at least three revolutions about the nucleus. 

\noindent
NGC2273 (CS) -- The bright ring at the center, described by \citet{vandriel91}, 
is clearly connected to dust lanes at larger radii. In the color map the 
emission line regions are not as prominent as in the $V$ image, and the 
dust lanes do not form a coherent spiral. The narrowband images in 
\citet{ferruit00} clearly trace the line emission in this ring and show it 
is a star formation ring, rather an extended narrow line region. 

\noindent
NGC2276 (CS) -- This is one of the galaxies in which we disputed the 
nuclear classification. There is only one obvious dust spiral arm in the 
$V-H$ color map, although this is mostly because the nucleus is near the 
edge of the $H$ image. The $V$ image clearly shows several additional 
dust spiral arms to the south. 

\noindent
NGC2336 (C) -- At large scales there is a clear, multiarm spiral, 
although within the central kiloparsec the dust is significantly less 
coherent. There are only a few, scattered dust lanes with low contrast 
that do not imply a sense of rotation and therefore this galaxy is not 
classified as a nuclear dust spiral. 

\noindent
NGC2460 (CS) -- This galaxy has many scattered nuclear spiral features 
that do not maintain sufficient coherence to be classified as anything 
but a chaotic spiral. 

\noindent
NGC2639 (GD) -- This is an excellent example of a grand design nuclear 
spiral, although the inclination of the host galaxy and dust geometry make the 
second spiral arm difficult to trace. As what few features present 
maintain the symmetry of the higher contrast arm, we classify this galaxy 
as a grand design spiral.  

\noindent
NGC2903 (C) -- The well-known `hot spot' galaxy has a large number of 
resolved circumnuclear star clusters, although there is only one dust lane 
with sufficient curvature to suggest spiral structure. Because there are no 
other, similar features, this is not classified as a nuclear dust spiral. 

\begin{figure}
\epsscale{0.7}
\plotone{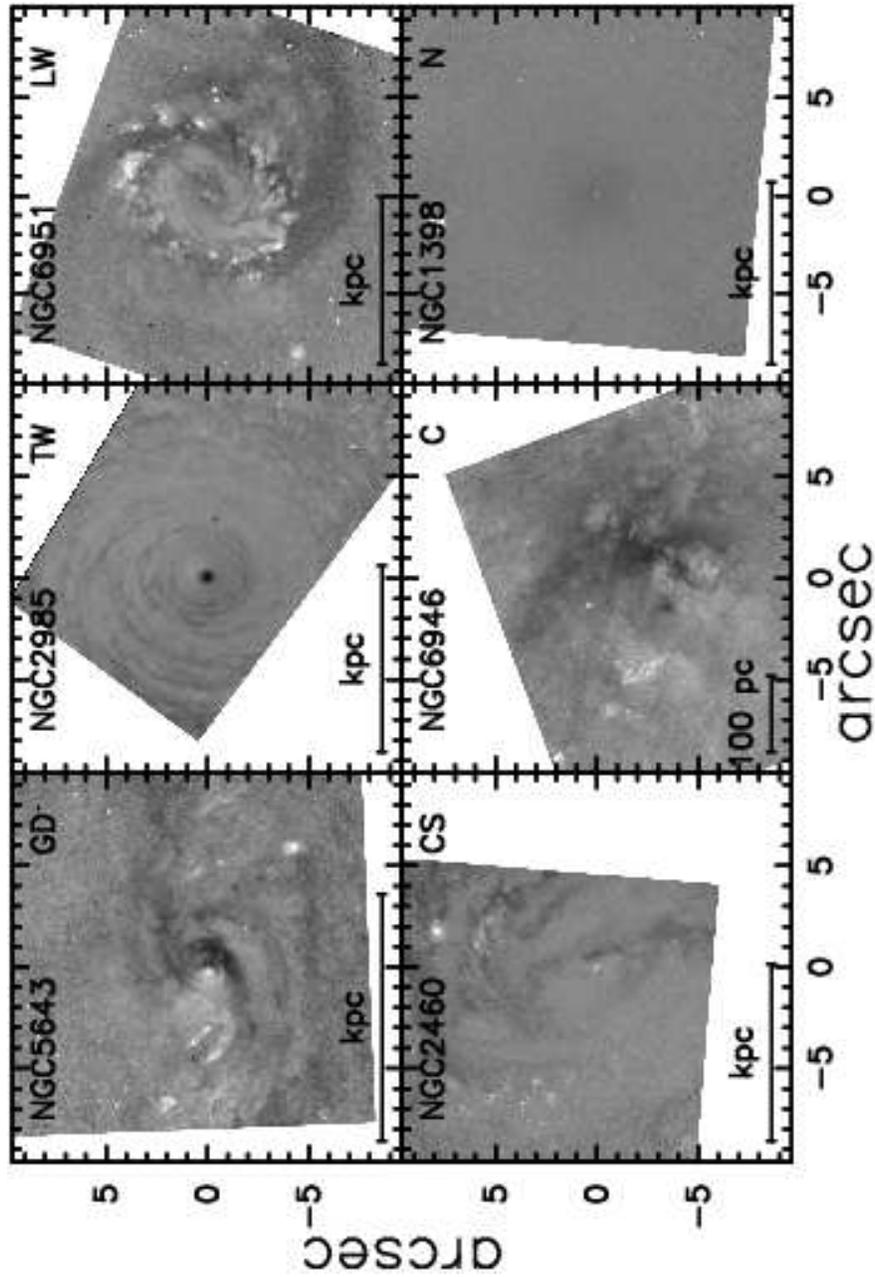}
\caption{Examples of each of the nuclear classifications described in 
\S\ref{sec:morph}. The colormaps are the same as those shown in 
Figure~\ref{fig:cmaps} for NGC\,5643 (GD: grand design nuclear spiral), 
NGC\,2985 (TW: tightly wound), NGC\,6951 (LW: loosely wound), NGC\,2460
(CS: chaotic spiral), NGC\,6946 (C: chaotic circumnuclear dust structure), 
and NGC\,1398 (N: no obvious circumnuclear dust structure). 
\label{fig:class} }
\end{figure}

\noindent 
NGC2985 (TW) -- This is one of the prototypes of the tightly wound nuclear 
spiral class and is also shown in Figure~\ref{fig:class}. While the nuclear 
spiral does appear to branch into additional dust lanes at larger radii, the 
main spiral arm remains coherent. 

\noindent 
NGC3032 (LW) -- This is one of the few examples of a nuclear dust spiral 
with associated star formation and there are several sites of star formation 
scattered throughout the circumnuclear disk. Given the high density of 
dust lanes, it is not possible to determine if these sites of star formation 
lead or trail the dust features, although the fact that they have the same 
pitch angle as the neighboring dust lanes suggests they are connected. 
The central region of the $V$ image is saturated, although even outside this 
saturated region the circumnuclear disk is very blue. 

\noindent 
NGC3079 (HI) -- This well-known LINER is too highly inclined for 
any structure in the circumnuclear dust to be reliably classified. 

\noindent 
NGC3081 (GD) -- This is one of the best examples of a resonance ring galaxy 
known and the nuclear ring described by \citet{buta98} lies between the 
two Inner Lindblad Resonances. This ring is well-resolved as a dust 
ring in the color map and interior to the ring the dust lanes form a
symmetric, two arm spiral. These dust lanes are somewhat unusual, however, 
as they appear to have nearly right angles where they meet the nuclear 
bar discussed by \citet{wozniak95} and \citet{mulchaey97b}. 
\citet{ferruit00} suggest that there is star formation in both the nuclear 
ring and the two spiral arms. 

\noindent 
NGC3145 (CS) -- While classified as a chaotic spiral, this galaxy could 
also meet the criteria of the loosely wound spiral class. There are several 
spiral dust lanes that stretch completely across the color map, although 
since they extend off the color map it is not possible to trace out how 
long they remain coherent. 

\noindent  
NGC3227 (C) -- The nucleus in $V$ is very saturated, although it is not so 
bright at $H$ that it overwhelms the circumnuclear region. While there is 
considerable dust present, even down to small scales, it does not take an 
obvious spiral form. 

\noindent 
NGC3300 (N) -- The colormap for this early-type galaxy is completely 
featureless, with the exception of a faint and blue nuclear point source. 

\noindent 
NGC3351 (C) -- The bright starburst ring dominates the appearance of the 
circumnuclear region. Within the ring there is a great deal of dust structure, 
although it does not take on a spiral form. 

\noindent 
NGC3362 (N) -- While bright, starforming spiral arms are prominent at large
scales, there are no obvious dust structures in the central kiloparsec. 
The only feature in the circumnuclear region is a blue structure in the 
colormap that is probably due to emission from the narrow line region. 

\noindent 
NGC3368 (CS) -- The dust lanes in this galaxy have very large pitch angles, 
although they are too incoherent to be classified as a loosely wound spiral. 

\noindent 
NGC3393 (CS) -- This galaxy has a spectacular, spiral-shaped ionization 
cone that appears morphologically similar to the one in Mrk\,573 
\citep{pogge97}. Unfortunately, the NICMOS observation just missed the 
nucleus and the color map only shows half of the circumnuclear region. 
Based on this half, and the $V$ image, there appears to be a nuclear spiral, 
although the contrast is insufficient to determine if the dust lanes and 
ionization cone are coupled as they are in Mrk\,573. 

\noindent 
NGC3458 (N) -- The circumnuclear region is completely featureless and 
only shows a very faint and blue nuclear point source. 

\noindent 
NGC3486 (LW) -- While this galaxy appears to have a weak grand design spiral 
at larger radii, in the central hundred parsecs these dust lanes fragment 
into a multiarm spiral. This galaxy was therefore classified as a loosely 
wound, rather than grand design type. 

\noindent 
NGC3516 (LW) -- This bright AGN has a number of spiral dust arms to the 
north of the nucleus, along with several large, diffuse emission line regions 
at larger radii. To the south there is relatively little dust structure 
in the circumnuclear region, although this may be due to inclination. 

\noindent 
NGC3627 (C) -- The high inclination of this relatively nearby galaxy makes
it difficult to discern the circumnuclear structure at distances much larger 
than a few hundred parsecs. The central region is dominated by several 
straight dust lanes that are parallel to the host galaxy semimajor 
axis. 

\noindent 
NGC3718 (C) -- The nucleus almost completely vanishes in the $V-H$ color 
map, as also shown for $V-I$ by \citet{pogge00}, and the dust structure 
around the nucleus does not imply the presence of any center of symmetry. 

\noindent 
NGC3786 (LW) -- Most of the visible dust is in a fragmented set of spiral 
arms to the north of the nucleus and on the near side of the galaxy. In 
the immediate vicinity of the nucleus the circumnuclear region is 
relatively blue, which may be due to scattered light from the nucleus or 
an emission line region. 

\noindent 
NGC3865 (CS) -- The lack of any dust on the far side of the galaxy disk 
makes it difficult to determine the coherence of the dust lanes, although 
the large amount of fragmentation at approximately a kiloparsec from the 
nucleus suggests that this spiral is chaotic. 

\noindent 
NGC3982 (GD) -- There is a grand design nuclear spiral in the central 
kiloparsec, although some of the other spiral dust lanes in the 
circumnuclear region have nearly comparable contrast. 

\noindent 
NGC4030 (TW) -- This galaxy has a spectacular, tightly wound nuclear 
spiral that matches the criteria for acoustic turbulence extremely well as 
the dust spirals fill the disk at all radii and decrease in contrast 
near the nucleus. This is one of the prototypes of the tightly wound 
nuclear spiral class. 

\noindent 
NGC4117 (LW) -- The main, prominent dust lane parallel to the semimajor axis 
dominates the circumnuclear dust, yet at low contrast this arm appears to 
make an entire revolution about the nucleus. This galaxy is relatively 
highly inclined. 

\noindent 
NGC4143 (LW) -- The circumnuclear dust structure is very low contrast, yet 
at least three dust spirals are still visible. While no circumnuclear 
star formation is visible, the LINER nucleus is visible as a faint, blue 
point source. 

\noindent
NGC4151 (LW) -- The bright nucleus in both $V$ and $H$ 
make it difficult to observe circumnuclear dust structure in the central 
hundred parsecs, yet at larger scales there are several spiral dust arms 
with large pitch angle. 

\noindent
NGC4253 (GD) -- This is a Seyfert galaxy with grand design nuclear dust 
spirals extending to the unresolved nuclear region, although due to the 
distance and bright nuclear point source, the smallest resolved radii 
correspond to about 300 parsecs. The dust lanes along the leading edge 
of the large scale bar, particularly to the east, show star formation 
in the dust lane along the bar. The presence of this star formation is 
characteristic of weak bars. 

\noindent
NGC4254 (CS) -- Although some spiral structure is present, it is very 
chaotic. There is some circumnuclear star formation scattered between the 
dust spirals, but they are not obvious associated with them. 

\noindent
NGC4258 (HI) -- The inclination of this galaxy is too high to classify
the circumnuclear dust structure. The near side of the galaxy is obscured 
by a great deal of dust, while there is no clear dust structure visible 
on the far side. It is also sufficiently nearby that the giant stars are 
resolved, which give the color map its mottled appearance.  The $V-I$ color 
map shown in \citet{pogge00} is less affected by resolved stars and 
indicates structure is present in the circumnuclear emission-line gas.

\noindent
NGC4260 (C) -- Dust structures are visible in the circumnuclear region, 
but they are not of sufficient contrast to classify easily. 
Those present do not imply any sense of rotation; we therefore classify this 
as a chaotic spiral. 

\noindent
NGC4303 (GD) -- The prominent grand design nuclear dust spirals connect to 
the dust lanes along the large-scale bar. As noted by \citet{colina00}, these 
two main dust spirals are composed of many smaller dust lanes and they
find that the star formation in the circumnuclear disk is due to 
gravitational instabilities.  

\noindent
NGC4314 (LW) -- \citet{benedict93} have extensively studied this galaxy and its 
bright, star forming ring. The ring includes a great deal of dust interior 
to the star formation. At larger scales there are two dust lanes that extend
to the large-scale bar, while interior to the ring there are a number of 
nuclear dust spiral arms with large pitch angle. 

\noindent
NGC4380 (CS) -- The dust spirals in this galaxy appear very flocculent. 
There are spiral arm fragments throughout the circumnuclear region, yet 
they continuously cross one another and reconnect. 

\noindent
NGC4388 (HI/C) -- While there is a copious amount of dust, the 
inclination is too great for a meaningful classification. 

\noindent
NGC4395 (C) -- This galaxy is almost too nearby to classify. At $H$ the 
brightest stars are resolved, while makes it difficult to trace the 
circumnuclear dust structure in the colormap. There are a few amorphous 
shapes, however, along with some emission coincident with the active 
nucleus and other star formation. 

\noindent
NGC4569 (C) -- There are clear dust structures present, yet no obvious spiral 
structure and this may be due to the moderate inclination of the galaxy. 
\citet{pogge00} also noted the chaotic circumnuclear dust in their $V-I$ 
color map. 
The round feature to the upper right in the NIR image and colormap are due 
to the NIC2 coronograph. 

\noindent
NGC4593 (TW) -- This galaxy has a tightly wound, one-arm nuclear spiral. 
It is somewhat unusual in that no other dust features outside of the one 
dust spiral are visible in the central kiloparsec. 
The circle to the right in the $H$ image and colormap is an artifact of the 
NIC2 coronograph. 

\noindent
NGC4725 (C) -- The nuclear dust structures are very low contrast in this 
galaxy and those visible do not form any obvious spiral pattern. 

\noindent
NGC4939 (C) -- The narrow line emission extends to the east and west, 
nearly perpendicular to the galaxy's semimajor axis. The straight dust 
lanes parallel to the semimajor axis extend across the frame, although the 
field of view is insufficient to see if they form any spiral structure. 

\noindent
NGC4941 (LW) -- This is a prototype of the loosely wound spiral class and 
also illustrates the effect of galaxy inclination on the relative 
contrast between dust features on the near and far sides of the galaxy. 

\noindent
NGC4968 (C) -- Any dust structure on the far side of the circumnuclear region 
is essentially washed out, and this makes it difficult to determine if the 
dust visible on the near side forms a spiral pattern. Southeast of the 
nucleus is a bluer region in the color map and this corresponds to the 
extended emission observed by \citet{ferruit00}, which may be the 
ionization cone. 

\noindent
NGC5005 (C) -- 
CO observations by \citet{sakamoto00} of this highly inclined LINER 
find $10^9 M_\odot$ of gas within the central five arcseconds and both 
the $V-H$ and the $V-I$ color map of \citet{pogge00} show that there is a 
large amount of chaotic, circumnuclear dust. The nucleus at $V$ is very 
indistinct, although the $H$ band image shows that the nucleus is 
approximately coincident with the northernmost emission feature at $V$. 
The peak at $H$ is still south of the location of the radio and 
CO peak shown by \citet{sakamoto00}. 

\noindent
NGC5033 (C) -- While this galaxy was considered to have a nuclear dust 
spiral by \citet{martini99}, inclination effects make it difficult to 
determine if these dust lanes actually form spiral structure. 

\noindent
NGC5054 (LW) -- There is some star formation in the circumnuclear disk, 
particularly to the south and west of the nucleus, but these star 
formation regions are not obviously associated with any of the many
nuclear dust spirals. 

\noindent
NGC5064 (TW) -- In spite of its relatively high inclination, the tightly 
wound nuclear dust spiral is apparent on both the near and far side of the 
circumnuclear disk. 

\noindent
NGC5135 (GD) -- There is a clear, grand design spiral at larger radii with 
some associated star formation along the leading edges of the dust arms. 
This star formation suggests that the bar is relatively weak
\citep{athanassoula92}. 
Within the central kiloparsec the spiral becomes much less ordered and the 
classification is consequently more uncertain. There also appears to be 
a great deal of circumnuclear star formation. 

\noindent
NGC5252 (CS) -- This colormap is relatively low signal to noise outside of 
the central kiloparsec, yet two spiral arm fragments are visible within 
the circumnuclear region. The structure map shown in \citet{pogge02} shows 
that this chaotic spiral has a flocculent form. There is also a small emission 
region to the immediate northeast of the nucleus. 

\noindent
NGC5256 (C) -- This chaotic, interacting system has no organized structure 
in the circumnuclear dust. The peak nuclear brightness at $V$ and $H$ are 
offset, which suggests the line of sight to the nucleus may be completely 
obscured at $V$. 

\noindent
NGC5273 (CS) -- The small nuclear spiral in this galaxy almost takes the 
form of a nuclear dust ring, although the arms do not quite form a 
circle. There are also some low contrast dust features at larger radii. 

\noindent
NGC5283 (CS) -- There is very little circumnuclear dust present and the 
central several hundred parsecs are dominated by emission regions. At 
slightly larger radii there are some dust structures to the south of the 
nucleus. 

\noindent
NGC5347 (GD) -- The two, symmetric dust lanes form a very well-defined 
grand design nuclear dust spiral, although the modest distance and Seyfert 
nucleus obscure these dust lanes before they make a full revolution about the 
nucleus. 

\noindent
NGC5383 (GD) -- This galaxy, studied in detail by \citet{sheth00}, has a 
grand design nuclear dust spiral that gradually decreases in prominence at 
smaller radii. There is a great deal of star formation associated with the 
dust lanes as they emerge from the inner edge of the large scale bar, 
particularly to the south, which suggests a weak bar \citep{athanassoula92}. 

\noindent
NGC5427 (LW) -- This galaxy is unusual as it appears to have star formation 
along the leading edges of a two arm, symmetric dust spiral. However, these 
arms are the continuation of the main disk, Sc grand design spiral pattern. 
Interior to these arms, at less than approximately 700 parsecs radius, there 
are four or five nuclear dust spirals and we therefore classify this 
galaxy as loosely wound, rather than grand design. 

\noindent
NGC5506 (HI/LW) --  This galaxy has too high an inclination for reliable 
classification, although the dust lanes present to the south of the nucleus 
suggest a dust spiral. 

\noindent
NGC5548 (TW) -- The extremely prominent nucleus of this Seyfert 1 at both 
$V$ and $H$ obscure the central five hundred parsecs with diffraction 
rings and other PSF artifacts. On larger scales, a tightly wound spiral is 
apparent and this spiral may change from a one arm to a two arm spiral at 
a radius of about a kiloparsec. 

\noindent
NGC5614 (TW) -- The tightly wound nuclear dust spiral in the central 
kiloparsec becomes less coherent at larger radii and branches into 
additional spiral dust lanes. 

\noindent
NGC5643 (GD) -- This well-defined, symmetric grand design nuclear dust spiral 
is another prototype of its class. While there are other dust spirals present 
in the circumnuclear region, the two main dust arms are wider and have 
greater contrast. The peak of nuclear brightness at $V$ and $H$ are slightly 
offset from one another, which is presumably due to dust. 

\noindent
NGC5674 (CS) -- There appears to be a blue, barlike feature oriented 
nearly east-west in the color map, however \citet{martini01} showed that 
this feature is not present at $H$ and is likely an artifact of the 
two, prominent dust lanes to the north and south of the nucleus. 

\noindent
NGC5691 (C) -- This galaxy has a great deal of distributed, circumnuclear 
star formation, although very few dust lanes. The nucleus, even at $H$, 
is not distinct. 

\noindent
NGC5695 (LW) -- This galaxy could arguably be also classified as a chaotic 
spiral as there are a number of dust lanes in the circumnuclear region with 
no connection to the curved dust lanes of the spiral. 

\noindent
NGC5929 (CS) -- Most of the circumnuclear dust is amorphous, although there 
are two dust spirals to the south that imply the same sense of rotation. 

\noindent
NGC5953 (TW) -- There are a large number of fragmented spiral dust arms with 
low pitch angle and unlike the other tightly wound nuclear dust spirals 
in the sample, these arms appear to have associated star formation. 

\noindent
NGC5970 (C) -- This may be a nuclear dust spiral observed at 
sufficiently high inclination that the spiral shape is not apparent. 
There are a number of straight dust lanes parallel to the host galaxy 
semimajor axis, yet they do not obvious wrap around the nuclear region. 

\noindent
NGC6221 (CS) -- The circumnuclear dust has a quite chaotic appearance, yet 
there are clear spiral arm fragments, particularly to the east. This galaxy 
is therefore one of our prototypes of the chaotic spiral class. 

\noindent
NGC6300 (C) -- This galaxy is difficult to classify as the nucleus falls in 
a corner of the $H$ image. There is a great deal of dust visible in the 
colormap, and the $V$ image shows that this dust extends to the south, yet 
it is dominated by a large cloud of dust rather than any filamentary structure. 
A two arm spiral is apparent at larger scales than shown in these frames. 

\noindent
NGC6384 (C) -- There is very little circumnuclear dust and the dust lanes 
are nearly straight. This may be due to inclination, but as only 
one of these dust lanes has any curvature it does not meet the requirements 
for a nuclear dust spiral. 

\noindent
NGC6412 (C) -- There is a very bright, radial filament of star formation 
extending nearly due north of the nucleus and about 600 parsecs in length.
The circumnuclear dust around it is very fragmented and it is not 
obviously associated with the star formation filament. 

\noindent
NGC6744 (C) -- Because the nucleus is near the corner of the $H$ image it is 
difficult to make out much structure in the circumnuclear region. 
There are several, distinct dust features, but they form no coherent 
structure. 

\noindent
NGC6814 (GD) -- The grand design nuclear spiral is the most obvious feature 
in the circumnuclear region, although there are several additional dust 
lanes at larger radii. This is unusual as most of the grand design spirals 
are wider and higher contrast at larger radii and become fragmented closer 
to the nucleus. 

\noindent
NGC6890 (GD) -- This is another prototype of the grand design nuclear spiral 
class and it illustrates the effect of inclination in enhancing and suppressing 
the dust contrast on the near and far sides of the galaxy, respectively. 

\noindent
NGC6946 (C) -- There is a great deal of dust in the circumnuclear region of 
this spiral galaxy, yet no spiral pattern. This galaxy is one of the 
prototypes of the chaotic class. 

\noindent
NGC6951 (LW) -- This galaxy has two dust lanes along its large scale bar 
and the southern one is clearly visible in the color map.  The dust lanes 
end at the circumnuclear starburst ring, however, and inside of the 
ring there appears to be a loosely wound spiral rather than a grand design. 
This galaxy is a prototype of the loosely wound nuclear spiral class. 

\noindent
NGC7096 (N) -- There may be two, very low contrast dust lanes in the 
central kiloparsec, although these features are much weaker than for 
the other galaxies with clear, circumnuclear dust structure. The nucleus 
is visible as a very faint, blue point source. 

\noindent
NGC7126 (TW) -- While the tightly wound nuclear dust spiral is lower contrast
on the far side of the galaxy, the dust lanes are still visible for 
several rotations about the nucleus. 

\noindent
NGC7130 (GD) -- This is a good example of a galaxy whose large scale bar 
is clearly transporting material to within several hundred kiloparsecs 
of the nucleus. The dust lanes from the large scale bar have some star 
formation on their leading edges, indicative of a weak bar, and there appear 
to be several knots of star formation in the very center. 

\noindent
NGC7177 (CS) -- Almost all of the circumnuclear dust structures are to the 
south of the nucleus and several of the dust lanes imply the same sense 
of rotation. 

\noindent
NGC7392 (LW) -- The symmetric, two arm nuclear dust spiral suggests appears 
similar to the grand design nuclear spiral class, yet these two dust arms 
are not higher contrast than other dust lanes in the circumnuclear 
region. 

\noindent
NGC7469 (TW) -- The most noticeable feature of this galaxy is the bright
circumnuclear starburst ring, followed by the prominent Seyfert~1 nucleus. 
Outside of the ring there is a tightly wound nuclear dust spiral, although 
it is difficult to determine if this structure continues inside the ring. 

\noindent
NGC7496 (C) -- There is a hint of a spiral dust arm to the southeast, yet 
the remainder of the circumnuclear region is sufficiently amorphous that 
we classify it as chaotic. 

\noindent
NGC7582 (HI/C) -- There is a great deal of star formation in the central 
kiloparsec, although the high inclination makes it difficult to determine 
whether or not the star formation is connected to the dust lanes. 

\noindent
NGC7674 (GD) -- There are two clear dust lanes entering the circumnuclear 
region from the large scale bar, yet the great distance and bright nuclear 
point source makes it difficult to resolve dust lanes in the central 
kiloparsec, although they appear to have a slight curve at the smallest 
radii observable. 

\noindent
NGC7682 (LW) -- It is difficult to reliably classify this galaxy due to its 
large distance. There are two weak dust arms that hint at a grand design 
structure, but due to their low contrast and absence of obvious symmetry 
we classify it as loosely wound. 
The ionization cones extending to the north and 
south are quite obvious in the colormap as two blue cones. 

\noindent
NGC7716 (CS) -- Because the nucleus is close to the edge of the $H$ image it 
is difficult to get a detailed picture of the circumnuclear dust morphology. 
There is a weak spiral arm that extends to the north, although to the south 
there are only a few scattered dust clumps. 

\noindent
NGC7743 (LW) -- There are several dust lanes with large pitch angle that 
imply the same sense of rotation, although these dust lanes do not 
appear to extend more than approximately 500 parsecs from the nucleus. 

\noindent
NGC7814 (HI) -- This nearly edge on galaxy has a striking dust lane 
across the nucleus that remains prominent in the NIR. 

\noindent
UGC12138 (GD) -- The two, curved dust lanes along the leading edges of the 
host galaxy bar curve inwards after half of a rotation about 
the circumnuclear region. Within the central several hundred parsecs the 
bright nuclear source obscures any smaller-scale structure. 

\noindent
UGC6100 (LW) -- There are spiral dust lanes within the central two kiloparsec
that form a coherent, loosely wound nuclear spiral. In the inner few 
hundred parsecs there is an emission line region that is probably associated 
with the narrow line region. 

\noindent
UM146 (LW) -- The signal to noise in the $H$ image makes it difficult to 
identify dust structures outside the central kiloparsec, but the circumnuclear 
region appears to be dominated by a two arm spiral. Because it is 
difficult to determine the symmetry of this spiral, it was not classified 
as a grand design nuclear spiral. The two arm spiral is more obvious in 
the structure map shown by \citet{pogge02}. 

{}

\clearpage

\clearpage

\begin{figure}
\epsscale{0.85}
\plotone{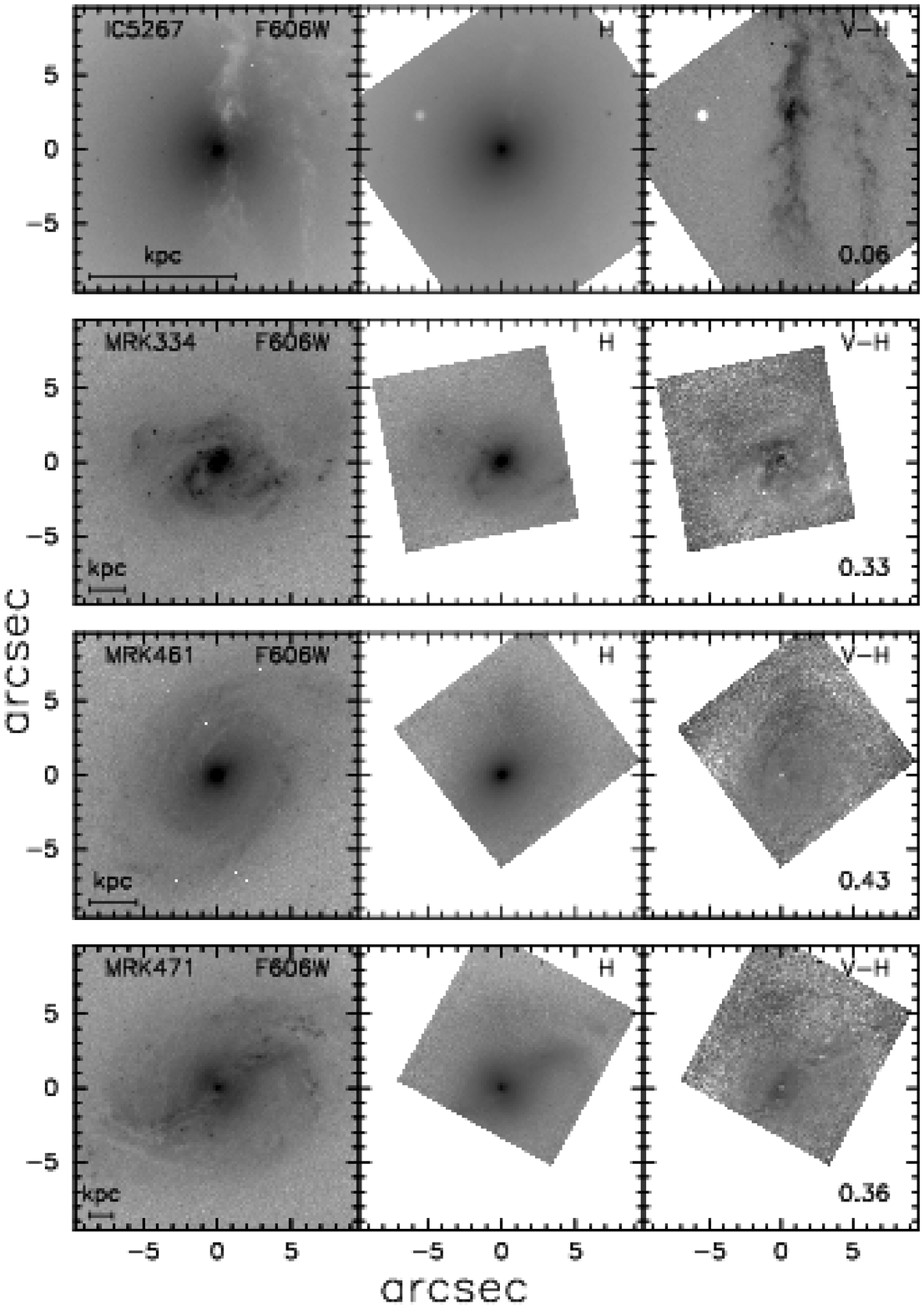}
\caption{Figure~\ref{fig:cmaps} -- {\it Continued}}
\end{figure}

\clearpage

\begin{figure}
\epsscale{0.85}
\plotone{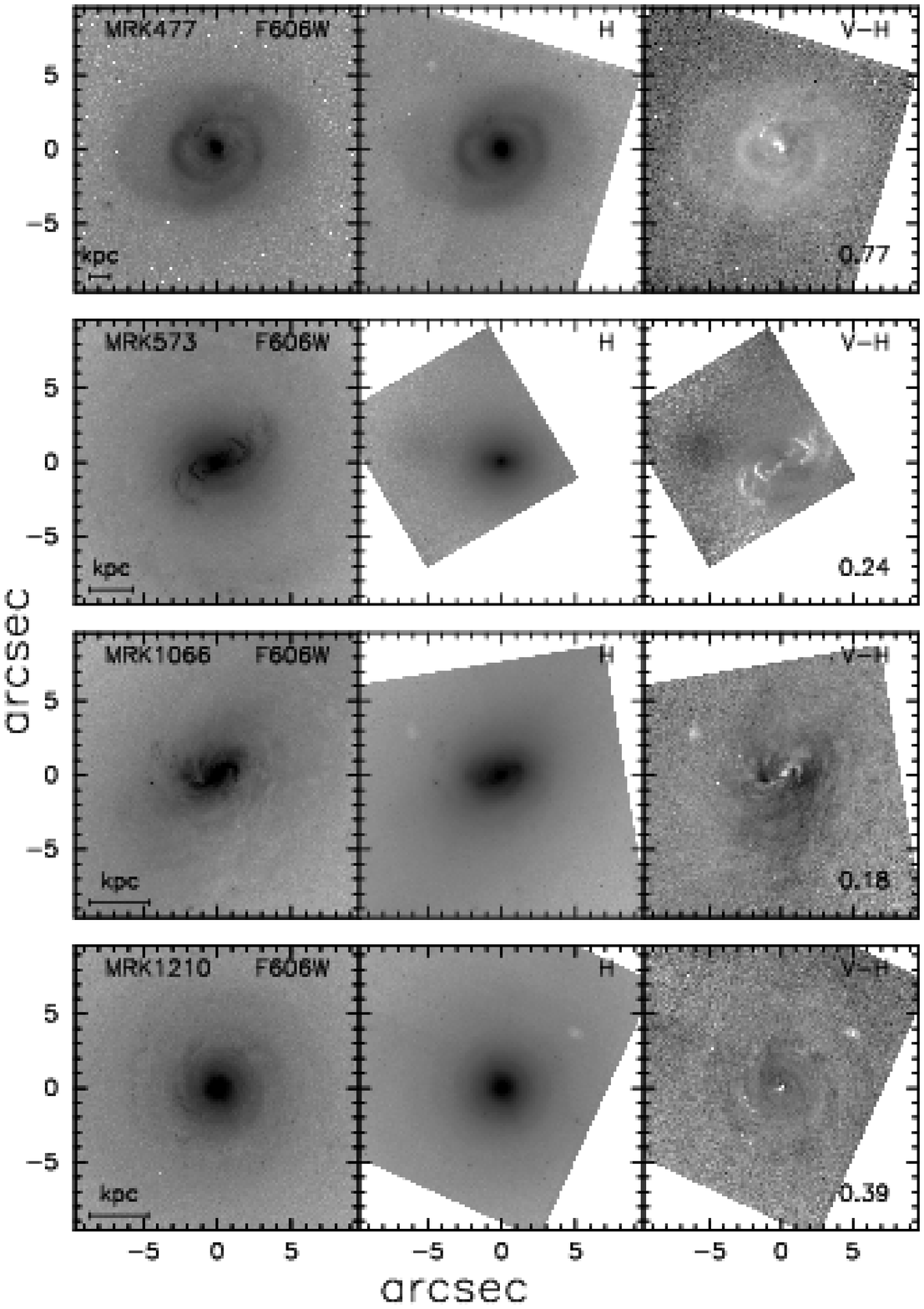}
\caption{Figure~\ref{fig:cmaps} -- {\it Continued}}
\end{figure}

\clearpage

\begin{figure}
\epsscale{0.85}
\plotone{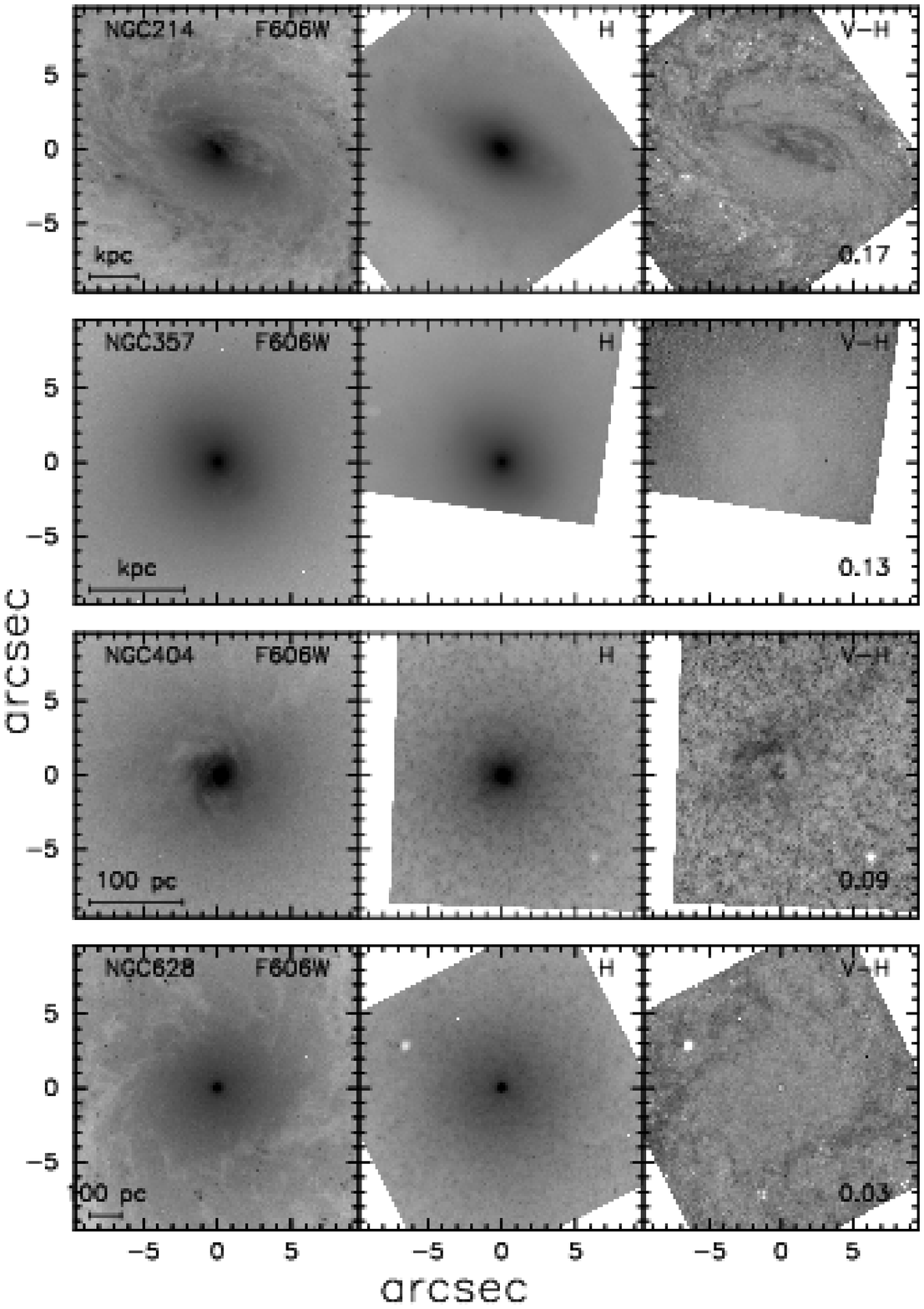}
\caption{Figure~\ref{fig:cmaps} -- {\it Continued}}
\end{figure}

\clearpage

\begin{figure}
\epsscale{0.85}
\plotone{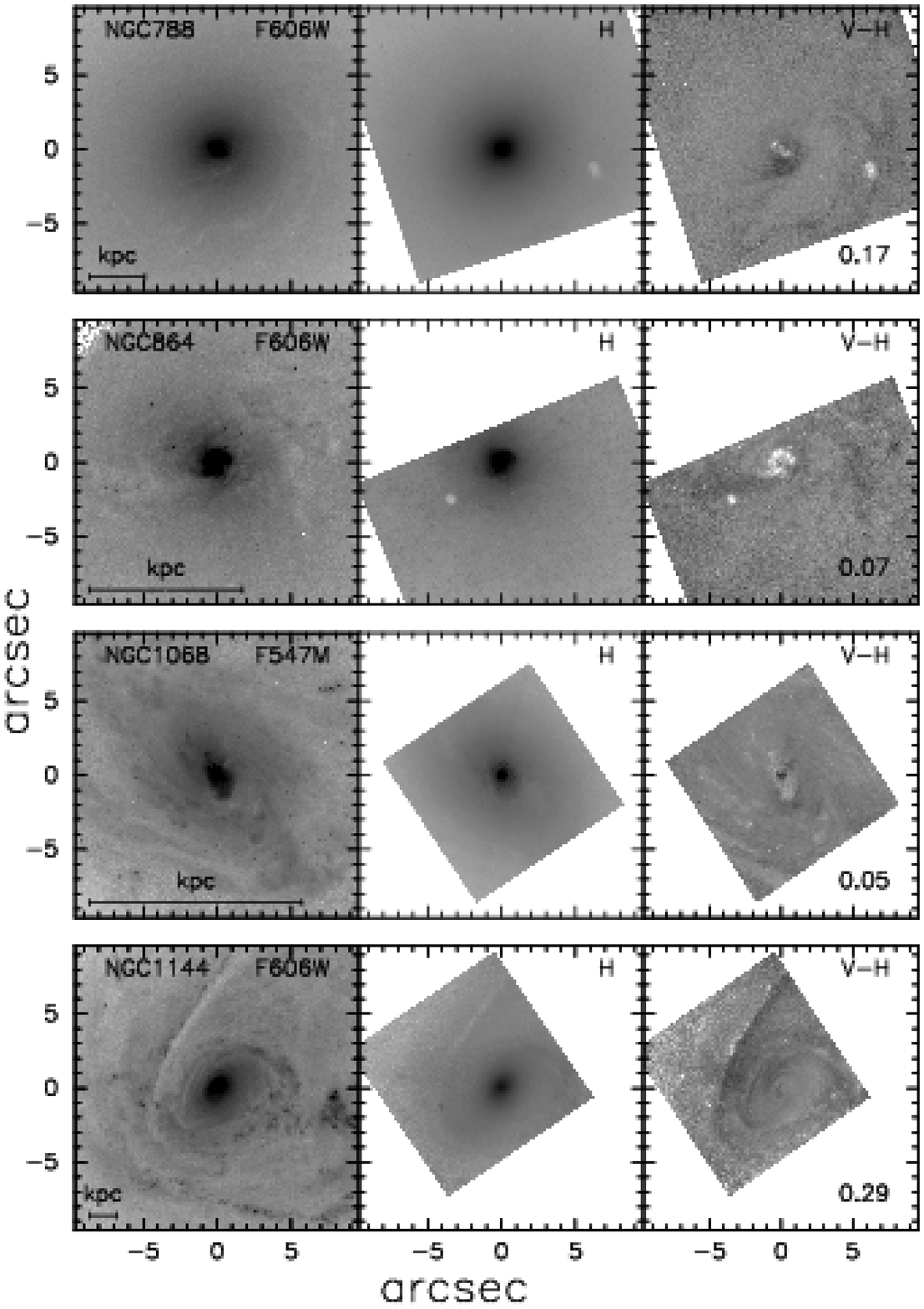}
\caption{Figure~\ref{fig:cmaps} -- {\it Continued}}
\end{figure}

\clearpage

\begin{figure}
\epsscale{0.85}
\plotone{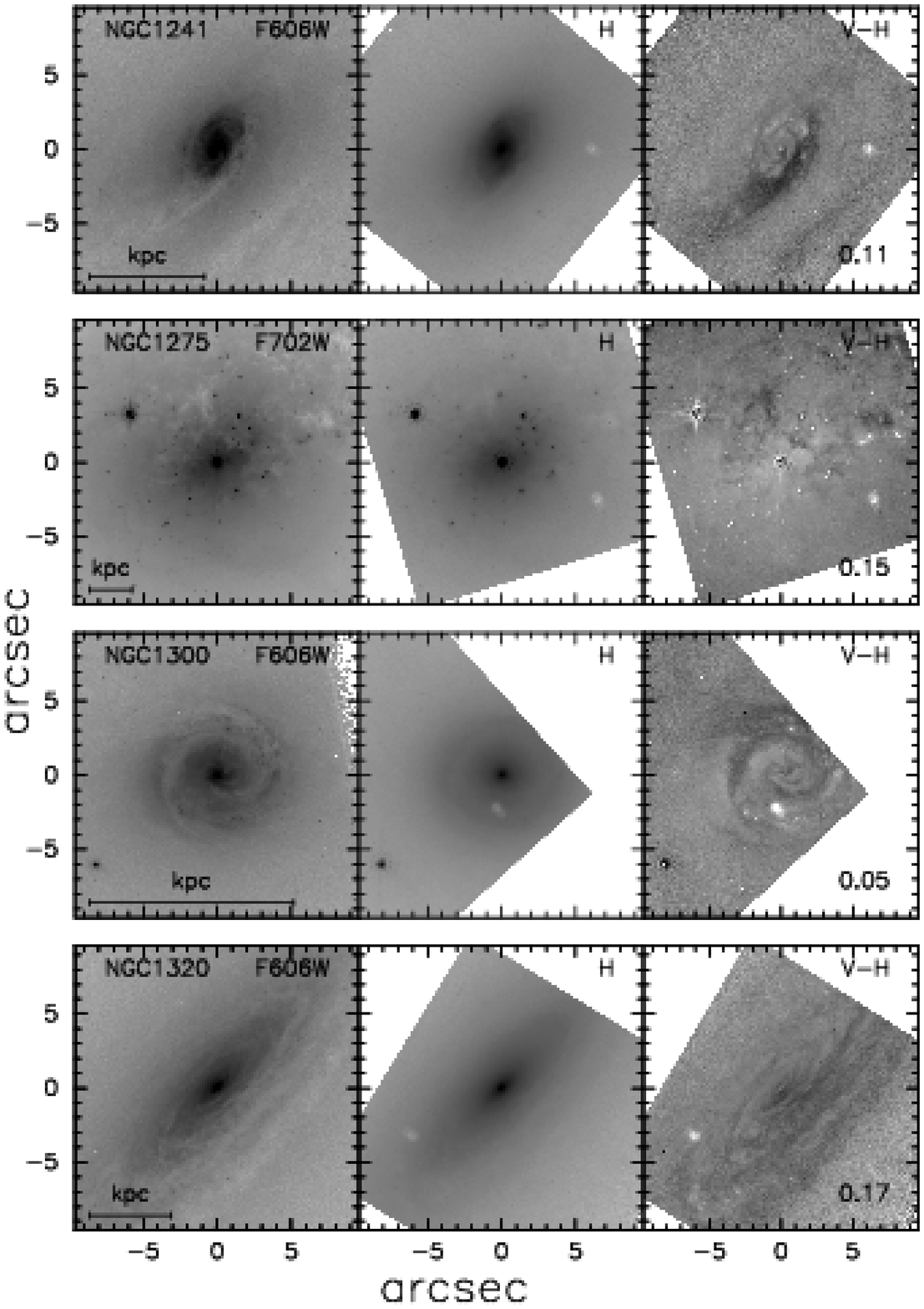}
\caption{Figure~\ref{fig:cmaps} -- {\it Continued}}
\end{figure}

\clearpage

\begin{figure}
\epsscale{0.85}
\plotone{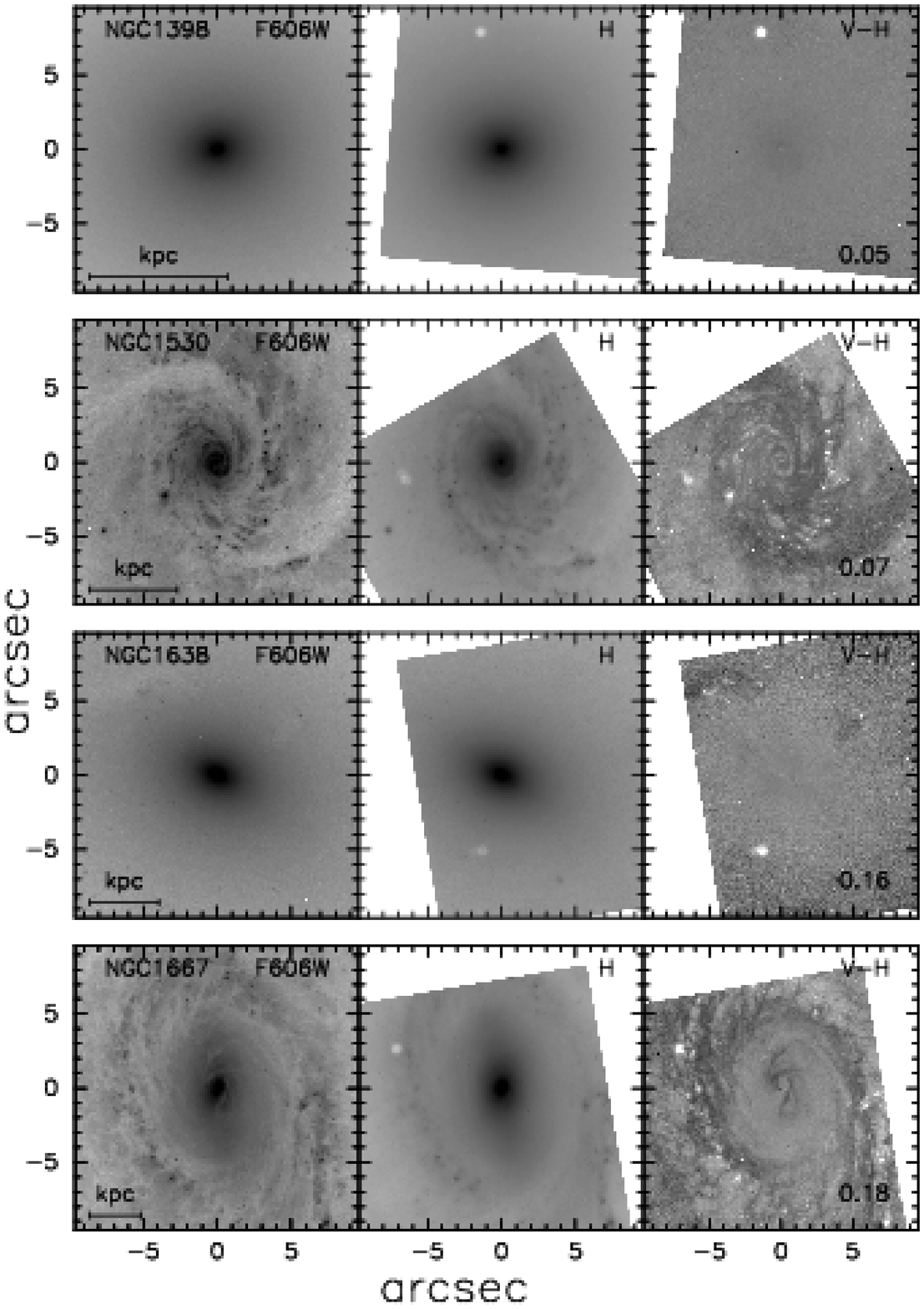}
\caption{Figure~\ref{fig:cmaps} -- {\it Continued}}
\end{figure}

\clearpage

\begin{figure}
\epsscale{0.85}
\plotone{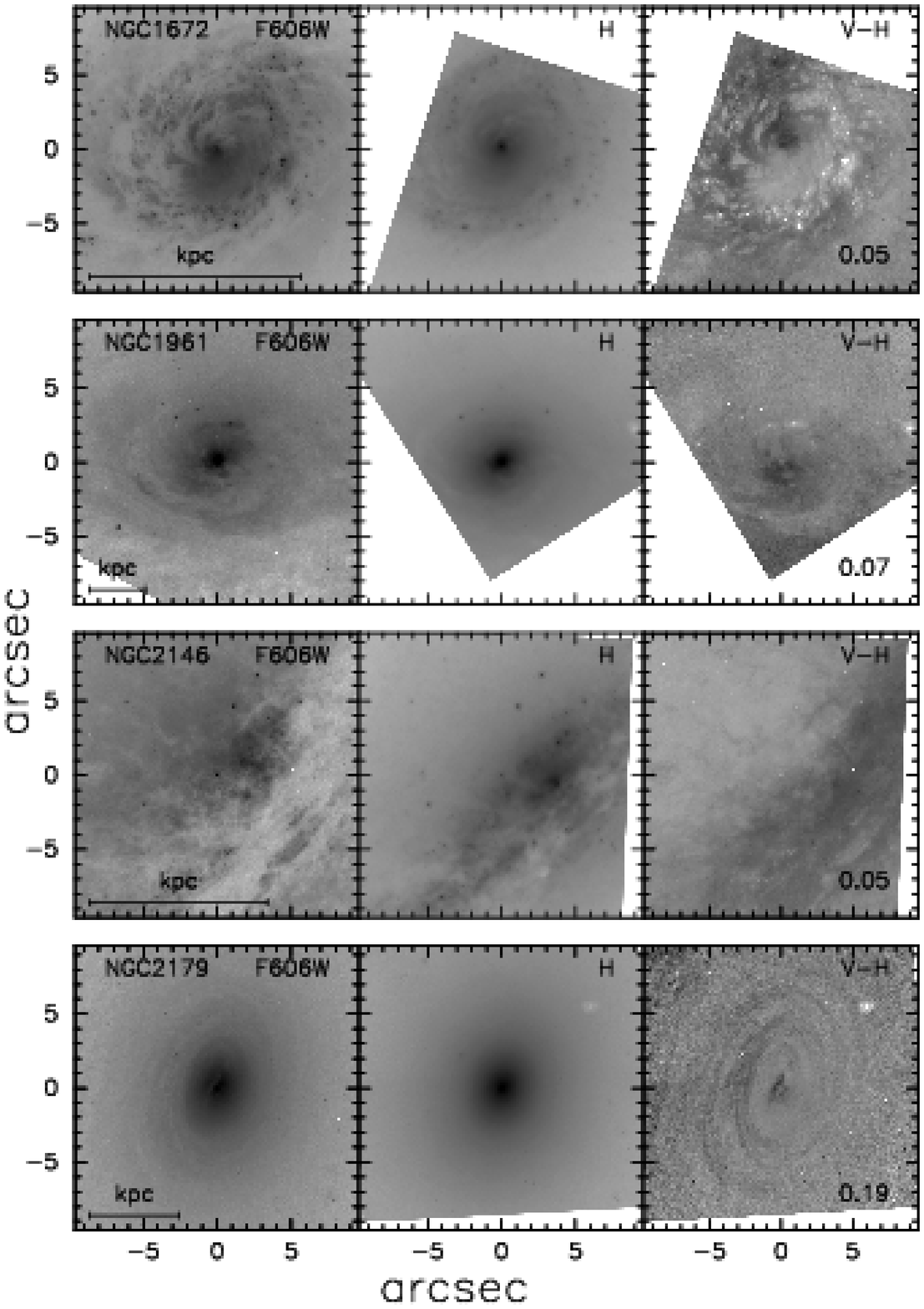}
\caption{Figure~\ref{fig:cmaps} -- {\it Continued}}
\end{figure}

\clearpage

\begin{figure}
\epsscale{0.85}
\plotone{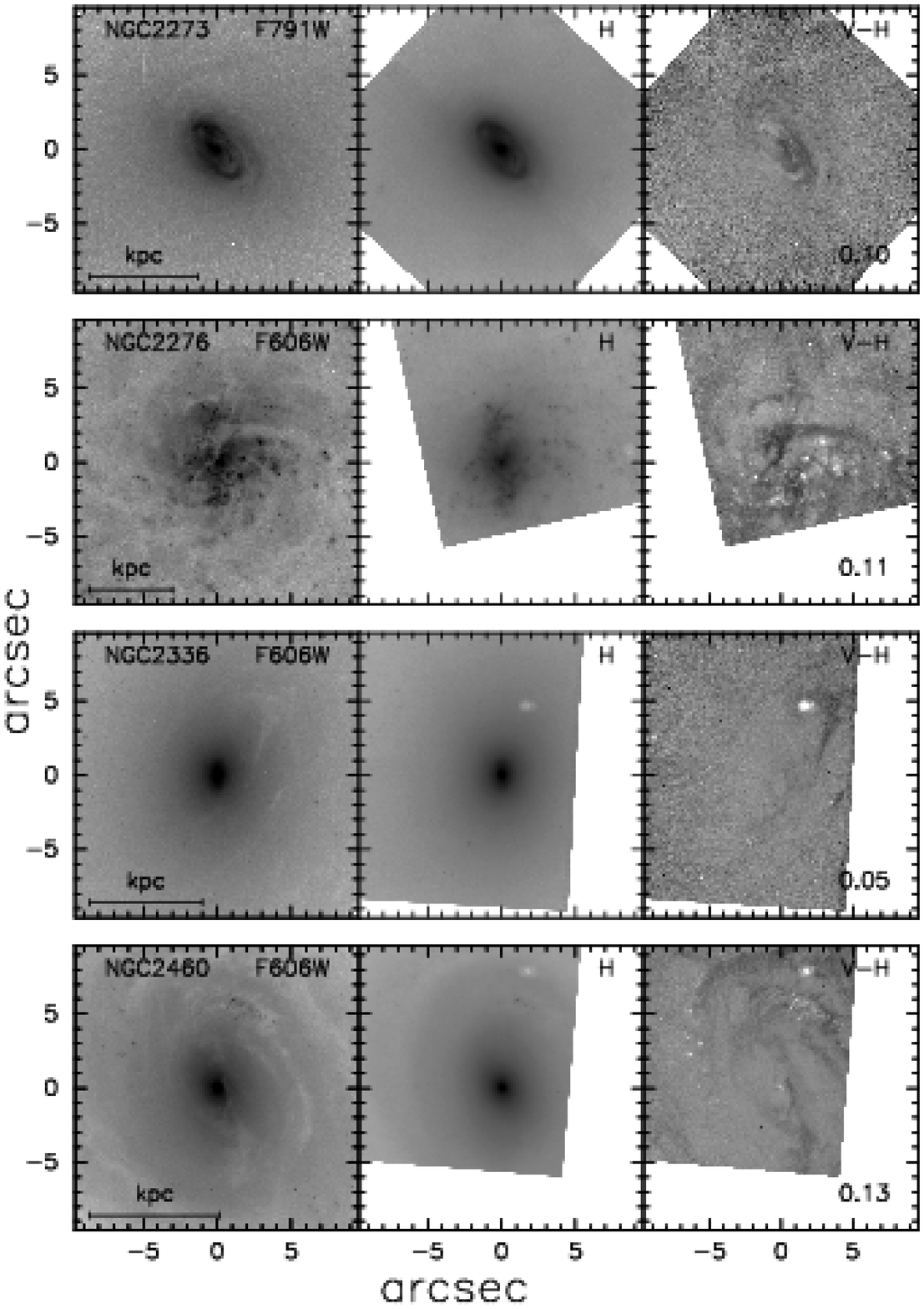}
\caption{Figure~\ref{fig:cmaps} -- {\it Continued}}
\end{figure}

\clearpage

\begin{figure}
\epsscale{0.85}
\plotone{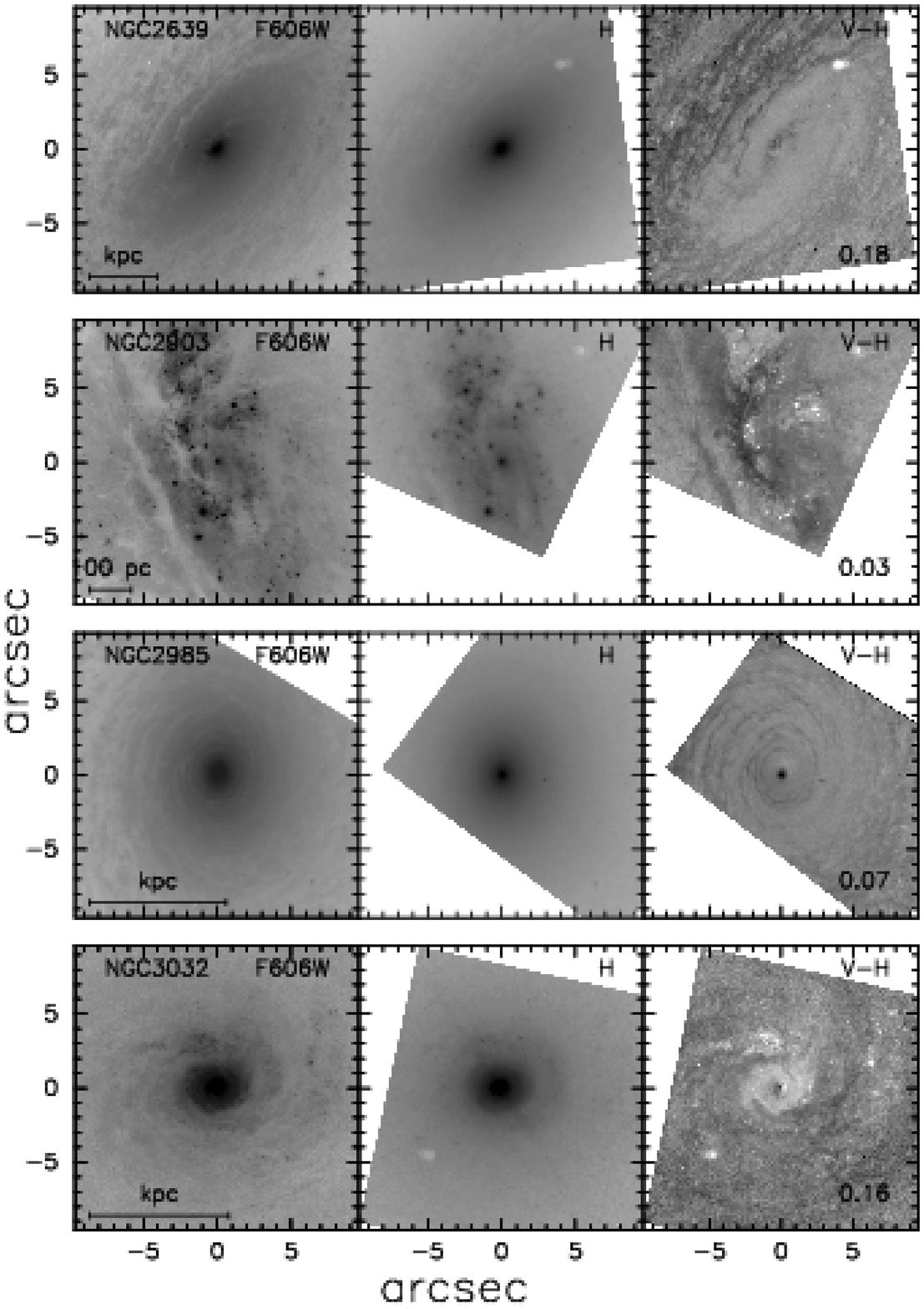}
\caption{Figure~\ref{fig:cmaps} -- {\it Continued}}
\end{figure}

\clearpage

\begin{figure}
\epsscale{0.85}
\plotone{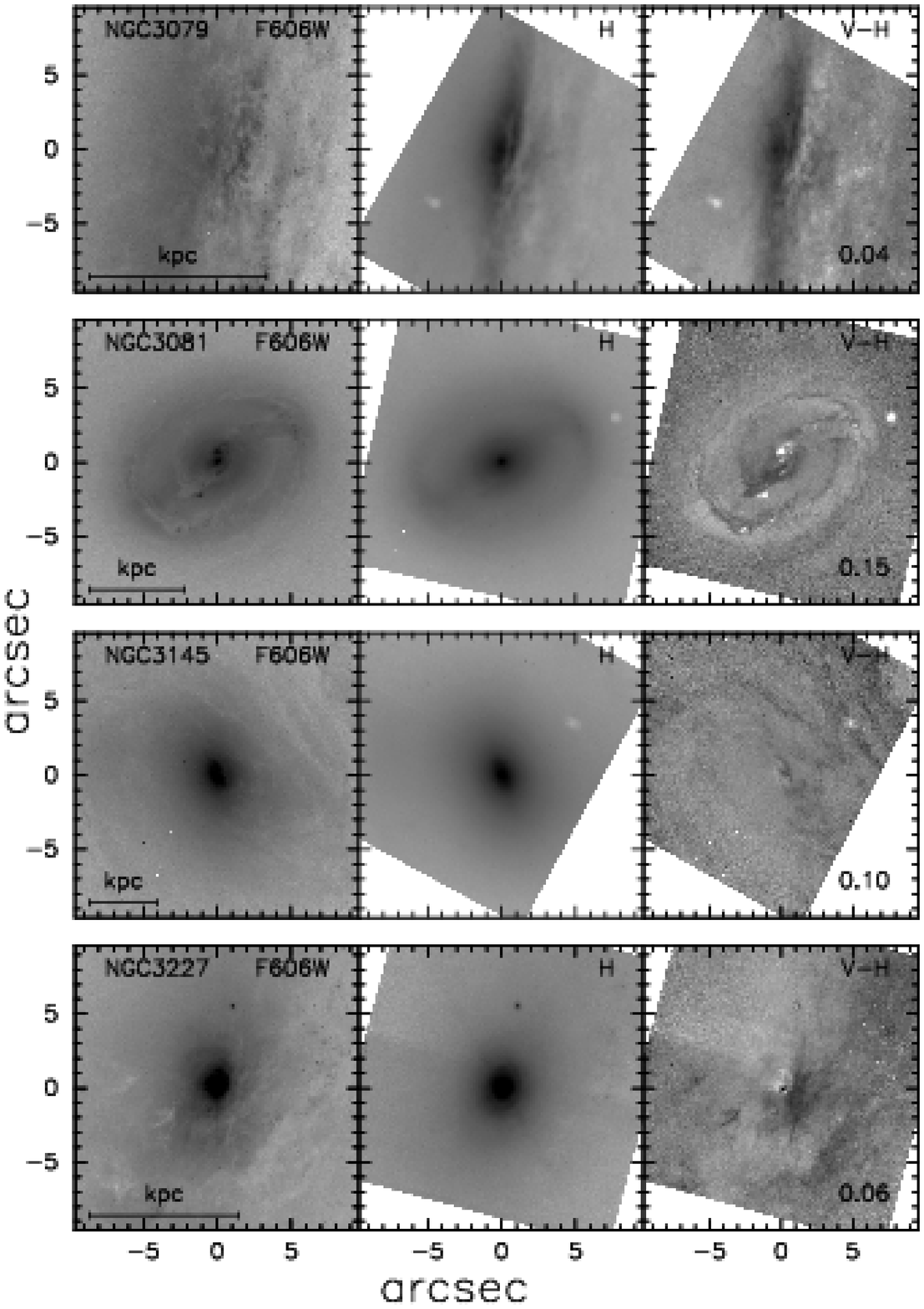}
\caption{Figure~\ref{fig:cmaps} -- {\it Continued}}
\end{figure}

\clearpage

\begin{figure}
\epsscale{0.85}
\plotone{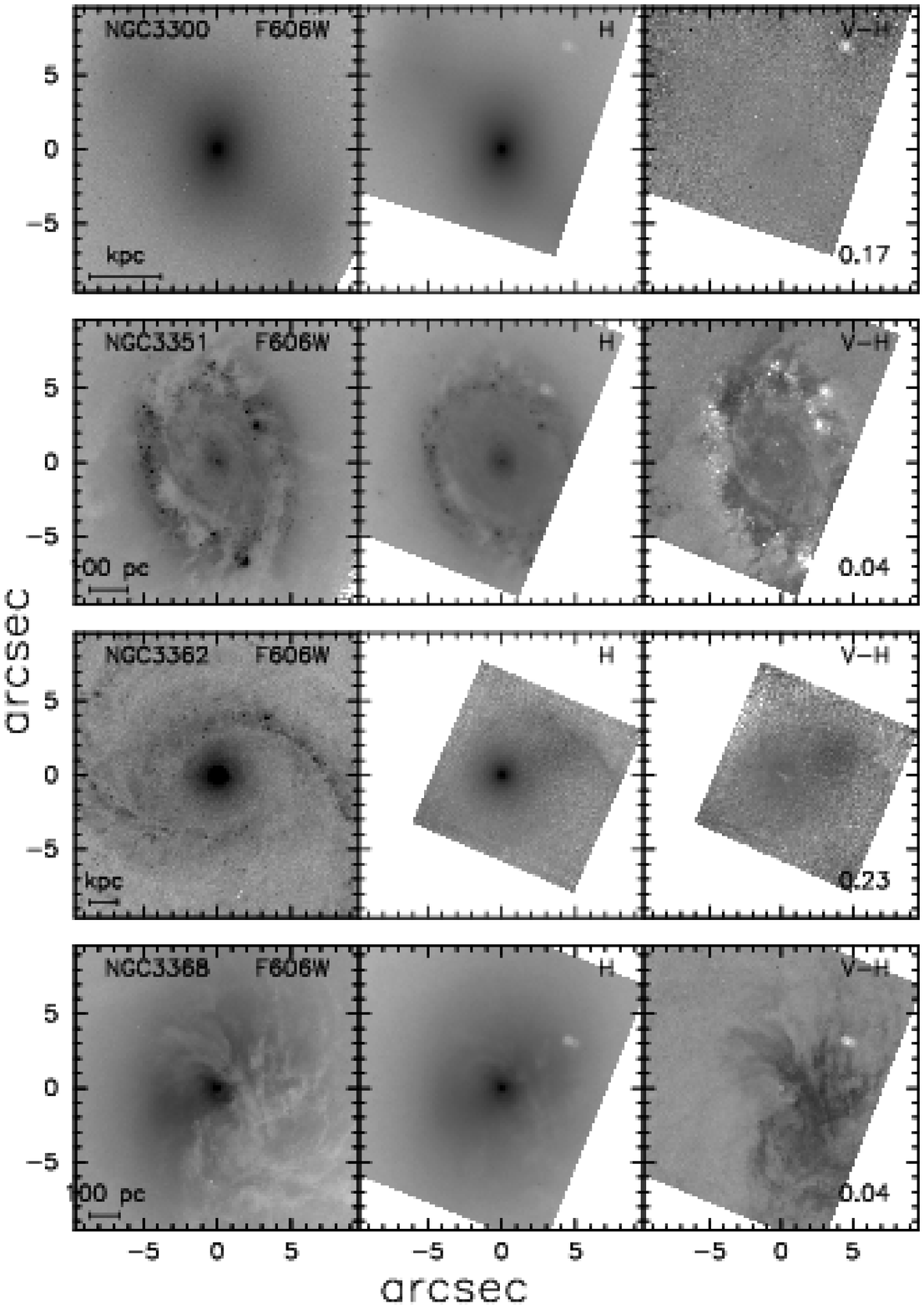}
\caption{Figure~\ref{fig:cmaps} -- {\it Continued}}
\end{figure}

\clearpage

\begin{figure}
\epsscale{0.85}
\plotone{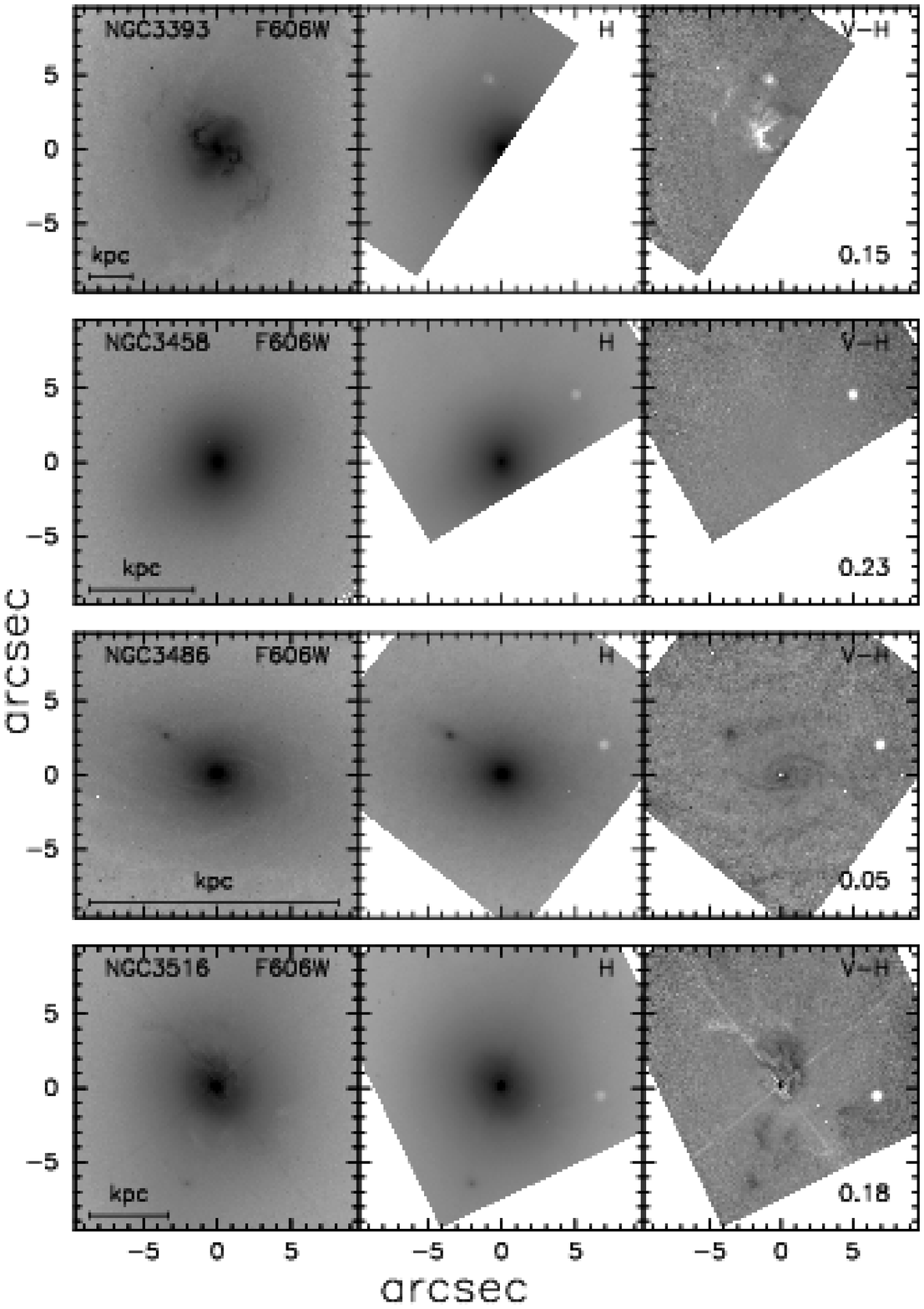}
\caption{Figure~\ref{fig:cmaps} -- {\it Continued}}
\end{figure}

\clearpage

\begin{figure}
\epsscale{0.85}
\plotone{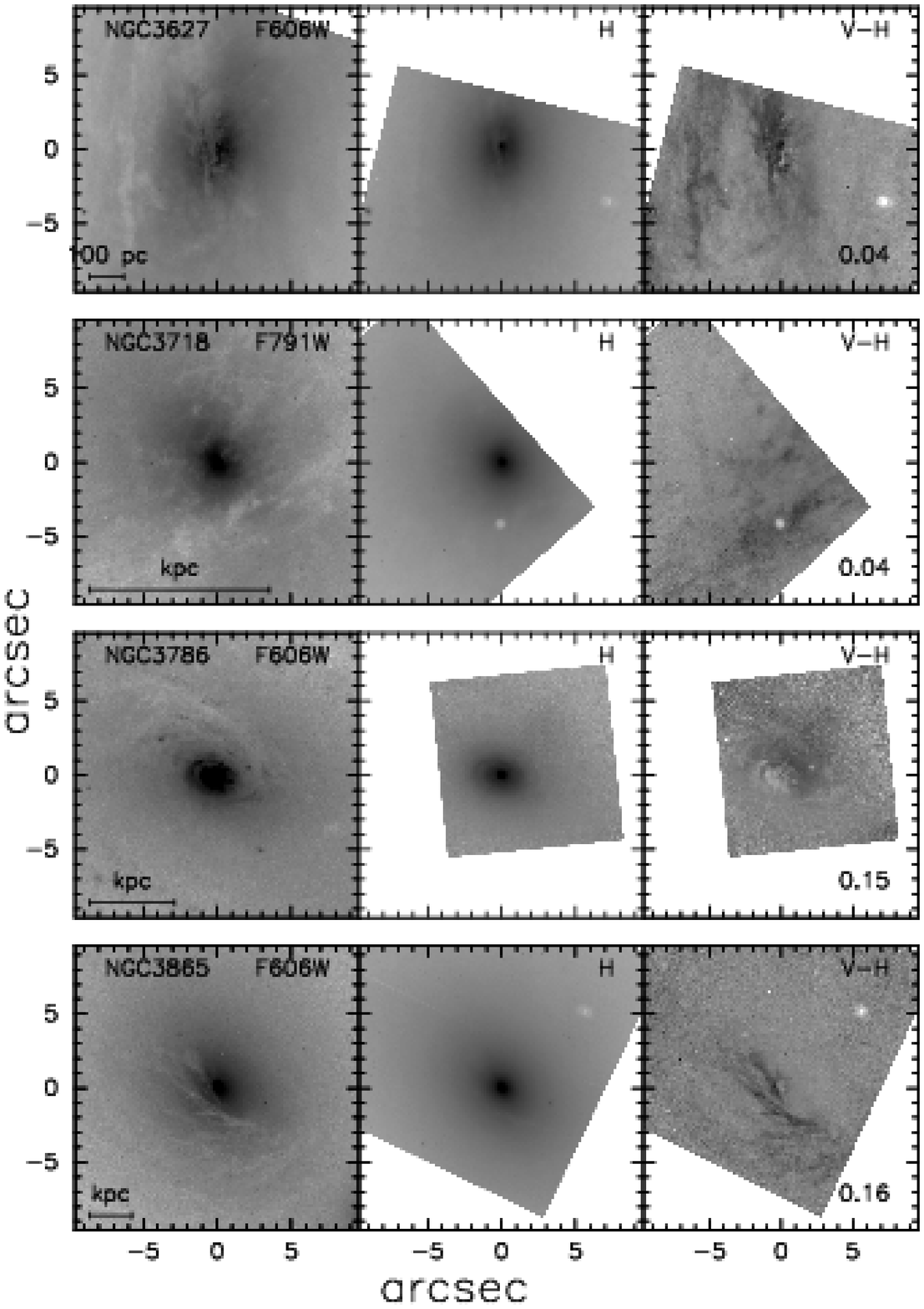}
\caption{Figure~\ref{fig:cmaps} -- {\it Continued}}
\end{figure}

\clearpage

\begin{figure}
\epsscale{0.85}
\plotone{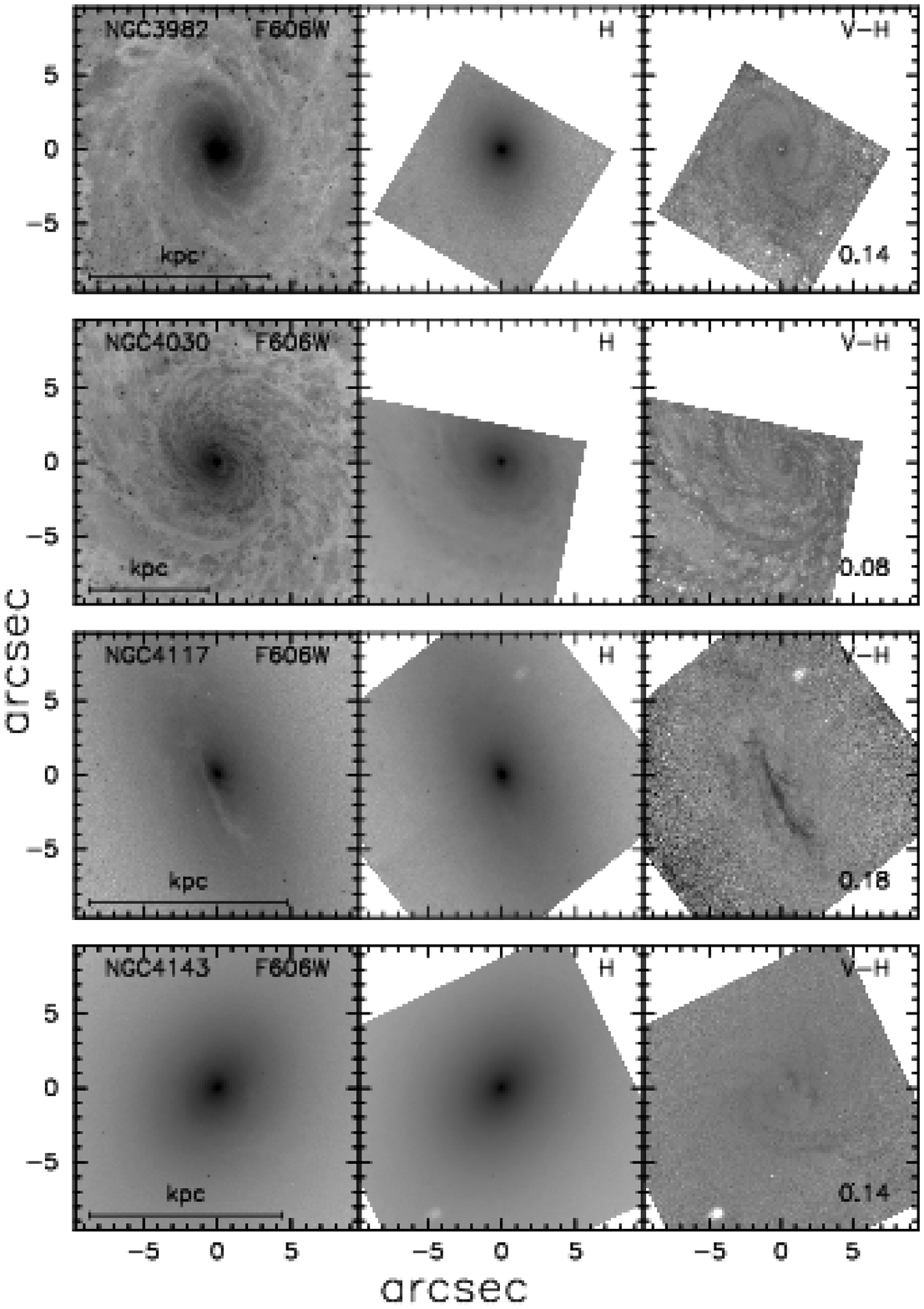}
\caption{Figure~\ref{fig:cmaps} -- {\it Continued}}
\end{figure}

\clearpage

\begin{figure}
\epsscale{0.85}
\plotone{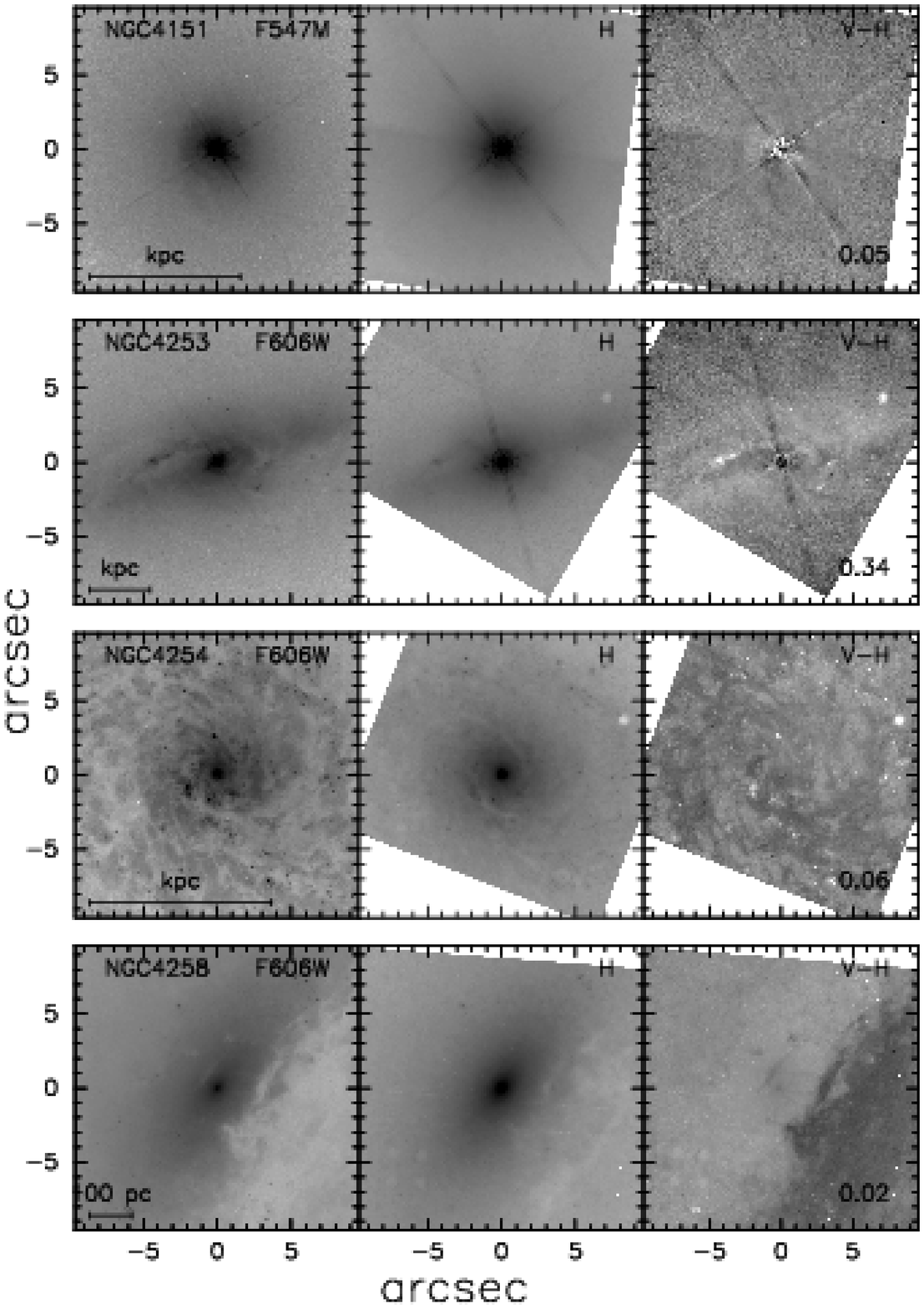}
\caption{Figure~\ref{fig:cmaps} -- {\it Continued}}
\end{figure}

\clearpage

\begin{figure}
\epsscale{0.85}
\plotone{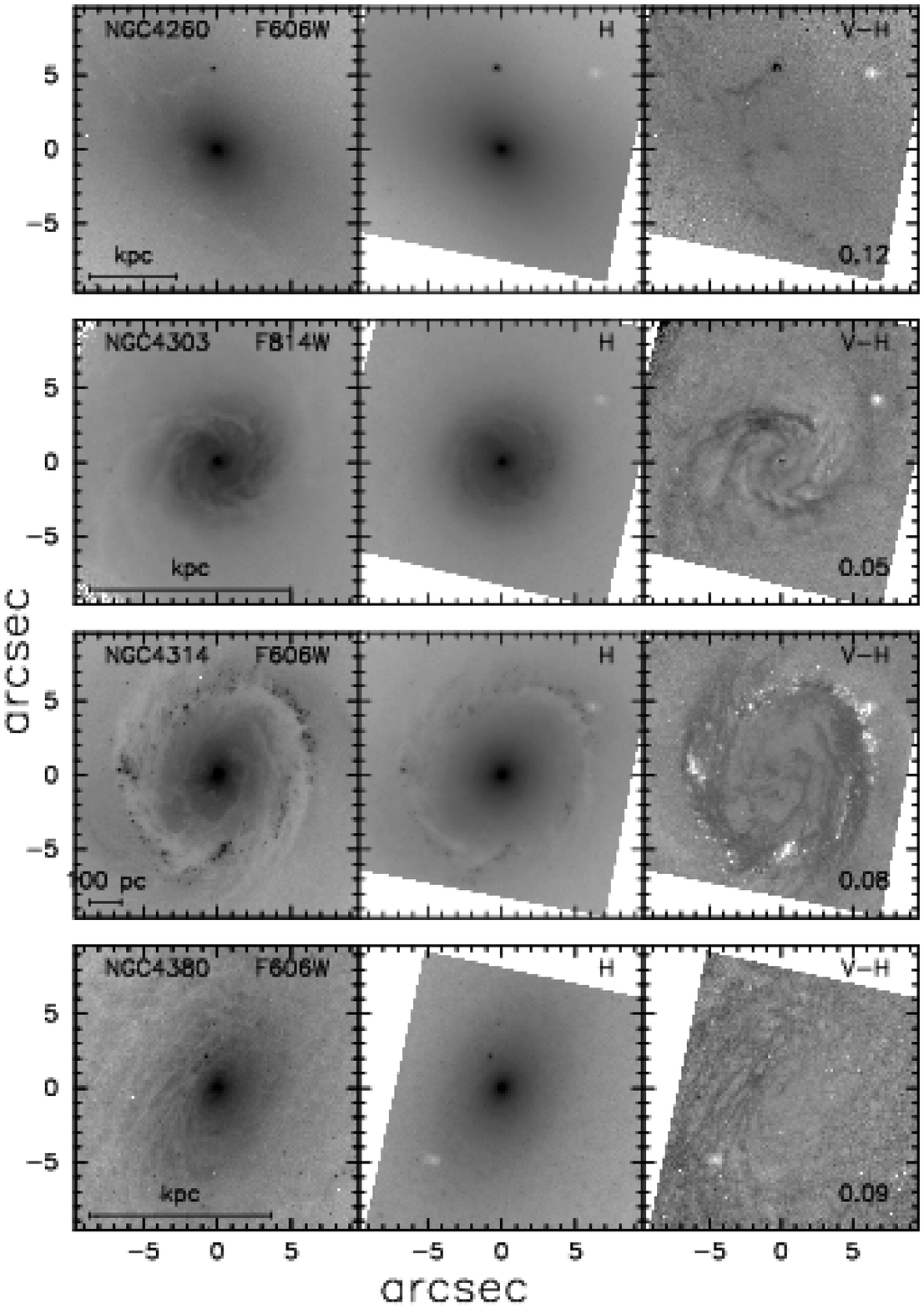}
\caption{Figure~\ref{fig:cmaps} -- {\it Continued}}
\end{figure}

\clearpage

\begin{figure}
\epsscale{0.85}
\plotone{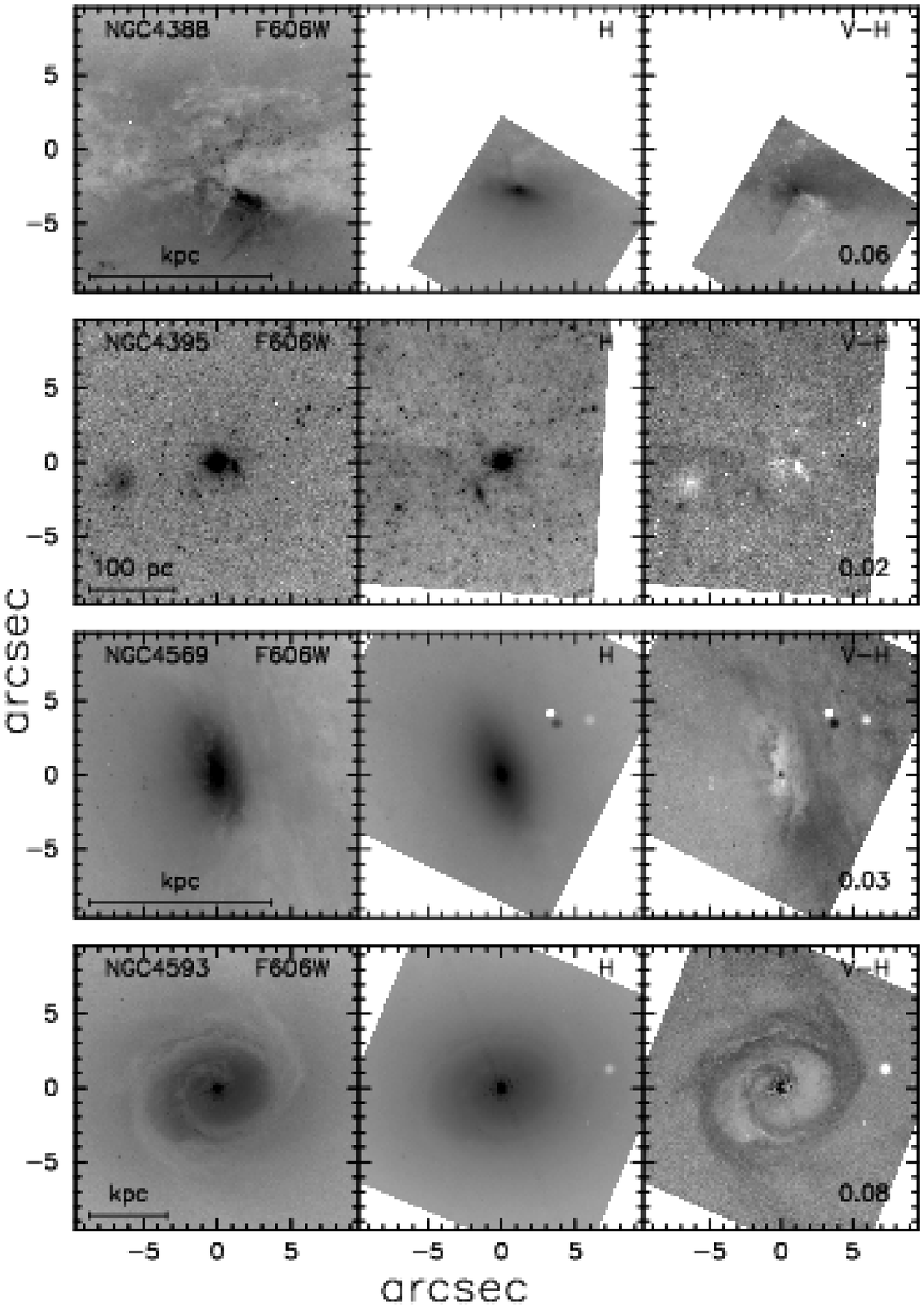}
\caption{Figure~\ref{fig:cmaps} -- {\it Continued}}
\end{figure}

\clearpage

\begin{figure}
\epsscale{0.85}
\plotone{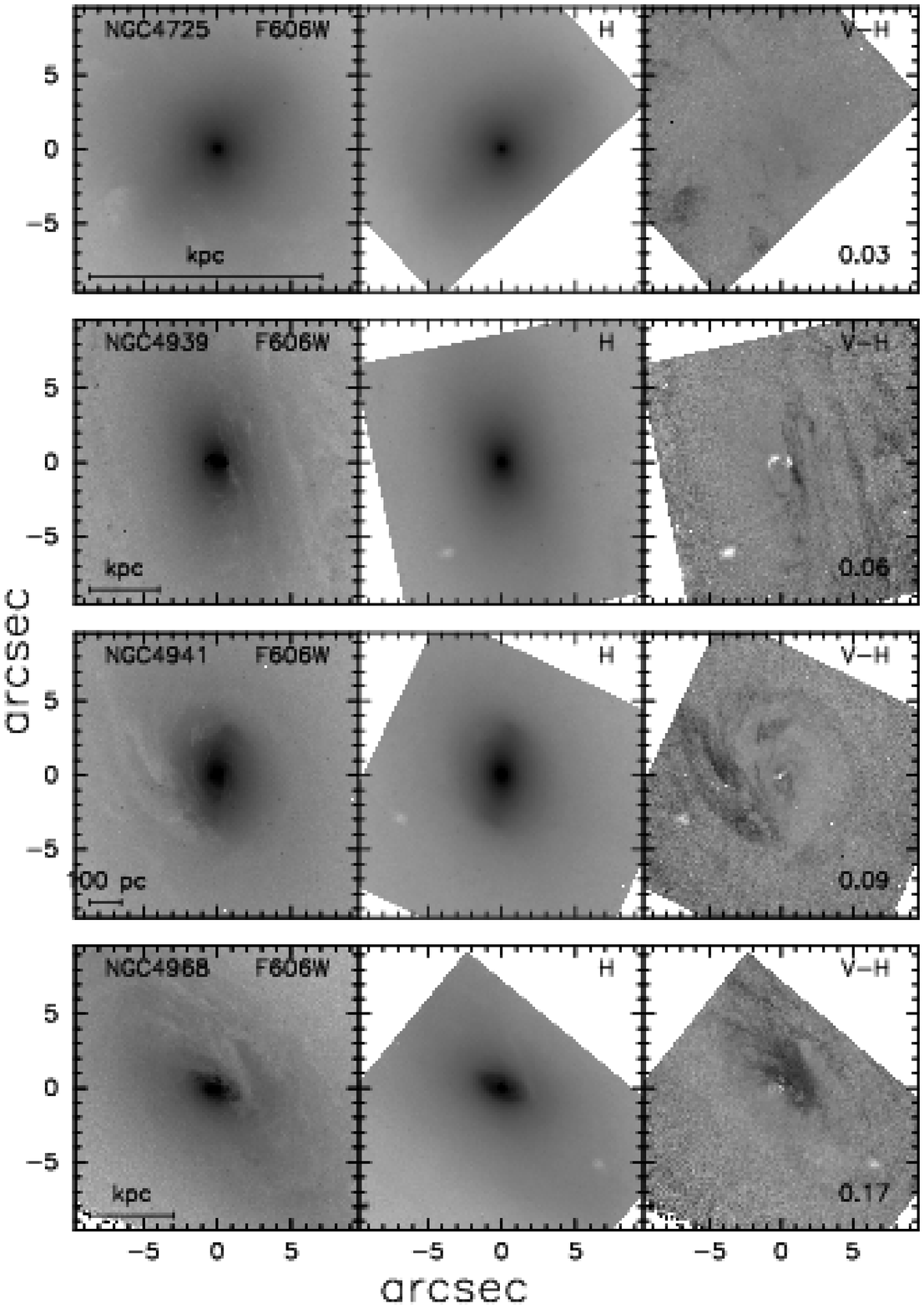}
\caption{Figure~\ref{fig:cmaps} -- {\it Continued}}
\end{figure}

\clearpage

\begin{figure}
\epsscale{0.85}
\plotone{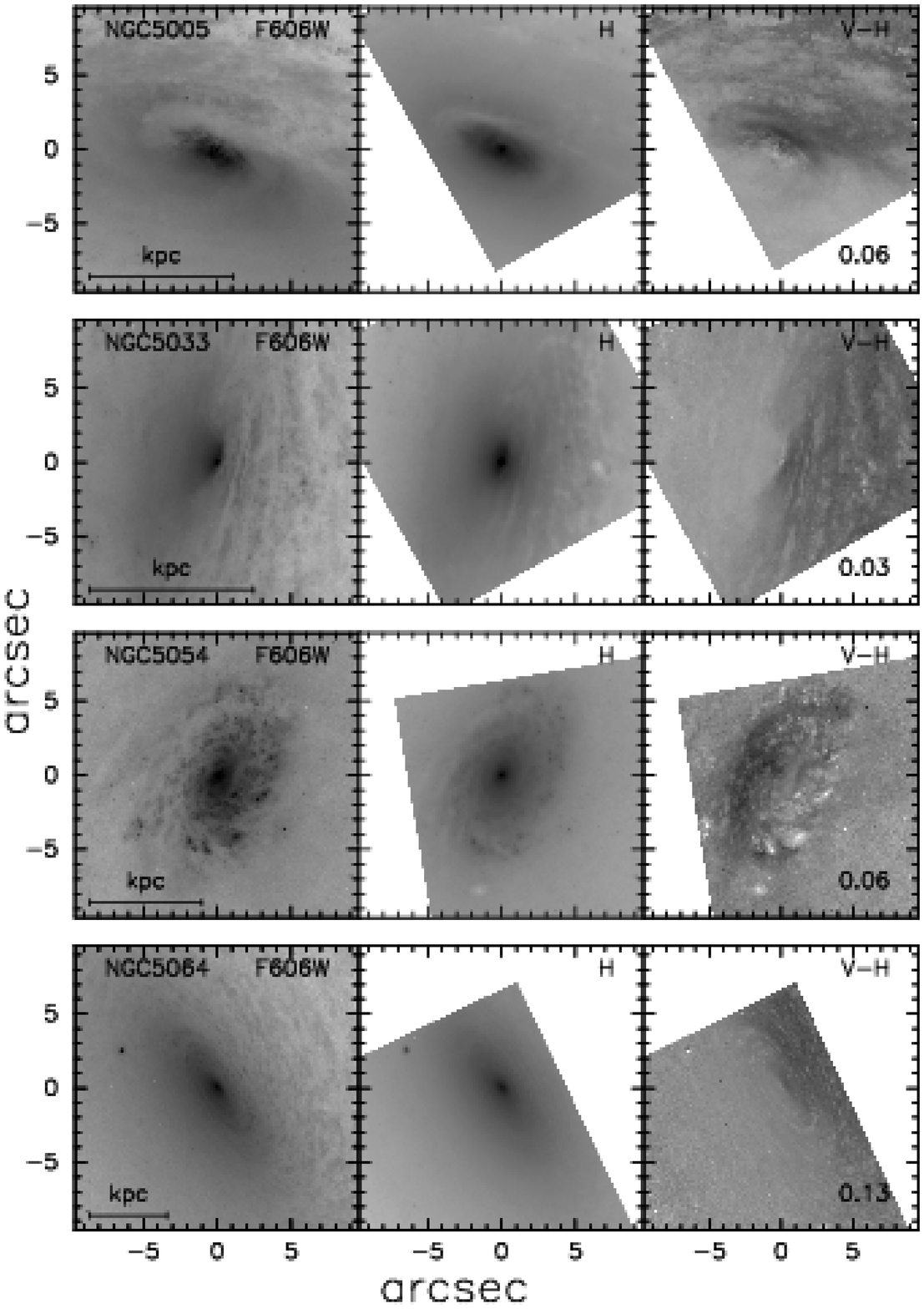}
\caption{Figure~\ref{fig:cmaps} -- {\it Continued}}
\end{figure}

\clearpage

\begin{figure}
\epsscale{0.85}
\plotone{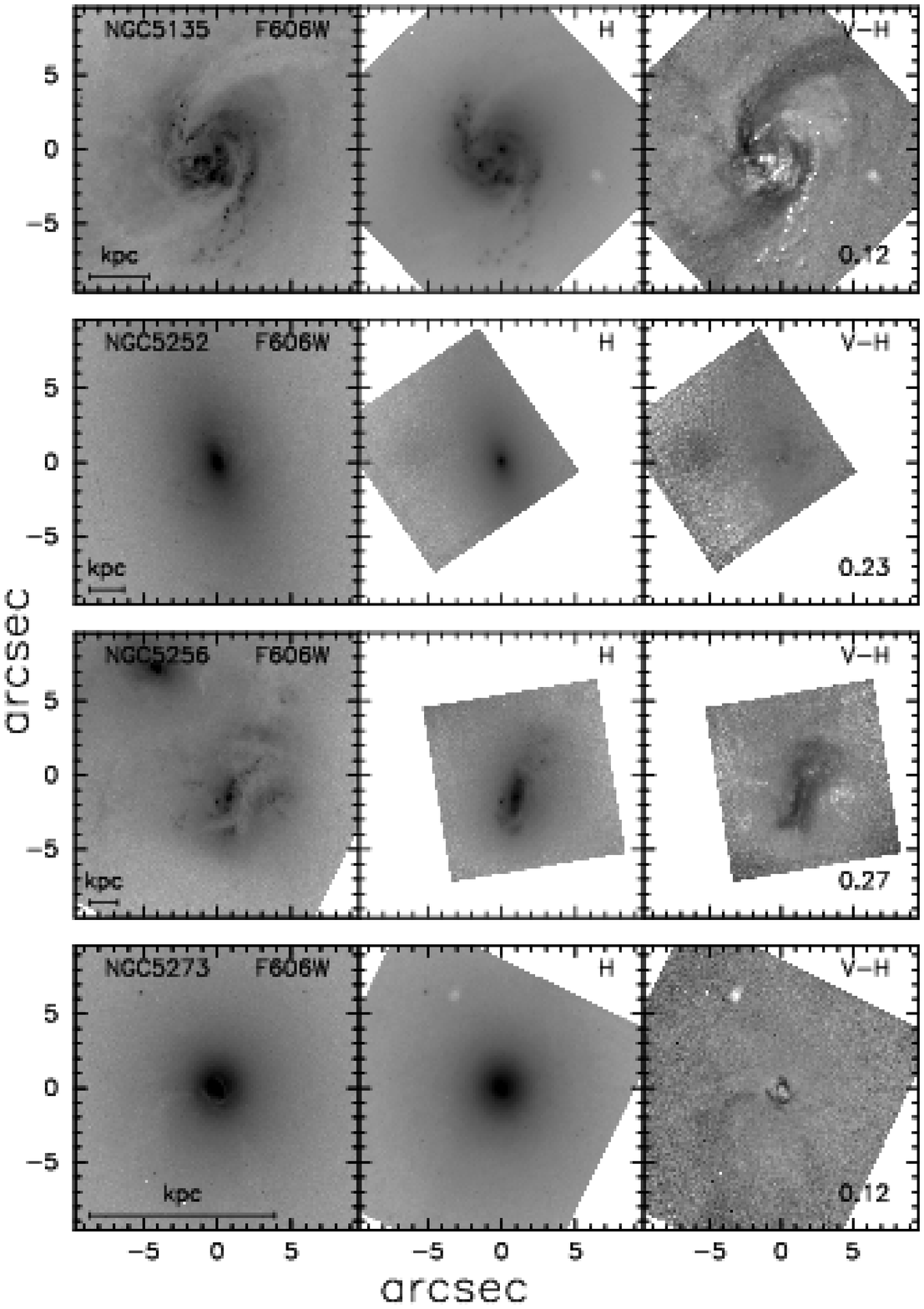}
\caption{Figure~\ref{fig:cmaps} -- {\it Continued}}
\end{figure}

\clearpage

\begin{figure}
\epsscale{0.85}
\plotone{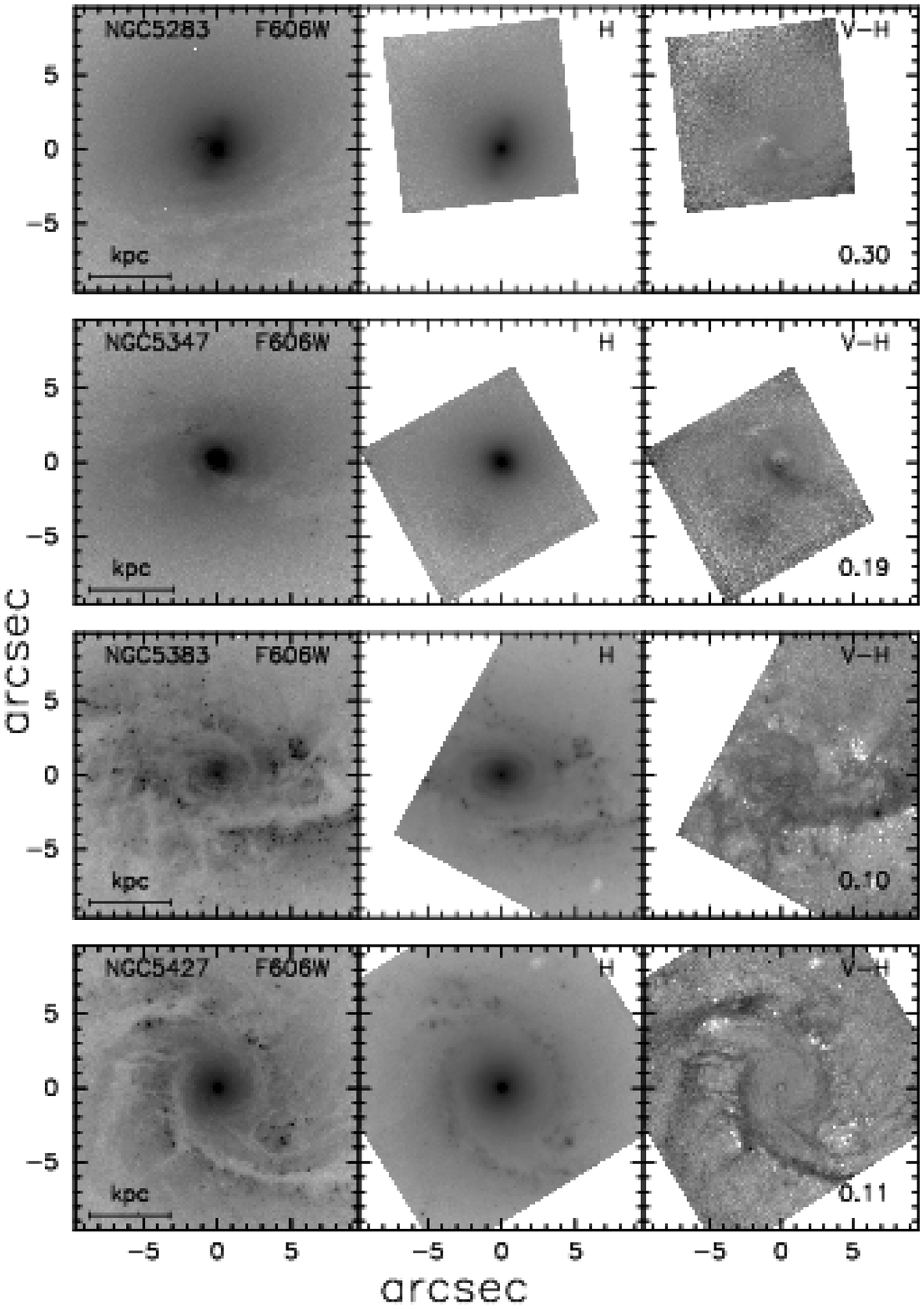}
\caption{Figure~\ref{fig:cmaps} -- {\it Continued}}
\end{figure}

\clearpage

\begin{figure}
\epsscale{0.85}
\plotone{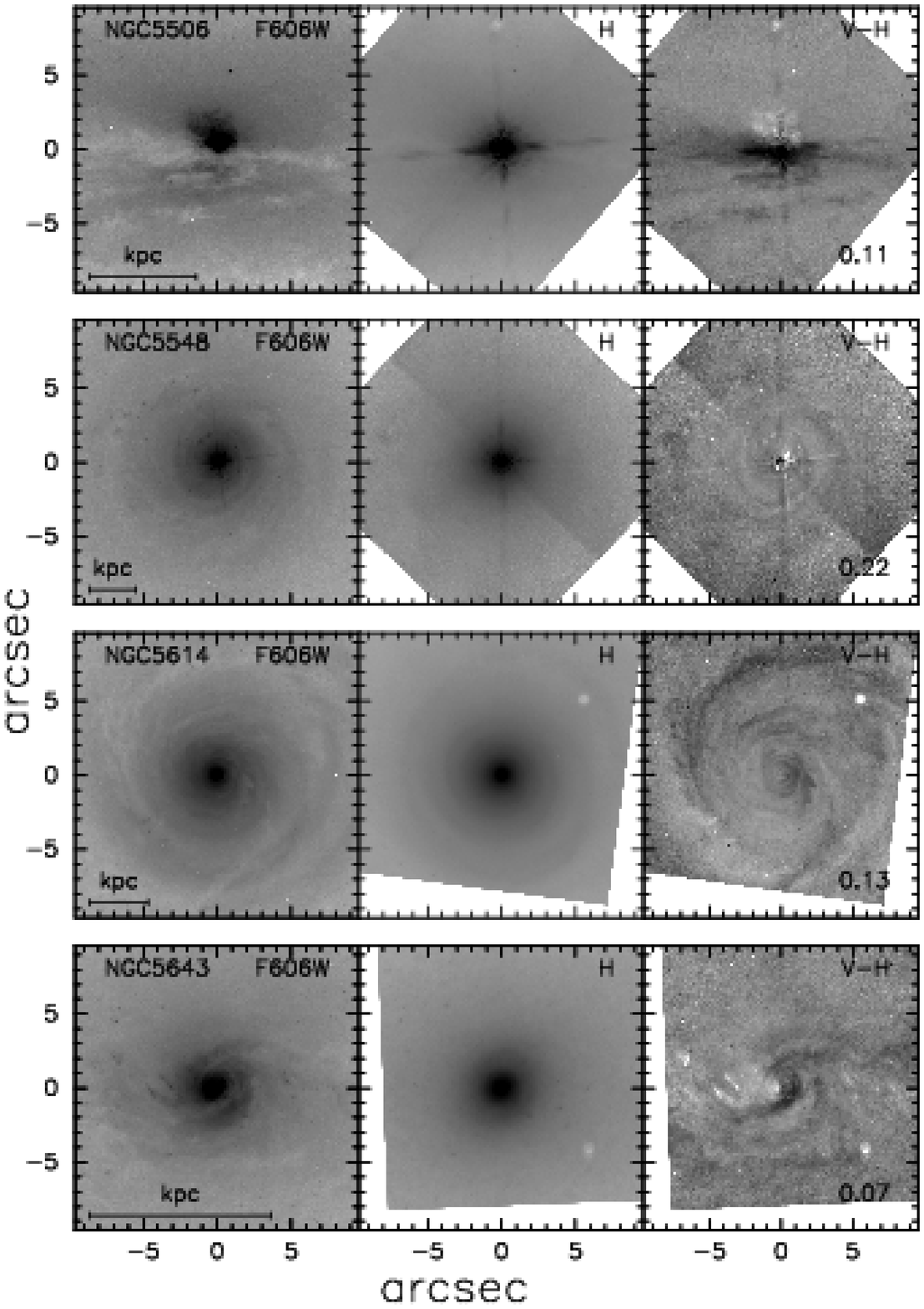}
\caption{Figure~\ref{fig:cmaps} -- {\it Continued}}
\end{figure}

\clearpage

\begin{figure}
\epsscale{0.85}
\plotone{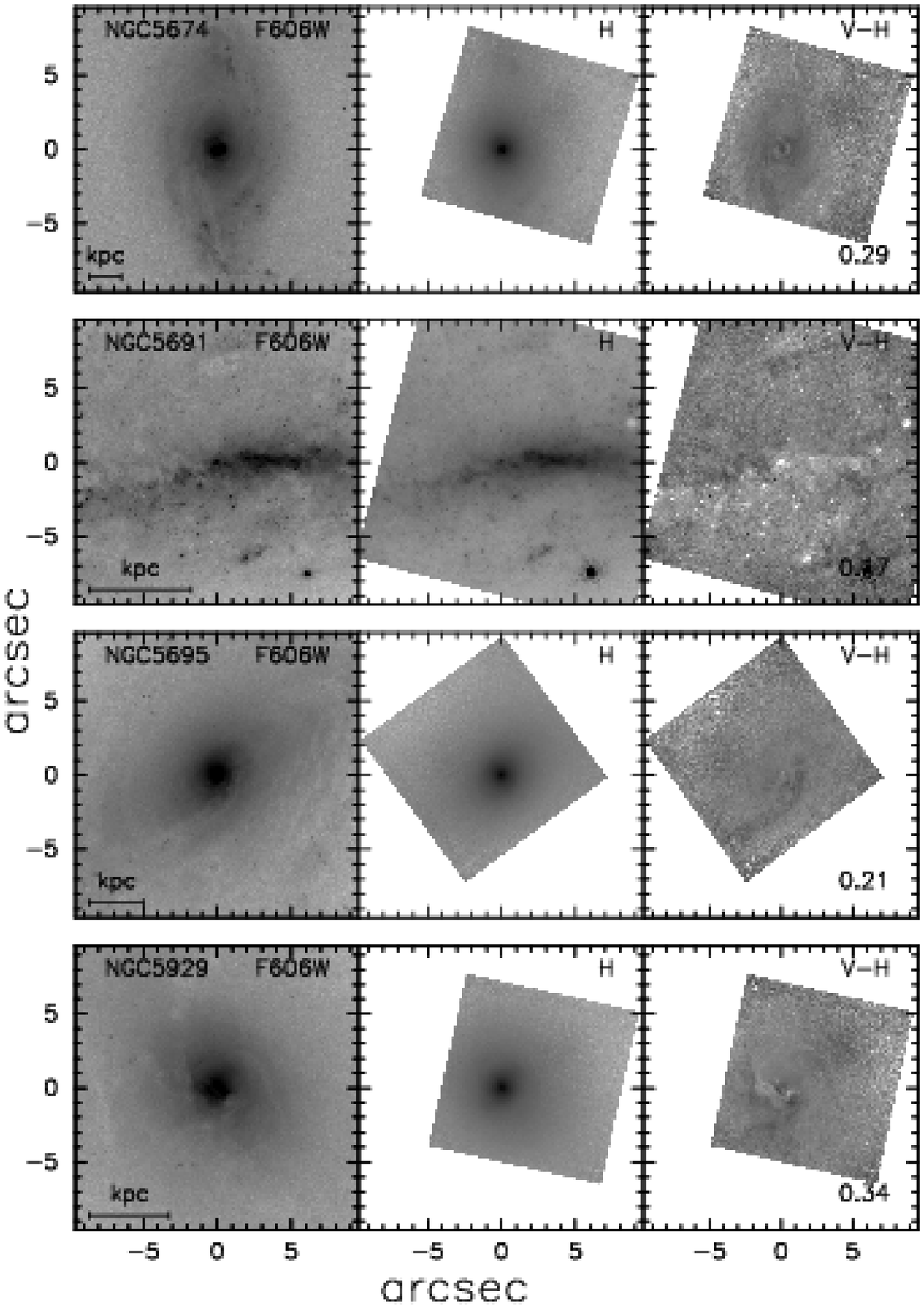}
\caption{Figure~\ref{fig:cmaps} -- {\it Continued}}
\end{figure}

\clearpage

\begin{figure}
\epsscale{0.85}
\plotone{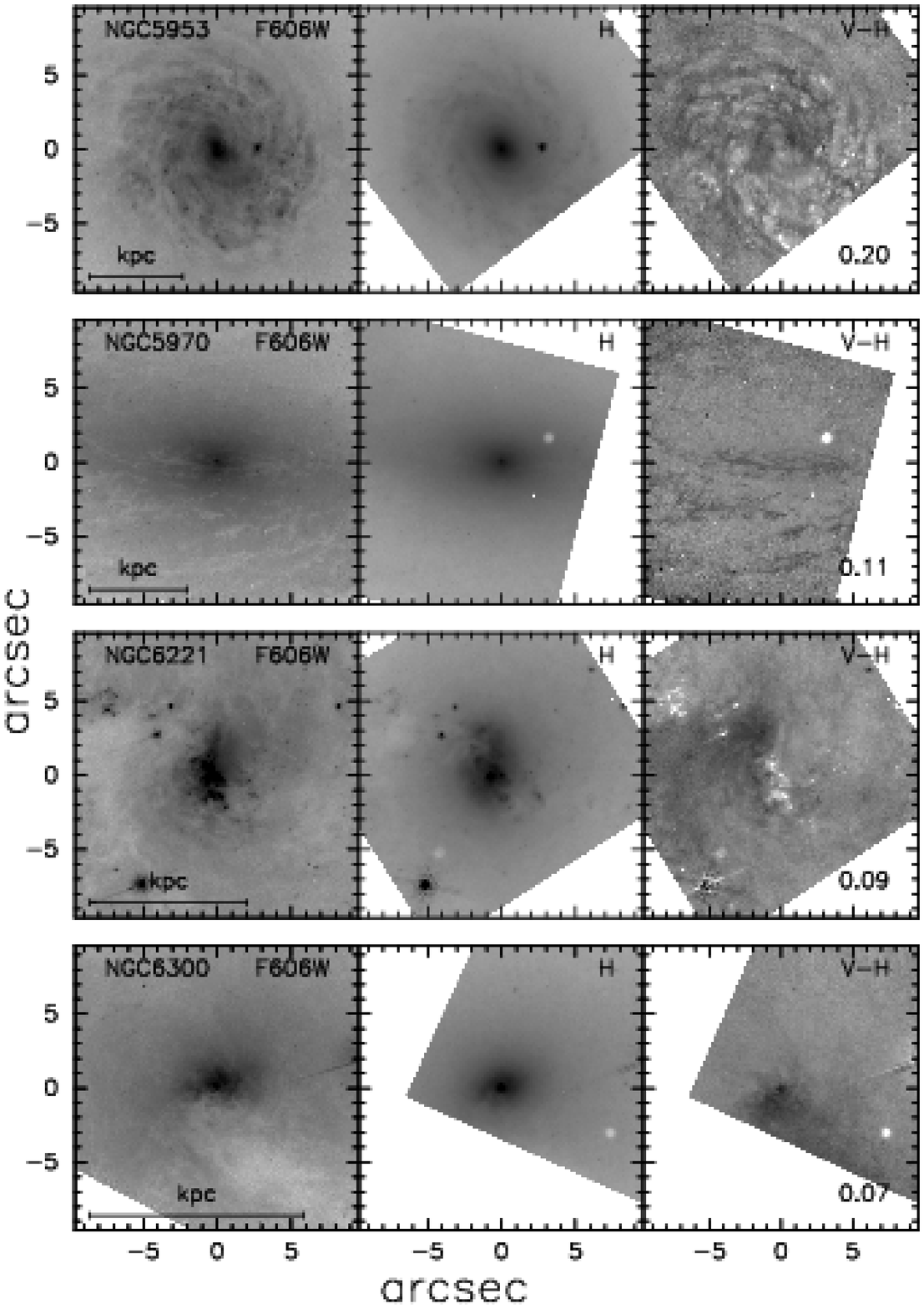}
\caption{Figure~\ref{fig:cmaps} -- {\it Continued}}
\end{figure}

\clearpage

\begin{figure}
\epsscale{0.85}
\plotone{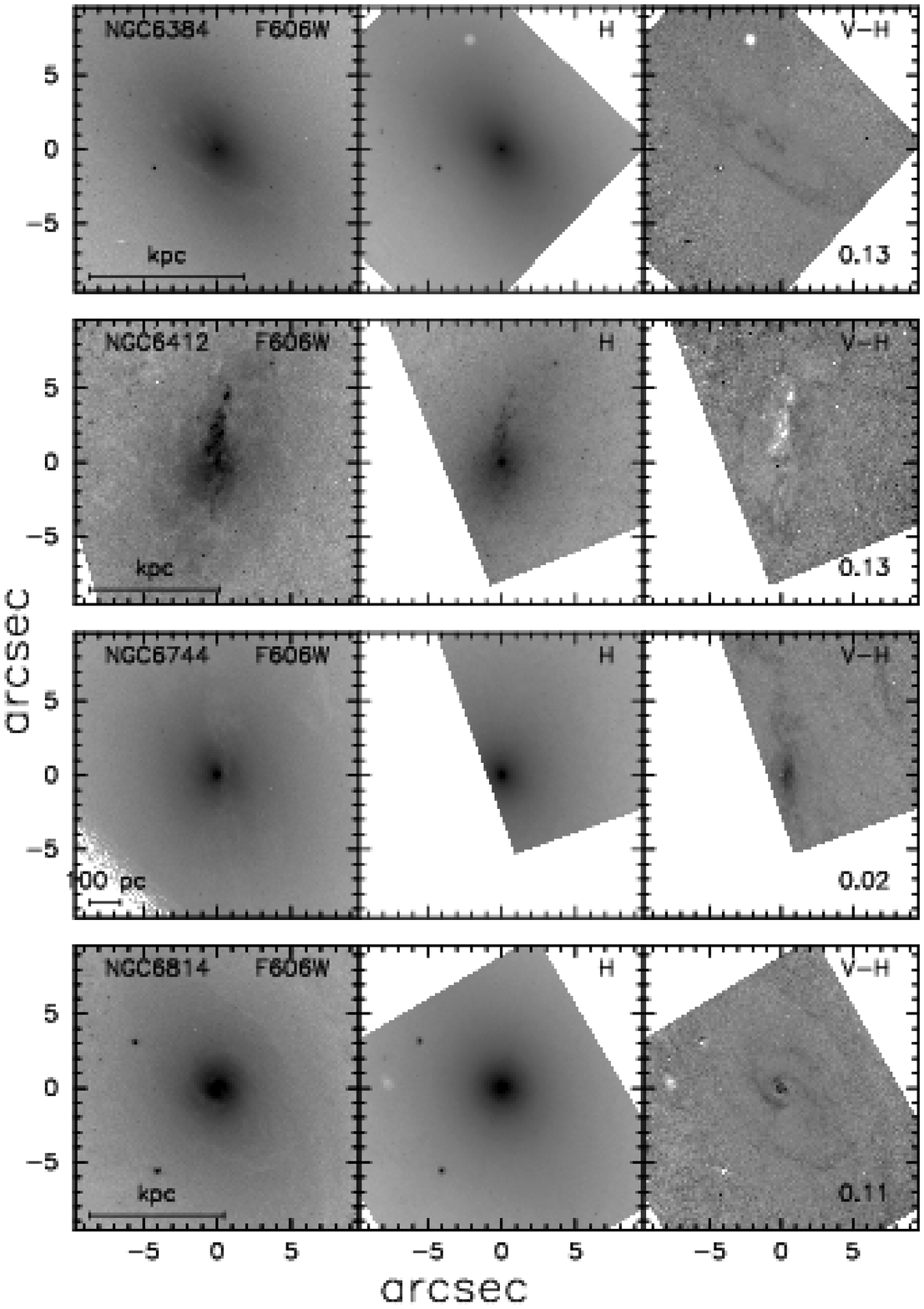}
\caption{Figure~\ref{fig:cmaps} -- {\it Continued}}
\end{figure}

\clearpage

\begin{figure}
\epsscale{0.85}
\plotone{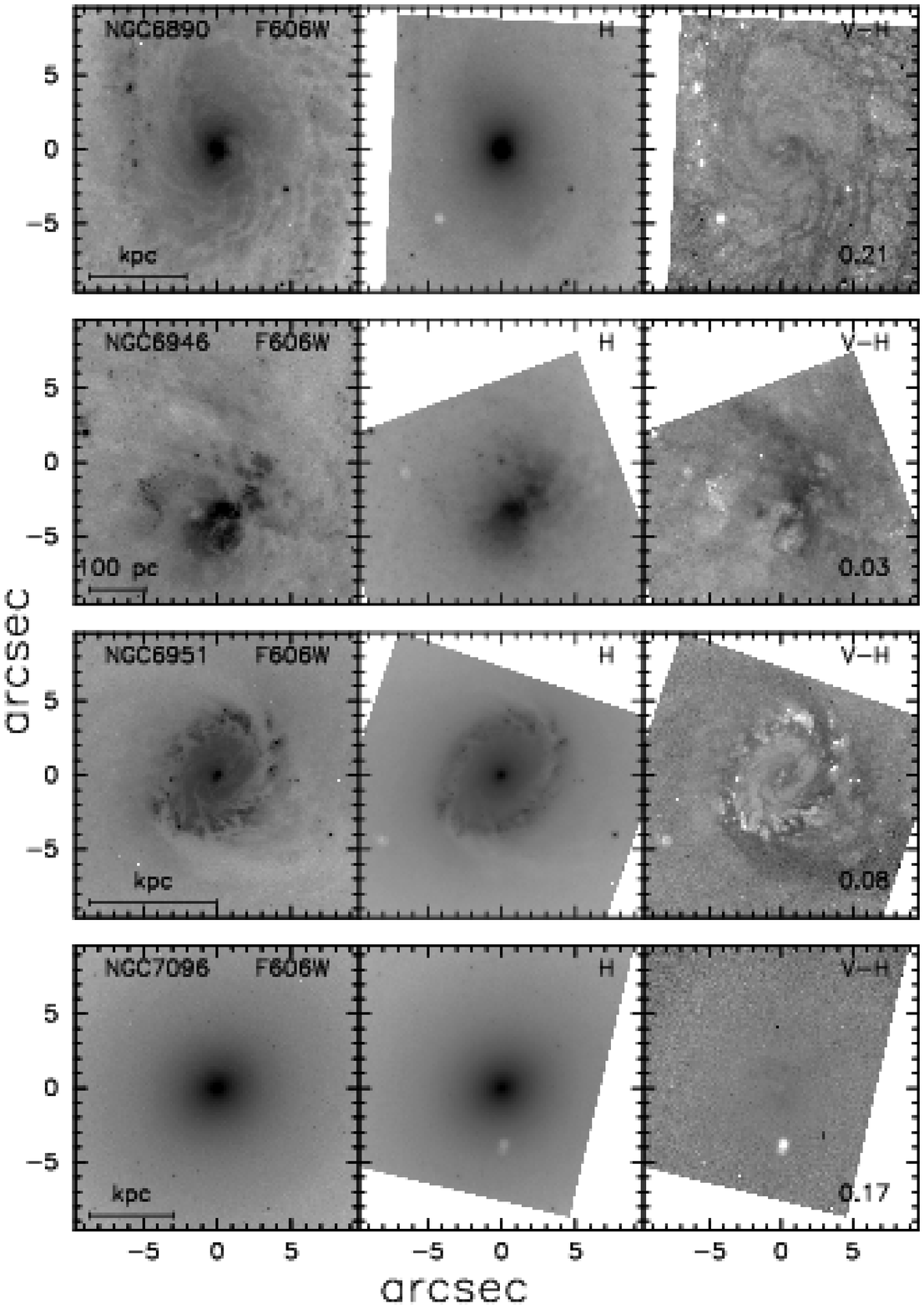}
\caption{Figure~\ref{fig:cmaps} -- {\it Continued}}
\end{figure}

\clearpage

\begin{figure}
\epsscale{0.85}
\plotone{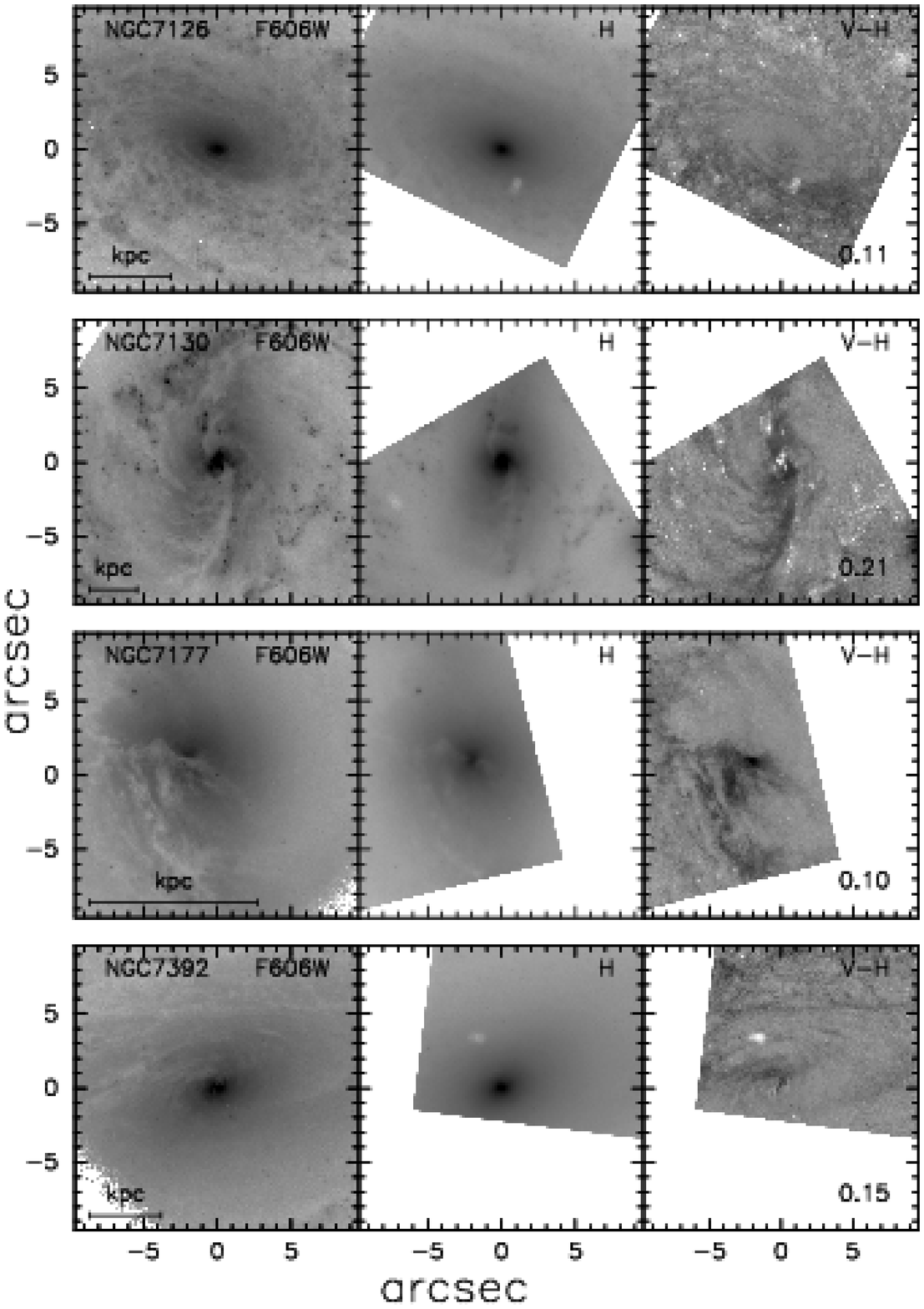}
\caption{Figure~\ref{fig:cmaps} -- {\it Continued}}
\end{figure}

\clearpage

\begin{figure}
\epsscale{0.85}
\plotone{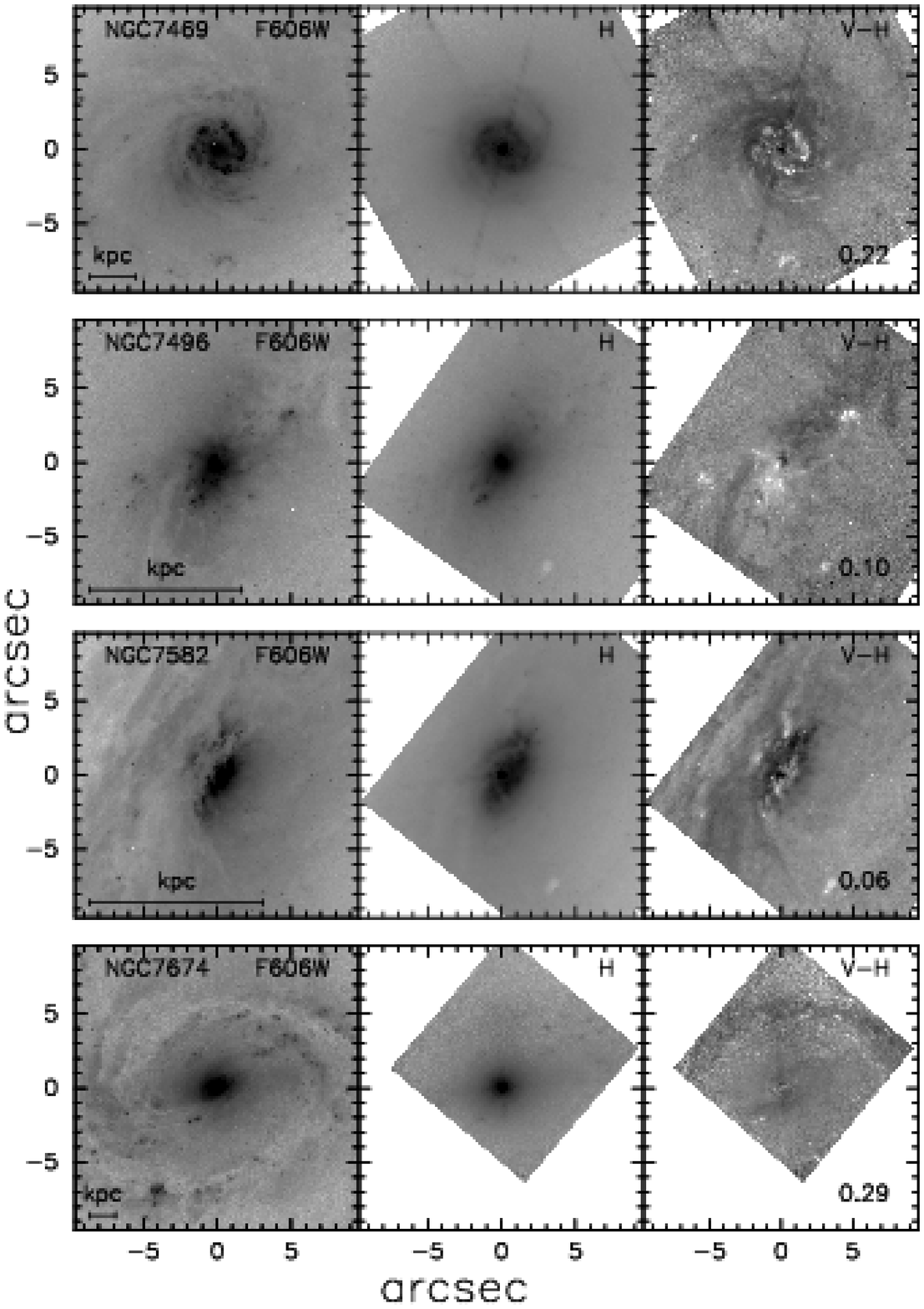}
\caption{Figure~\ref{fig:cmaps} -- {\it Continued}}
\end{figure}

\clearpage

\begin{figure}
\epsscale{0.85}
\plotone{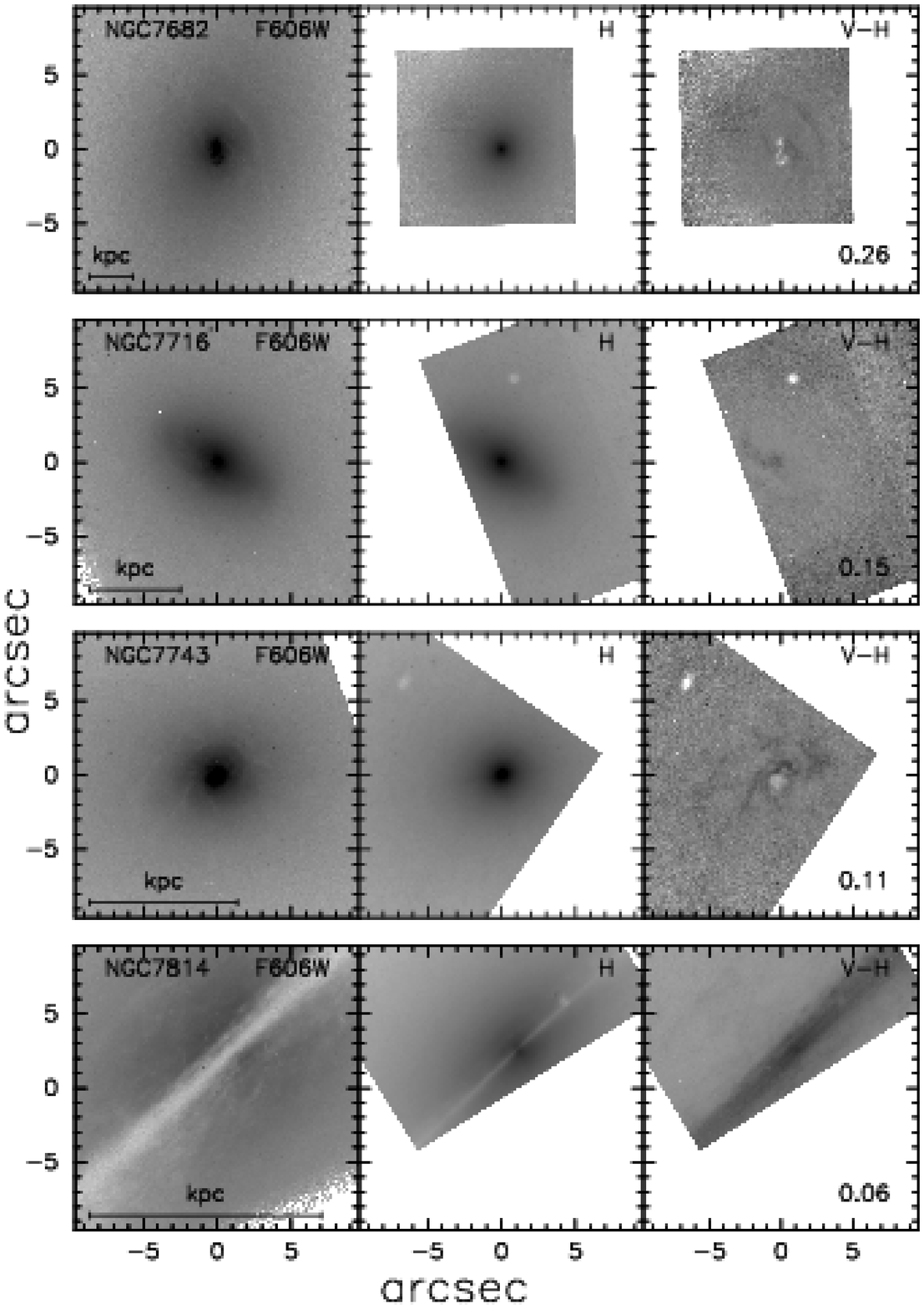}
\caption{Figure~\ref{fig:cmaps} -- {\it Continued}}
\end{figure}

\clearpage

\begin{figure}
\epsscale{0.85}
\plotone{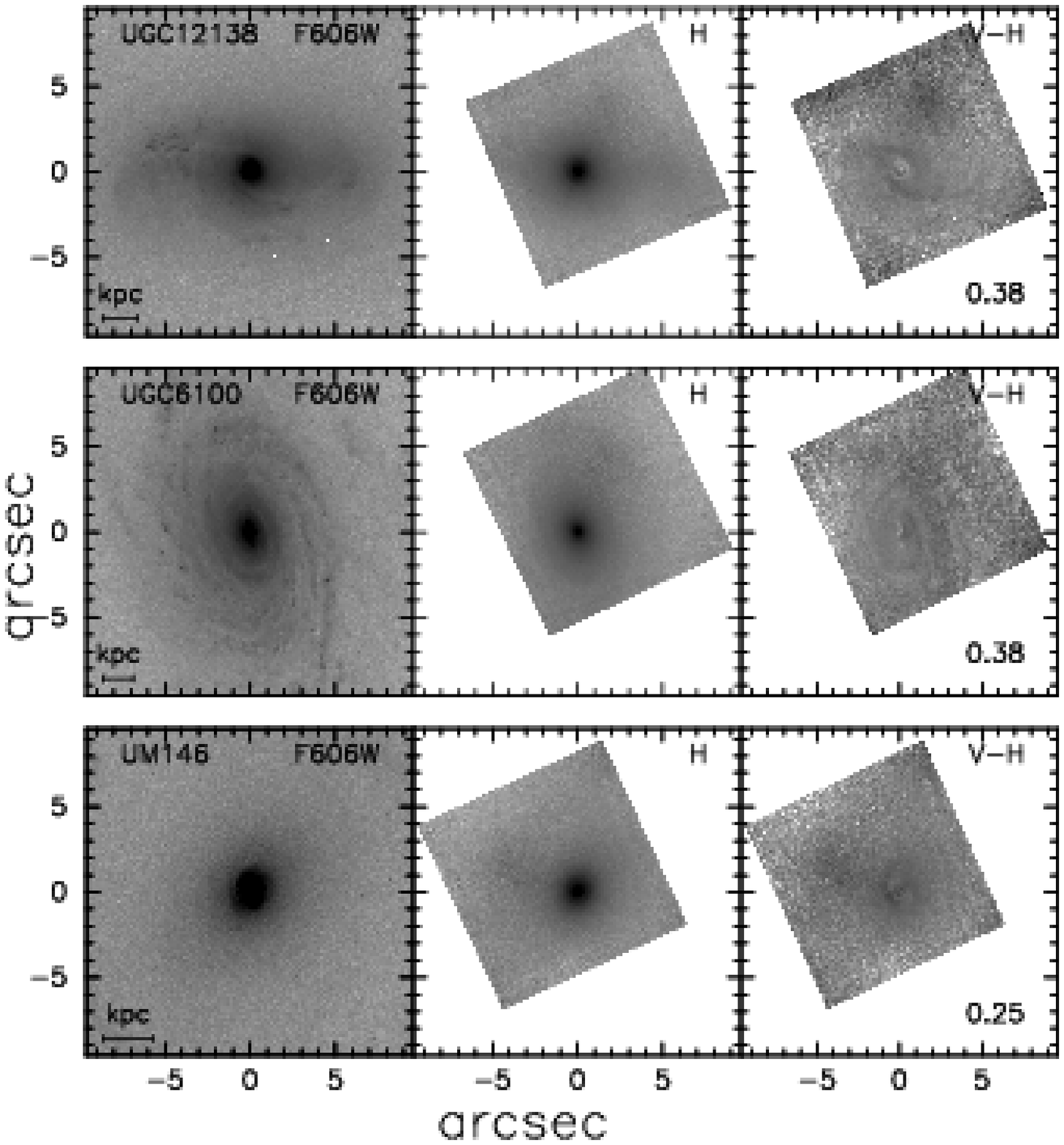}
\caption{Figure~\ref{fig:cmaps} -- {\it Continued}}
\end{figure}

\clearpage

\clearpage

\clearpage

\clearpage

\clearpage

\begin{center}
\begin{deluxetable}{llcccc} 
\tabletypesize{\scriptsize}
\tablecolumns{6} 
\tablenum{1}
\tablewidth{3.5truein}
\tablecaption{Observations \label{tbl:basic} } 
\tablehead{ 
\colhead{Name} & 
\colhead{Aliases} &  
\colhead{V Prop ID} & 
\colhead{V Filter} &  
\colhead{H Prop ID} \\
}
\startdata
ESO 137-G34 & - & 5479 & F606W & 7330 \\
ESO 138-G1  & - & 5479 & F606W & 7330 \\ 
IC 2560 & - & 8597 & F606W & 7330 \\ 
IC 5063 & PKS 2048-572 & 5479 & F606W & 7330 \\
IC 5267 & - & 8597 & F606W & 7330 \\
Mrk 334 & UGC 6 & 5479 & F606W & 7867 \\ 
Mrk 461 & UGC 8718 & 8597 & F606W & 7867 \\
Mrk 471 & UGC 9214 & 5479 & F606W & 7867 \\ 
Mrk 477 & IZw92 & 8597 & F606W & 7330 \\
Mrk 573 & UGC 1214 & 5479 & F606W & 7330 \\
Mrk 1066 & UGC 2456 & 5479 & F606W & 7330 \\
Mrk 1210 & UGC 4203 & 5479 & F606W & 7330 \\ 
NGC 214 & - & 8597 & F606W & 7330 \\
NGC 357 & - & 8597 & F606W & 7330 \\
NGC 404 & - & 8597 & F606W & 7330 \\ 
NGC 628 & M74 & 8597 & F606W & 7330 \\
NGC 788 & - & 5479 & F606W & 7330 \\ 
NGC 864 & - & 8597 & F606W & 7330 \\
NGC 1068 & M77 & 8597 & F606W & 7215 \\
NGC 1144 & - & 5479 & F606W & 7867 \\
NGC 1241 & - & 5479 & F606W & 7330 \\ 
NGC 1275 & 3C84,PersA & 6228 & F702W & 7330 \\
NGC 1300 & - & 8597 & F606W & 7330 \\
NGC 1320 & Mrk 607 & 5479 & F606W & 7330 \\
NGC 1398 & - & 8597 & F606W & 7330 \\
NGC 1530 & - & 8597 & F606W & 7330 \\
NGC 1638 & - & 8597 & F606W & 7330 \\
NGC 1667 & - & 5479 & F606W & 7330 \\
NGC 1672 & - & 8597 & F606W & 7330 \\
NGC 1961 & - & 8597 & F606W & 7330 \\
NGC 2146 & - & 8597 & F606W & 7330 \\
NGC 2179 & - & 8597 & F606W & 7330 \\
NGC 2273 & Mrk 620 & 6419 & F791W & 7172 \\
NGC 2276 & - & 8597 & F606W & 7330 \\
NGC 2336 & - & 8597 & F606W & 7330 \\
NGC 2460 & - & 8597 & F606W & 7330 \\
NGC 2639 & - & 5479 & F606W & 7330 \\
NGC 2903 & NGC 2905 & 8597 & F606W & 7330 \\
NGC 2985 & - & 5479 & F606W & 7330 \\
NGC 3032 & - & 5479 & F606W & 7330 \\
NGC 3079 & - & 8597 & F606W & 7330 \\
NGC 3081 & - & 5479 & F606W & 7330 \\
NGC 3145 & - & 8597 & F606W & 7330 \\
NGC 3227 & - & 5479 & F606W & 7172 \\ 
NGC 3300 & - & 8597 & F606W & 7330 \\
NGC 3351 & M95 & 8597 & F606W & 7330 \\
NGC 3362 & - & 5479 & F606W & 7867 \\ 
NGC 3368 & - & 8597 & F606W & 7330 \\ 
NGC 3393 & - & 5479 & F606W & 7330 \\
NGC 3458 & - & 8597 & F606W & 7330 \\
NGC 3486 & - & 8597 & F606W & 7330 \\
NGC 3516 & - & 5479 & F606W & 7330 \\ 
NGC 3627 & M66 & 8597 & F606W & 7330 \\ 
NGC 3718 & - & 6436 & F791W & 7330 \\ 
NGC 3786 & Mrk 744 & 5479 & F606W & 7867 \\ 
NGC 3865 & - & 8597 & F606W & 7330 \\
NGC 3982 & - & 5479 & F606W & 7330 \\  
NGC 4030 & - & 6359 & F606W & 7330 \\
NGC 4117 & - & 8597 & F606W & 7330 \\ 
NGC 4143 & - & 8597 & F606W & 7330 \\
NGC 4151 & - & 5433 & F547M & 7215 \\
NGC 4253 & Mrk 766 & 5479 & F606W & 7330 \\ 
NGC 4254 & M99 & 8597 & F606W & 7330 \\
NGC 4258 & M106 & 8597 & F606W & 7230 \\
NGC 4260 & - & 8597 & F606W & 7330 \\
NGC 4303 & M61 & 9042 & F814W & 7330 \\
NGC 4314 & - & 8597 & F606W & 7330 \\
NGC 4380 & - & 8597 & F606W & 7330 \\
NGC 4388 & Arp 120 & 8597 & F606W & 7867 \\
NGC 4395 & - & 8597 & F606W & 7330 \\
NGC 4569 & - & 8597 & F606W & 7331 \\
NGC 4593 & Mrk 1330 & 5479 & F606W & 7330 \\
NGC 4725 & - & 8597 & F606W & 7330 \\
NGC 4939 & - & 5479 & F606W & 7330 \\
NGC 4941 & - & 8597 & F606W & 7330 \\
NGC 4968 & - & 5479 & F606W & 7330 \\
NGC 5005 & - & 8597 & F606W & 7330 \\
NGC 5033 & - & 8597 & F606W & 7330 \\
NGC 5054 & - & 8597 & F606W & 7330 \\
NGC 5064 & - & 8597 & F606W & 7330 \\
NGC 5135 & - & 5479 & F606W & 7330 \\
NGC 5252 & - & 5479 & F606W & 7330 \\
NGC 5256 & Mrk 266SW & 5479 & F606W & 7867 \\
NGC 5273 & - & 8597 & F606W & 7330 \\
NGC 5283 & Mrk 270 & 5479 & F606W & 7867 \\
NGC 5347 & - & 5479 & F606W & 7330 \\  
NGC 5383 & - & 8597 & F606W & 7330 \\
NGC 5427 & - & 5479 & F606W & 7330 \\ 
NGC 5506 & Mrk 1376 & 5479 & F606W & 7330 \\ 
NGC 5548 & - & 5479 & F606W & 7172 \\ 
NGC 5614 & - & 8597 & F606W & 7330 \\
NGC 5643 & - & 8597 & F606W & 7330 \\
NGC 5674 & - & 5479 & F606W & 7867 \\ 
NGC 5691 & - & 8597 & F606W & 7330 \\
NGC 5695 & Mrk 686 & 5479 & F606W & 7867 \\
NGC 5929 & - & 5479 & F606W & 7330 \\  
NGC 5953 & - & 5479 & F606W & 7330 \\
NGC 5970 & - & 8597 & F606W & 7330 \\
NGC 6221 & - & 5479 & F606W & 7330 \\ 
NGC 6300 & - & 5479 & F606W & 7330 \\ 
NGC 6384 & - & 6359 & F606W & 7330 \\ 
NGC 6412 & - & 8597 & F606W & 7330 \\
NGC 6744 & - & 8597 & F606W & 7330 \\
NGC 6814 & - & 5479 & F606W & 7330 \\ 
NGC 6890 & - & 8597 & F606W & 7330 \\ 
NGC 6946 & - & 8597 & F606W & 7330 \\
NGC 6951 & - & 8597 & F606W & 7330 \\
NGC 7096 & - & 8597 & F606W & 7330 \\
NGC 7126 & - & 8597 & F606W & 7330 \\
NGC 7130 & IC 5135 & 5479 & F606W & 7330 & \\
NGC 7177 & - & 8597 & F606W & 7330 \\
NGC 7392 & - & 8597 & F606W & 7330 \\
NGC 7469 & - & 5479 & F606W & 7219 \\
NGC 7496 & - & 8597 & F606W & 7330 \\
NGC 7582 & - & 8597 & F606W & 7330 \\
NGC 7674 & Mrk 533 & 5479 & F606W & 7867 \\
NGC 7682 & - & 5479 & F606W & 7867 \\
NGC 7716 & - & 8597 & F606W & 7330 \\
NGC 7743 & - & 8597 & F606W & 7330 \\
NGC 7814 & - & 8597 & F606W & 7330 \\
UGC 12138& 2237+07 & 5479 & F606W & 7867 \\
UGC 6100 & A1058+45 & 5479 & F606W & 7867 \\
UM 146   & UGC 1395 & 5479 & F606W & 7867 \\
\enddata
\end{deluxetable}
\end{center}

\clearpage

\begin{center}
\begin{deluxetable}{lcccccccc}
\tabletypesize{\scriptsize}
\tablecolumns{9}
\tablenum{2}
\tablewidth{5.5truein}
\tablecaption{Literature Data and Sample Assignments \label{tbl:morph} }
\tablehead{
\colhead{Name} & 
\colhead{D} & 
\colhead{References} & 
\colhead{pc/$''$}& 
\colhead{Bar} & 
\colhead{Note} & 
\colhead{References} & 
\colhead{AGN} & 
\colhead{References} \\ 
}
\startdata
ESO 137-G34& 33.8&7 & 164 & B & 1 & 5 & S2 & 27 \\ 
ESO 138-G1 & 33.8&7 & 164 & 0 & 4 & 5 & S2 & 26 \\ 
IC 2560  & 34.4 & 7 & 167 & B & 1 & 1 & S2 & 23 \\ 
IC 5063  & 43.8 & 7 & 212 & 0 & 5 & 6 & S2 & 10,13,18 \\ 
IC 5267  & 21.0 & 1 & 102 & 0 & 5 & 6 & No & 24 \\ 
Mrk 334 & 88.4 & 7 & 429 & 0 & 5 & 4 & S1.8 & 20,29 \\ 
Mrk 461 & 65.6 & 7 & 318 & 0 & 5 & 4 & S2 & 28,29 \\ 
Mrk 471 &137.3 & 7 & 666 & B & 1 & 1 & S1.8 & 7,29 \\ 
Mrk 477 &153.5 & 7 & 744 & ? & 3 & . & S2 & 7 \\ 
Mrk 573 & 71.0 & 7 & 344 & X & 2 & 1 & S2 & 9,29 \\ 
Mrk 1066 & 50.9 & 7 & 247 & B & 1 & 1 & S2 & 17 \\ 
Mrk 1210 & 52.3 & 7 & 254 & 0 & 5 & 8 & S2 & 27 \\ 
NGC 214 & 63.9 & 7 & 310 & X & 2 & 1 & No & 16 \\ 
NGC 357 & 32.1 & 1 & 156 & B & 1 & 6 & No & 24 \\ 
NGC 404 &  3.3 & 6 & 16  & 0 & 5 & 8 & No(L2) & 30 \\ 
NGC 628 &  9.7 & 1 & 47  & 0 & 5 & 8 & No & 30 \\ 
NGC 788 & 55.7 & 7 & 270 & 0 & 5 & 6 & S2 & 14 \\ 
NGC 864 & 20.0 & 1 & 97  & B & 1 & 6 & No & 30 \\ 
NGC 1068 & 14.4 & 1 & 70  & B & 1 & 2 & S1.8 & 1,29,30 \\ 
NGC 1144 & 116.4& 7 & 564 & B & 1 & 1 & S2 & 14,29 \\ 
NGC 1241 & 26.6 & 1 & 129 & B & 1 & 1 & S2 & 20 \\ 
NGC 1275 & 73.0 & 7 & 354 & 0 & 5 & 3 & S1.5 & 1,30 \\ 
NGC 1300 & 15.0 & 2 &  73 & B & 1 & 1 & No & 2 \\ 
NGC 1320 & 37.7 & 1 & 183 & 0 & 4 & 1 & S2 & 22 \\
NGC 1398 & 22.2 & 2 & 108 & B & 1 & 1 & No & 2,24 \\ 
NGC 1530 & 36.6 & 1 & 177 & B & 1 & 1 & No & 2 \\ 
NGC 1638 & 44.3 & 7 & 215 & 0 & 5 & 6 & No & 32 \\ 
NGC 1667 & 60.3 & 7 & 292 & B & 1 & 6 & S2 & 10,18 \\ 
NGC 1672 & 14.5 & 1 & 70  & B & 1 & 1 & S2 & 24 \\ 
NGC 1961 & 55.2 & 7 & 268 & 0 & 5 & 8 & No(L2) & 30 \\ 
NGC 2146 & 17.2 & 1 & 83  & B & 1 & 1 & No & 30 \\ 
NGC 2179 & 34.2 & 1 & 166 & 0 & 4 & 1 & No & 10 \\ 
NGC 2273 & 28.4 & 1 & 138 & B & 1 & 6 & S2 & 14 \\ 
NGC 2276 & 36.8 & 1 & 178 & X & 2 & 1 & No & 30 \\ 
NGC 2336 & 27.2 & 2 & 132 & X & 2 & 1 & S2 & 30 \\ 
NGC 2460 & 23.6 & 1 & 114 & B & 1 & 8 & No & 29 \\ 
NGC 2639 & 45.6 & 7 & 221 & B & 1 & 7 & S1 & 14 \\ 
NGC 2903 &  7.4 & 2 & 36  & X & 2 & 1 & No & 2,30 \\ 
NGC 2985 & 22.4 & 1 & 109 & 0 & 4 & 1 & No(T1.9) & 30 \\ 
NGC 3032 & 22.0 & 6 & 107 & X & 2 & 1 & No & 32 \\ 
NGC 3079 & 17.3 & 2 & 84  & B & 1 & 1 & S2 & 28,30 \\ 
NGC 3081 & 32.5 & 1 & 158 & D & 1 & 6 & S2 & 10,18 \\ 
NGC 3145 & 45.5 & 7 & 221 & B & 1 & 1 & No & 19 \\ 
NGC 3227 & 20.6 & 1 & 100 & B & 1 & 6 & S1.5 & 1,29 \\
NGC 3300 & 42.9 & 1 & 208 & X & 2 & 1 & No & 10 \\ 
NGC 3351 & 10.1 & 4 & 49  & B & 1 & 1 & No & 30 \\ 
NGC 3362 & 108.7& 7 & 527 & X & 2 & 1 & S2 & 14,29 \\ 
NGC 3368 & 10.4 & 6 & 50  & X & 2 & 1 & No & 30 \\ 
NGC 3393 & 73.0 & 7 & 354 & B & 1 & 6 & S2 & 6 \\ 
NGC 3458 & 29.5 & 1 & 143 & X & 2 & 1 & No & 10 \\ 
NGC 3486 & 12.3 & 2 & 60  & X & 2 & 1 & S2 & 30 \\ 
NGC 3516 & 38.9 & 1 & 189 & B & 1 & 4 & S1.2 & 1,30 \\ 
NGC 3627 & 11.1 & 5 & 54  & X & 2 & 1 & S2 & 30 \\ 
NGC 3718 & 17.0 & 1 & 82  & B & 1 & 1 & L1.9 & 30 \\ 
NGC 3786 & 36.1 & 7 & 175 & B & 1 & 8 & S1.8 & 14 \\ 
NGC 3865 & 73.1 & 7 & 354 & 0 & 5 & 6 & No & 10 \\ 
NGC 3982 & 17.0 & 1 & 82  & B & 1 & 6 & S2 & 10,18 \\ 
NGC 4030 & 25.9 & 1 & 126 & 0 & 5 & 8 & No & 24 \\ 
NGC 4117 & 15.5 & 3 & 75  & 0 & 5 & 8 & S2 & 14 \\ 
NGC 4143 & 15.9 & 6 & 77  & X & 2 & 1 & L1.9 & 30 \\ 
NGC 4151 & 20.3 & 1 & 98  & ? & 3 & 4,6 & S1 & 1,28 \\ 
NGC 4253 & 50.7 & 7 & 246 & B & 1 & 4 & S1 & 5 \\ 
NGC 4254 & 16.8 & 1 & 81  & 0 & 5 & 8 & No & 30 \\ 
NGC 4258 &  7.3 & 6 & 35  & X & 2 & 1 & S1.9 & 2,28,30 \\ 
NGC 4260 & 35.1 & 1 & 170 & B & 1 & 1 & No & 8 \\ 
NGC 4303 & 15.2 & 1 & 74  & X & 2 & 1 & L & 30 \\ 
NGC 4314 &  9.7 & 1 & 47  & B & 1 & 1 & No(L2) & 2,30 \\ 
NGC 4380 & 16.8 & 1 & 81  & 0 & 4 & 1 & No & 30 \\ 
NGC 4388 & 16.8 & 1 & 81  & 0 & 5 & 4 & S1.9 & 10,14,18,30 \\ 
NGC 4395 &  3.6 & 1 & 17  & 0 & 5 & 4 & S1.8 & 28,30 \\ 
NGC 4569 & 16.8 & 1 & 81  & X & 2 & 1 & No & 30 \\ 
NGC 4593 & 39.5 & 1 & 192 & B & 1 & 6 & S1 & 9 \\ 
NGC 4725 & 13.2 & 6 & 64  & X & 2 & 1 & S2 & 30 \\ 
NGC 4939 & 44.3 & 1 & 215 & 0 & 5 & 8 & S2 & 24 \\ 
NGC 4941 &  9.7 & 2 & 47  & X & 2 & 1 & S2 & 15,20 \\ 
NGC 4968 & 36.2 & 7 & 176 & X & 2 & 1 & S2 & 23 \\ 
NGC 5005 & 21.3 & 1 & 103 & X & 2 & 1 & L1.9 & 2,30 \\ 
NGC 5033 & 18.7 & 1 & 91  & 0 & 5 & 4 & S1.5 & 15,29,30 \\ 
NGC 5054 & 27.3 & 1 & 132 & 0 & 5 & 8 & No & 24 \\ 
NGC 5064 & 39.5 & 1 & 192 & 0 & 5 & 8 & L  & 24,31 \\ 
NGC 5135 & 51.6 & 7 & 250 & B & 1 & 6 & S2 & 10,18 \\ 
NGC 5252 & 90.7 & 7 & 440 & 0 & 5 & 4 & S1.9 & 6,14,29 \\ 
NGC 5256 & 111.0& 7 & 538 & 0 & 5 & 4 & S2 & 17,29 \\ 
NGC 5273 & 16.5 & 6 & 80  & 0 & 5 & 6 & S1.5 & 14,29,30 \\ 
NGC 5283 & 38.2 & 7 & 185 & 0 & 5 & 8 & S2 & 5,29 \\ 
NGC 5347 & 36.7 & 1 & 178 & B & 1 & 6 & S2 & 10,18,29 \\ 
NGC 5383 & 37.8 & 1 & 183 & B & 1 & 1 & No & 30 \\ 
NGC 5427 & 38.1 & 1 & 185 & 0 & 5 & 6 & S2 & 20 \\ 
NGC 5506 & 28.7 & 1 & 139 & ? & 3 & . & S1.9 & 4 \\ 
NGC 5548 & 66.4 & 7 & 322 & 0 & 5 & 6 & S1 & 3 \\ 
NGC 5614 & 52.5 & 7 & 255 & 0 & 5 & 8 & No & 21 \\ 
NGC 5643 & 16.9 & 1 & 82  & B & 1 & 6 & S2 & 10,18 \\ 
NGC 5674 & 98.1 & 7 & 476 & B & 1 & 1 & S1.9 & 14,29 \\ 
NGC 5691 & 30.2 & 1 & 146 & B & 1 & 6 & No & 10 \\ 
NGC 5695 & 56.9 & 7 & 276 & B & 1 & 4 & S2 & 17,29 \\ 
NGC 5929 & 38.5 & 1 & 187 & 0 & 5 & 8 & S2 & 14,29 \\ 
NGC 5953 & 33.0 & 1 & 160 & 0 & 4 & 1 & S2 & 21 \\ 
NGC 5970 & 31.6 & 1 & 153 & B & 1 & 1 & No(T2/L2:) & 30 \\ 
NGC 6221 & 19.4 & 1 & 94  & B & 1 & 1 & S2 & 11 \\ 
NGC 6300 & 14.3 & 1 & 69  & B & 1 & 6 & S2 & 10,18 \\ 
NGC 6384 & 19.8 & 2 & 96  & X & 2 & 1 & No(T2) & 30 \\ 
NGC 6412 & 23.5 & 1 & 114 & B & 1 & 6 & No & 30 \\ 
NGC 6744 & 10.4 & 1 & 50  & X & 2 & 1 & L  & 24,31 \\ 
NGC 6814 & 22.8 & 1 & 111 & B & 1 & 6 & S1.5 & 3,24 \\ 
NGC 6890 & 31.8 & 1 & 154 & B & 1 & 6 & S2 & 10,18 \\ 
NGC 6946 &  5.5 & 1 &  27 & X & 2 & 1 & No & 30 \\ 
NGC 6951 & 24.1 & 1 & 117 & B & 1 & 6 & S2 & 2,30 \\ 
NGC 7096 & 36.7 & 1 & 178 & 0 & 5 & 8 & No & 10 \\ 
NGC 7126 & 37.5 & 1 & 182 & 0 & 4 & 1 & No & 10,24 \\ 
NGC 7130 & 64.7 & 7 & 314 & B & 1 & 6 & S2 & 10,18 \\ 
NGC 7177 & 18.2 & 1 & 88  & X & 2 & 1 & No(T2) & 30 \\ 
NGC 7392 & 43.0 & 7 & 208 & B & 1 & 9 & No & 25 \\ 
NGC 7469 & 66.9 & 7 & 324 & B & 1 & 4 & S1.2 & 1,3 \\ 
NGC 7496 & 20.1 & 1 & 97  & B & 1 & 6 & S2 & 24 \\ 
NGC 7582 & 17.6 & 1 & 85  & B & 1 & 1 & S2 & 12 \\ 
NGC 7674 & 118.5& 7 & 575 & B & 1 & 4 & S2 & 17,29 \\ 
NGC 7682 & 70.8 & 7 & 343 & B & 1 & 1 & S2 & 14,29 \\ 
NGC 7716 & 33.7 & 1 & 163 & B & 1 & 6 & No & 24 \\ 
NGC 7743 & 20.7 & 6 & 100 & B & 1 & 1 & S2 & 15,28,30 \\ 
NGC 7814 & 13.2 & 6 & 64  & 0 & 4 & 1 & No(L2::) & 30 \\ 
UGC 12138& 102.8& 7 & 498 & B & 1 & 1 & S1.8 & 14,29 \\ 
UGC 6100 & 117.6& 7 & 570 & 0 & 5 & 4 & S2 & 14,29 \\ 
UM 146   & 71.6 & 7 & 347 & B & 1 & 4 & S1.9 & 14 \\ 
\enddata
\tablecomments{
Literature data and references for the distance (col.~2), derived 
projected spatial scale (col.~4), bar (col.~5), and nuclear activity 
class (col.~8) for each galaxy in the sample. 
Distance references: (1) \citet{tully88}; (2) \citet{tully92}; 
(3) \citet{tully96}; (4) \citet{graham97}; (5) \citet{saha99}; 
(6) \citet{tonry01}; (7) \citet{yahil77}. 
Bar Notes: (1) Bar in RC3 or other survey; (2) weak bar in RC3; 
(3) Bar classification unknown; (4) unbarred at visible wavelengths, \eg RC3; 
(5) unbarred in near-infrared image. All galaxies with flags 1 and 2 were 
included in the barred sample, those with class 4 or 5 were included in the 
unbarred sample. Galaxies in class 3 were not included in the bar discussion. 
Bar references: (1) RC3; (2); \citet{scoville88}; (3) \citet{poulain92}; 
(4) \citet{mcleod95}; (5) \citet{mulchaey96}; (6) \citet{mulchaey97b}; 
(7) \citet{marquez99}; (8) \citet{laine02}; (9) \citet{buta95}. 
AGN references: (1) \citet{seyfert43}; (2); \citet{burbidge62}; 
(3) \citet{anderson70}; (4) \citet{wilson76}; (5) \citet{markarian77}; 
(6) \citet{bohuski78}; (7) \citet{debruyn78}; (8) \citet{eastmond78}; 
(9) \citet{koski78}; (10) \citet{sandage78}; (11) \citet{phillips79}; 
(12) \citet{ward80}; (13) \citet{veron81}; (14) \citet{huchra82}; 
(15) \citet{stauffer82}; (16) \citet{huchra83}; (17) \citet{osterbrock83}; 
(18) \citet{phillips83}; (19) \citet{dacosta84}; (20) \citet{dahari85}; 
(21) \citet{keel85}; (22) \citet{derobertis86}; (23) \citet{fairall86}; 
(24) \citet{veron86}; (25) \citet{bica87}; (26) \citet{fairall88}; 
(27) \citet{hewitt91}; (28) \citet{huchra92}; (29) \citet{osterbrock93}; 
(30) \citet{ho97b}; (31) \citet{vaceli97}; (32) \citet{trager98}. 
All galaxies with ``No'' in column~8 are considered to be inactive. 
}
\end{deluxetable}
\end{center}

\clearpage

\begin{center}
\begin{deluxetable}{llcccccccc}
\tabletypesize{\scriptsize}
\tablecolumns{10}
\tablenum{3}
\tablewidth{5.5truein}
\tablecaption{Morphological Classifications \label{tbl:nc} }
\tablehead{
\colhead{Name}   &
\colhead{RC3}   &
\colhead{T}   &
\colhead{$B_T$}   &
\colhead{$M_B$}   &
\colhead{$v$}   &
\colhead{D$_{25}$}   &
\colhead{20$''$frac}   &
\colhead{R$_{25}$}   &
\colhead{Nuc Class} \\
}
\startdata
ESO 137-G34& .S?.... & 0.0 & 10.1 & -22.8 & 2747 & 0.98 & 0.35 & 0.45 & HI/C \\ 
ESO 138-G1 & .E?.... & -3.0 & 13.3 & -19.5 & 2740 & 1.02 & 0.32 & 0.30 & C \\ 
IC 2560 & PSBR3*. & 3.3 & 11.9 & -20.8 & 2925 & 1.50 & 0.11 & 0.20 & LW \\ 
IC 5063 & .LAS+*. & -0.8 & 12.6 & -20.7 & 3402 & 1.33 & 0.16 & 0.17 & C \\ 
IC 5267 & .SAS0.. & 0.0 & 11.3 & -20.3 & 1713 & 1.72 & 0.06 & 0.13 & C \\ 
Mrk 334 & .P..... & 0.0 & 14.4 & -20.4 & 6605 & 0.99 & 0.34 & 0.14 & LW \\ 
Mrk 461 & .S..... & 0.0 & 14.3 & -19.7 & 4894 & 0.87 & 0.45 & 0.13 & LW \\ 
Mrk 471 & .SB.1.. & 1.0 & 14.2 & -21.5 & 10263 & 0.95 & 0.37 & 0.19 & GD \\ 
Mrk 477 & .S...*P & 0.0 & 15.2 & -20.7 & 11379 & 0.62 & 0.80 & 0.11 & LW \\ 
Mrk 573 & RLXT+*. & -1.0 & 13.6 & -20.7 & 5178 & 1.13 & 0.25 & 0.01 & GD \\ 
Mrk 1066 & RLBS+.. & -1.0 & 13.0 & -20.4 & 3605 & 1.24 & 0.19 & 0.25 & LW \\ 
Mrk 1210 & .S?.... & 1.0 & 14.2 & -19.4 & 4046 & 0.91 & 0.41 & 0.00 & TW \\ 
NGC 214 & .SXR5.. & 5.0 & 12.6 & -21.4 & 4533 & 1.27 & 0.18 & 0.13 & LW \\ 
NGC 357 & .SBR0*. & 0.0 & 12.6 & -19.9 & 2541 & 1.38 & 0.14 & 0.14 & N \\ 
NGC 404 & .LAS-*. & -3.0 & 10.9 & -16.7 & -48 & 1.54 & 0.10 & 0.00 & C \\ 
NGC 628 & .SAS5.. & 5.0 & 9.8 & -20.2 & 656 & 2.02 & 0.03 & 0.04 & N \\ 
NGC 788 & .SAS0*. & 0.0 & 12.8 & -20.9 & 4078 & 1.28 & 0.17 & 0.12 & LW \\ 
NGC 864 & .SXT5.. & 5.0 & 11.3 & -20.2 & 1550 & 1.67 & 0.07 & 0.12 & LW \\ 
NGC 1068 & RSAT3.. & 3.0 & 9.5 & -21.3 & 1109 & 1.85 & 0.05 & 0.07 & CS \\ 
NGC 1144 & .RING.B & -5.0 & 13.2 & -22.1 & 8641 & 1.04 & 0.30 & 0.21 & TW \\ 
NGC 1241 & .SBT3.. & 3.0 & 12.1 & -20.0 & 4030 & 1.45 & 0.12 & 0.22 & GD \\ 
NGC 1275 & .P..... & 99.0 & 12.6 & -21.7 & 5260 & 1.34 & 0.15 & 0.12 & C \\ 
NGC 1300 & .SBT4.. & 4.0 & 10.8 & -20.1 & 16 & 1.79 & 0.05 & 0.18 & GD \\ 
NGC 1320 & .S..1*/ & 1.0 & 13.2 & -19.7 & 2716 & 1.28 & 0.17 & 0.47 & HI/TW \\ 
NGC 1398 & PSBR2.. & 2.0 & 10.4 & -21.3 & 1407 & 1.85 & 0.05 & 0.12 & N \\ 
NGC 1530 & .SBT3.. & 3.0 & 11.4 & -21.3 & 2461 & 1.66 & 0.07 & 0.28 & GD \\ 
NGC 1638 & .LXT0?. & -2.3 & 12.8 & -20.4 & 3320 & 1.30 & 0.17 & 0.13 & N \\ 
NGC 1667 & .SXR5.. & 5.0 & 12.4 & -21.5 & 4547 & 1.25 & 0.19 & 0.11 & LW \\ 
NGC 1672 & .SBS3.. & 3.0 & 10.2 & -20.6 & 1350 & 1.82 & 0.05 & 0.08 & LW \\ 
NGC 1961 & .SXT5.. & 5.0 & 11.0 & -22.7 & 3930 & 1.66 & 0.07 & 0.19 & TW \\ 
NGC 2146 & .SBS2P. & 2.0 & 10.6 & -20.6 & 893 & 1.78 & 0.06 & 0.25 & C \\ 
NGC 2179 & .SAS0.. & 0.0 & 12.8 & -19.8 & 2798 & 1.23 & 0.20 & 0.16 & TW \\ 
NGC 2273 & .SBR1*. & 0.5 & 12.0 & -20.2 & 1840 & 1.51 & 0.10 & 0.12 & CS \\ 
NGC 2276 & .SXT5.. & 5.0 & 11.8 & -21.1 & 2417 & 1.45 & 0.12 & 0.02 & CS \\ 
NGC 2336 & .SXR4.. & 4.0 & 10.6 & -21.6 & 2200 & 1.85 & 0.05 & 0.26 & C \\ 
NGC 2460 & .SAS1.. & 1.0 & 12.3 & -19.6 & 1451 & 1.39 & 0.14 & 0.12 & CS \\ 
NGC 2639 & RSAR1*\$ & 1.0 & 12.2 & -21.1 & 3336 & 1.26 & 0.18 & 0.22 & GD \\ 
NGC 2903 & .SXT4.. & 4.0 & 9.1 & -20.2 & 556 & 2.10 & 0.03 & 0.32 & C \\ 
NGC 2985 & PSAT2.. & 2.0 & 11.0 & -20.8 & 1322 & 1.66 & 0.07 & 0.10 & TW \\ 
NGC 3032 & .LXR0.. & -2.0 & 12.8 & -18.9 & 1533 & 1.30 & 0.17 & 0.05 & LW \\ 
NGC 3079 & .SBS5./ & 7.0 & 10.4 & -20.8 & 1114 & 1.90 & 0.04 & 0.74 & HI \\ 
NGC 3081 & RSXR0.. & 0.0 & 12.6 & -20.0 & 2367 & 1.32 & 0.16 & 0.11 & GD \\ 
NGC 3145 & .SBT4.. & 4.0 & 11.8 & -21.5 & 3652 & 1.49 & 0.11 & 0.29 & CS \\ 
NGC 3227 & .SXS1P. & 1.0 & 11.2 & -20.4 & 1152 & 1.73 & 0.06 & 0.17 & C \\ 
NGC 3300 & .LXRO*\$ & -2.0 & 13.0 & -20.2 & 3075 & 1.28 & 0.17 & 0.28 & N \\ 
NGC 3351 & .SBR3.. & 3.0 & 10.3 & -19.3 & 778 & 1.87 & 0.04 & 0.17 & C \\ 
NGC 3362 & .SX.5.. & 5.0 & 13.2 & -22.0 & 8318 & 1.15 & 0.24 & 0.11 & N \\ 
NGC 3368 & .SXT2.. & 2.0 & 9.8 & -20.3 & 897 & 1.88 & 0.04 & 0.16 & CS \\ 
NGC 3393 & PSBT1*. & 1.0 & 12.6 & -21.8 & 5750 & 1.34 & 0.15 & 0.04 & CS \\ 
NGC 3458 & .LX..*. & -2.0 & 13.2 & -19.2 & 1818 & 1.14 & 0.24 & 0.20 & N \\ 
NGC 3486 & .SXR5.. & 5.0 & 10.8 & -19.7 & 682 & 1.85 & 0.05 & 0.13 & LW \\ 
NGC 3516 & RLBS0*. & -2.0 & 12.1 & -20.8 & 2540 & 1.24 & 0.19 & 0.11 & LW \\ 
NGC 3627 & .SXS3.. & 3.0 & 9.1 & -20.6 & 727 & 1.96 & 0.04 & 0.34 & C \\ 
NGC 3718 & .SBS1P. & 1.0 & 11.2 & -20.0 & 994 & 1.91 & 0.04 & 0.31 & C \\ 
NGC 3786 & .SXT1P. & 1.0 & 13.0 & -19.8 & 2737 & 1.34 & 0.15 & 0.23 & LW \\ 
NGC 3865 & .SXT3P* & 3.0 & 12.6 & -21.8 & 5702 & 1.31 & 0.16 & 0.14 & CS \\ 
NGC 3982 & .SXR3*. & 3.0 & 11.7 & -19.5 & 1188 & 1.37 & 0.14 & 0.06 & GD \\ 
NGC 4030 & .SAS4.. & 4.0 & 11.2 & -20.9 & 1460 & 1.62 & 0.08 & 0.14 & TW \\ 
NGC 4117 & .L..0* & -2.3 & 14.0 & -16.9 & 943 & 1.25 & 0.19 & 0.31 & LW \\ 
NGC 4143 & .LXS0.. & -2.0 & 11.9 & -19.1 & 985 & 1.36 & 0.15 & 0.20 & LW \\ 
NGC 4151 & PSXT2*. & 2.0 & 10.7 & -20.8 & 970 & 1.80 & 0.05 & 0.15 & LW \\ 
NGC 4253 & PSBS1*. & 1.0 & 13.8 & -19.7 & 3836 & 0.98 & 0.35 & 0.06 & GD \\ 
NGC 4254 & .SAS5.. & 5.0 & 10.1 & -21.0 & 2407 & 1.73 & 0.06 & 0.06 & CS \\ 
NGC 4258 & .SXS4.. & 4.0 & 8.5 & -20.8 & 449 & 2.27 & 0.02 & 0.41 & HI \\ 
NGC 4260 & .SBS1.. & 1.0 & 12.3 & -20.4 & 1958 & 1.43 & 0.12 & 0.30 & C \\ 
NGC 4303 & .SXT4.. & 4.0 & 10.1 & -20.8 & 1585 & 1.81 & 0.05 & 0.05 & GD \\ 
NGC 4314 & .SBT1.. & 1.0 & 11.2 & -18.8 & 963 & 1.62 & 0.08 & 0.05 & LW \\ 
NGC 4380 & .SAT3*\$ & 3.0 & 12.1 & -19.2 & 970 & 1.54 & 0.10 & 0.26 & CS \\ 
NGC 4388 & .SAS3*/ & 3.0 & 10.8 & -20.3 & 2535 & 1.75 & 0.06 & 0.64 & HI/C \\ 
NGC 4395 & .SAS9*. & 9.0 & 10.6 & -17.2 & 318 & 2.12 & 0.03 & 0.08 & C \\ 
NGC 4569 & .SXT2.. & 2.0 & 9.8 & -21.3 & -245 & 1.98 & 0.03 & 0.34 & C \\ 
NGC 4593 & RSBT3.. & 3.0 & 11.4 & -21.6 & 2492 & 1.59 & 0.09 & 0.13 & TW \\ 
NGC 4725 & .SXR2P. & 2.0 & 9.8 & -20.8 & 1206 & 2.03 & 0.03 & 0.15 & C \\ 
NGC 4939 & .SAS4.. & 4.0 & 11.2 & -22.0 & 3111 & 1.74 & 0.06 & 0.29 & C \\ 
NGC 4941 & RSXR2*. & 2.0 & 11.7 & -18.3 & 1111 & 1.56 & 0.09 & 0.27 & LW \\ 
NGC 4968 & PLX.0.. & -2.0 & 13.4 & -19.8 & 2957 & 1.27 & 0.18 & 0.34 & C \\ 
NGC 5005 & .SXT4.. & 4.0 & 10.2 & -21.5 & 1022 & 1.76 & 0.06 & 0.32 & C \\ 
NGC 5033 & .SAS5.. & 5.0 & 10.2 & -21.1 & 892 & 2.03 & 0.03 & 0.33 & C \\ 
NGC 5054 & .SAS4.. & 4.0 & 11.1 & -21.1 & 1741 & 1.71 & 0.06 & 0.24 & LW \\ 
NGC 5064 & PSA.2*. & 2.5 & 11.7 & -21.3 & 3002 & 1.39 & 0.14 & 0.34 & TW \\ 
NGC 5135 & .SBS2.. & 2.0 & 12.4 & -21.3 & 4112 & 1.41 & 0.13 & 0.15 & GD \\ 
NGC 5252 & .L..... & -2.0 & 13.9 & -20.9 & 6926 & 1.14 & 0.24 & 0.21 & CS \\ 
NGC 5256 & .P..... & 99.0 & 0.0 & -35.2 & 8239 & 1.08 & 0.28 & 0.05 & C \\ 
NGC 5273 & .LAS0.. & -2.0 & 12.4 & -18.7 & 1089 & 1.44 & 0.12 & 0.04 & CS \\ 
NGC 5283 & .L...?. & -2.0 & 14.1 & -18.8 & 2697 & 1.03 & 0.31 & 0.04 & CS \\ 
NGC 5347 & PSBT2.. & 2.0 & 13.1 & -19.7 & 2335 & 1.23 & 0.20 & 0.10 & GD \\ 
NGC 5383 & PSBT3*P & 3.0 & 11.9 & -20.9 & 2250 & 1.50 & 0.11 & 0.07 & GD \\ 
NGC 5427 & .SAS5P. & 5.0 & 11.7 & -21.2 & 2618 & 1.45 & 0.12 & 0.07 & LW \\ 
NGC 5506 & .S..1P/ & 1.0 & 12.3 & -20.0 & 1815 & 1.45 & 0.12 & 0.52 & HI/LW \\ 
NGC 5548 & PSAS0.. & 0.0 & 12.8 & -21.3 & 4981 & 1.16 & 0.23 & 0.05 & TW \\ 
NGC 5614 & .SAR2P. & 2.0 & 12.4 & -21.2 & 3892 & 1.39 & 0.14 & 0.08 & TW \\ 
NGC 5643 & .SXT5.. & 5.0 & 10.2 & -20.9 & 1199 & 1.66 & 0.07 & 0.06 & GD \\ 
NGC 5674 & .SX.5.. & 5.0 & 13.6 & -21.4 & 7442 & 1.04 & 0.30 & 0.02 & CS \\ 
NGC 5691 & .SXS1*P & 1.0 & 12.5 & -19.9 & 1870 & 1.27 & 0.18 & 0.12 & C \\ 
NGC 5695 & .S?.... & 3.0 & 13.4 & -20.4 & 4209 & 1.19 & 0.22 & 0.15 & LW \\ 
NGC 5929 & .S..2*P & 2.0 & 14.1 & -18.8 & 2504 & 0.98 & 0.35 & 0.03 & CS \\ 
NGC 5953 & .SA.1*P & 1.0 & 13.3 & -19.3 & 1965 & 1.21 & 0.21 & 0.08 & TW \\ 
NGC 5970 & .SBR5.. & 5.0 & 11.8 & -20.7 & 1963 & 1.46 & 0.12 & 0.17 & C \\ 
NGC 6221 & .SBS5.. & 5.0 & 9.8 & -21.7 & 1482 & 1.55 & 0.09 & 0.16 & CS \\ 
NGC 6300 & .SBT3.. & 3.0 & 10.2 & -20.6 & 1110 & 1.65 & 0.07 & 0.18 & C \\ 
NGC 6384 & .SXR4.. & 4.0 & 10.6 & -20.9 & 1663 & 1.79 & 0.18 & 0.05 & C \\ 
NGC 6412 & .SAS5.. & 5.0 & 12.1 & -19.8 & 1324 & 1.40 & 0.13 & 0.06 & C \\ 
NGC 6744 & .SXR4.. & 4.0 & 8.8 & -21.2 & 841 & 2.30 & 0.02 & 0.19 & C \\ 
NGC 6814 & .SXT4.. & 4.0 & 11.3 & -20.5 & 1563 & 1.48 & 0.11 & 0.03 & GD \\ 
NGC 6890 & .SAT3.. & 3.0 & 12.8 & -19.7 & 2419 & 1.19 & 0.22 & 0.10 & GD \\ 
NGC 6946 & .SXT6.. & 6.0 & 7.8 & -20.9 & 52 & 2.06 & 0.03 & 0.07 & C \\ 
NGC 6951 & .SXT4.. & 4.0 & 10.7 & -21.2 & 1426 & 1.59 & 0.09 & 0.08 & LW \\ 
NGC 7096 & .SAS1.. & 1.0 & 12.6 & -20.2 & 3100 & 1.27 & 0.18 & 0.06 & N \\ 
NGC 7126 & .SAT5.. & 5.0 & 12.4 & -20.5 & 3054 & 1.45 & 0.12 & 0.34 & TW \\ 
NGC 7130 & .S..1P. & 1.0 & 12.9 & -21.2 & 4842 & 1.18 & 0.22 & 0.04 & GD \\ 
NGC 7177 & .SXR3.. & 3.0 & 11.5 & -19.8 & 1150 & 1.49 & 0.11 & 0.19 & CS \\ 
NGC 7392 & .SAS4.. & 4.0 & 12.2 & -20.8 & 3128 & 1.33 & 0.16 & 0.23 & LW \\ 
NGC 7469 & PSXT1.. & 1.0 & 12.6 & -21.5 & 4790 & 1.17 & 0.23 & 0.14 & TW \\ 
NGC 7496 & .SBS3.. & 3.0 & 11.8 & -19.7 & 1649 & 1.52 & 0.10 & 0.04 & C \\ 
NGC 7582 & PSBS2.. & 2.0 & 10.8 & -20.4 & 1575 & 1.70 & 0.07 & 0.38 & HI/C \\ 
NGC 7674 & .SAR4P. & 4.0 & 13.6 & -21.8 & 8662 & 1.05 & 0.30 & 0.04 & GD \\ 
NGC 7682 & .SBR2.. & 2.0 & 13.7 & -20.6 & 5107 & 1.09 & 0.27 & 0.05 & LW \\ 
NGC 7716 & .SXR3*. & 3.0 & 12.5 & -20.1 & 2571 & 1.33 & 0.16 & 0.08 & CS \\ 
NGC 7743 & RLBS+.. & -1.0 & 12.2 & -19.4 & 1658 & 1.48 & 0.11 & 0.07 & LW \\ 
NGC 7814 & .SAS2*/ & 2.0 & 11.0 & -19.6 & 1054 & 1.74 & 0.06 & 0.38 & HI \\ 
UGC 12138 & .SB.1.. & 1.0 & 13.8 & -21.3 & 7487 & 0.92 & 0.40 & 0.08 & GD \\ 
UGC 6100 & .S..1?. & 1.0 & 14.0 & -21.3 & 8778 & 0.92 & 0.40 & 0.18 & LW \\ 
UM 146 & .SAT3.. & 3.0 & 13.9 & -20.4 & 5208 & 1.10 & 0.26 & 0.10 & LW \\ 
\enddata
\tablecomments{
Morphological data, including our nuclear classifications (col.~10), for all 
123 galaxies in the sample. Most of these data were obtained from the RC3 
catalog, including the morphological classification (col.~2), $T$ type 
(col.~3), total $B$ magnitude (col.~4), diameter $D_{25}$ (col.~7), and 
axis ratio $R_{25}$ (col.~9). The absolute $B$ magnitude (col.~5) was derived 
based on the distance given in Table~\ref{tbl:morph}, the heliocentric 
velocity $v$ (col.~6) is from either the RC3 or NED, and $20''$ frac 
(col.~8) is the fraction of the galaxy diameter $D_{25}$ contained within 
the $20''$ field of view shown in Figures~\ref{fig:cmaps} and 
\ref{fig:convtest}. 
The classification codes in column~10 are: GD: grand design nuclear spiral; 
TW: tightly wound spiral; LW: loosely wound spiral; CS: chaotic spiral; 
C: chaotic circumnuclear dust; N: no circumnuclear dust structure. 
These classes are defined in \S\ref{sec:morph}. Galaxies 
with axis ratios $R_{25} > 0.35$ are classified as HI (high inclination), 
along with a nuclear classification if possible. Galaxies with 
$R_{25} \geq 0.30$ are still too highly inclined for 
accurate classifications and are not included in the upper left panels of 
Figures~\ref{fig:agnhist} and \ref{fig:barhist}. 
}
\end{deluxetable}
\end{center}

\clearpage

\begin{center}
\begin{deluxetable}{llcccl}
\tabletypesize{\scriptsize}
\tablecolumns{6}
\tablenum{4}
\tablewidth{5.0truein}
\tablecaption{Ring Properties \label{tbl:ring} }
\tablehead{
\colhead{Name}   &
\colhead{RC3}   &
\colhead{$R$ [$''$]}   &
\colhead{$R$ [pc]}   &
\colhead{$D(ring)/D_{25}$}   &
\colhead{Note}  \\
}
\startdata
ESO 138-G1&.E?....	& 6.2	& 1100	& 0.20  & smooth in $V$ and $H$ \\
Mrk 477	& .S...*P	& 2.8	& 2100	& 0.22 	& smooth in $V$ and $H$ \\
NGC 864	& .SXT5..	& 0.6	&   60	& 0.004	& very small, starburst \\
NGC 1300 & .SBT4..	& 6.8 	&  500	& 0.037	& smooth in $V$ and $H$, dusty\\
NGC 1667 & .SXR5..	& 7.3	& 2100 	& 0.14 	& stars, incl. young clusters \\
NGC 1672 & .SBS3..	& 5.1	&  360	& 0.026	& stars, incl. young clusters \\
NGC 2273 & .SBR1*.	& 2.3	&  310	& 0.012	& stars, incl. young clusters \\
NGC 3081 & RSXR0..	& 4.5	&  700	& 0.072	& smooth in $V$ and $H$, dusty\\
NGC 3351 & .SBR3..	& 5.6	&  220	& 0.025	& stars, incl. young clusters \\
NGC 4314 & .SBT1..	& 5.6   &  260	& 0.045 & stars, incl. young clusters \\
NGC 5427 & .SAS5P.	& 5.6 	& 1000	& 0.066	& stars, incl. young clusters \\
NGC 6890 & .SAT3..	& 7.9	& 1200	& 0.17	& stars, incl. young clusters \\
NGC 6951 & .SXT4..	& 3.9	&  460	& 0.033 & stars, incl. young clusters \\
NGC 7469 & PSXT1..	& 1.4	&  450	& 0.032 & very bright SB ring\\
\enddata
\tablecomments{
Properties of the rings shown in Figure~\ref{fig:cmaps}. Columns~1 and 2 
list the name and morphology for each galaxy from Table~\ref{tbl:morph}. The 
angular (col.~3) and physical (col.~4) radius is listed next, followed by 
the ratio of the ring size to the galaxy diameter (col.~5). A note on the 
appearance of each ring is listed in the last column. 
}
\end{deluxetable}
\end{center}

\end{document}